\documentclass[acmsmall]{acmart}
\AtBeginDocument{%
  \providecommand\BibTeX{{%
    \normalfont B\kern-0.5em{\scshape i\kern-0.25em b}\kern-0.8em\TeX}}}

\usepackage{todonotes}
\usepackage{subfigure}
\usepackage{bbding}

\begin{document}

\title{The Serverless Computing Survey: A Technical Primer for Design Architecture}

\author{Zijun Li}
\email{lzjzx1122@sjtu.edu.cn}

\author{Linsong Guo}
\email{gls1196@sjtu.edu.cn}

\author{Jiagan Cheng}
\email{chengjiagan@sjtu.edu.cn}

\author{Quan Chen}
\email{chen-quan@cs.sjtu.edu.cn}
\affiliation{%
  \institution{Department of Computer Science and Engineering, Shanghai Jiao Tong University}
  %\streetaddress{P.O. Box 1212}
  \country{China}
}

\author{BingSheng He}
\email{hebs@comp.nus.edu.sg}
\affiliation{%
  \institution{National University of Singapore, Department of Computer Science}
  %\streetaddress{P.O. Box 1212}
  \country{Singapore}
}

\author{Minyi Guo}
\email{guo-my@cs.sjtu.edu.cn}
\affiliation{%
  \institution{Department of Computer Science and Engineering, Shanghai Jiao Tong University}
  %\streetaddress{P.O. Box 1212}
  \country{China}
}

% \affiliation{%

% \affiliation{%

% \affiliation{%

% \affiliation{%

% \affiliation{%



\renewcommand{\shortauthors}{Li, Guo, Cheng and Quan, et al.}

\begin{abstract}
  The development of cloud infrastructures inspires the emergence of cloud-native computing. As the most promising architecture for deploying microservices, serverless computing has recently attracted more and more attention in both industry and academia. Due to its inherent scalability and flexibility, serverless computing becomes attractive and more pervasive for ever-growing Internet services. 
  Despite the momentum in the cloud-native community, the existing challenges and compromises still wait for more advanced research and solutions to further explore the potentials of the serverless computing model.
  As a contribution to this knowledge, this article surveys and elaborates the research domains in the serverless context by decoupling the architecture into four stack layers: Virtualization, Encapsule, System Orchestration, and System Coordination. 
  We highlight the key implications and limitations of these works in each layer, and make suggestions for potential challenges to the field of future serverless computing.

  {\bf Note: This paper has been accepted by ACM Computing Surveys (CSUR), and the current e-print version is our major revision. For a complete view, please visit ACM CSUR.}

  %many aspects of serverless computing are still unclear and ignored, leading to misunderstandings for research. 
\end{abstract}

  


\maketitle

\section{Introduction}
\subsection{Definition of Serverless Computing}
Traditional Infrastructure-as-a-Service (IaaS) deployment mode demands a long-term running server for sustainable service delivery. However, this exclusive allocation needs to retain resources regardless of whether the user application is running or not. Consequently, it results in low resource utilization in current data centers by only about 10\% on average, especially for an online service with a diurnal pattern. The contradiction attracts the development of a platform-managed on-demand service model to attain higher resource utilization and lower cloud computing costs. To this end, serverless computing was put forward, and most large cloud vendors such as Amazon, Google, Microsoft, IBM, and Alibaba have already offered such elastic computing services.

In the following, we will first review the definition given in Berkeley View~\cite{DBLP:journals/corr/abs-1902-03383}, and then we will give a broader definition. We believe that a narrow perception of the FaaS-based serverless model may weaken its advancement. So far, there is no formal definition of serverless computing. The common acknowledged definitions from Berkeley View~\cite{DBLP:journals/corr/abs-1902-03383} are presented as follows:
\begin{itemize}
  \item $Serverless\ Computing= FaaS (Function$-$as$-$a$-$Service)$ + $BaaS (Backend$-$as$-$a$-$Service)$. One fallacy is that `Serverless' is interchangeable with `FaaS', 
  which is revealed in a recent interview~\cite{DBLP:journals/jss/LeitnerWSH19}. 
  To be precise, they both are essential to serverless computing. $FaaS$ model enables the function isolation and invocation, while $BaaS$ provides overall backend support for online services.
  \item In the $FaaS$ model (aka Lambda paradigm), an application is sliced into functions or function-level microservices~\cite{DBLP:journals/corr/abs-1902-03383,DBLP:journals/csur/BuyyaSCCSVGJVNT19,DBLP:conf/micro/ShahradBW19,DBLP:journals/tpds/HoseinyFarahabady18,DBLP:journals/internet/EykTTVUI18,DBLP:conf/usenix/WangLZRS18}. The function identifier, the language runtime, the memory limit of one instance, and the function code blob URI (Uniform Resource Identifier) together define the existence of a 
  function~\cite{DBLP:conf/icdcsw/McGrathB17}. 
  \item The $BaaS$ covers a wide range of services that any application relies on can be categorized into it. For example, the cloud
  storage (Amazon S3 and DynamoDB), the message bus system for passing (Google cloud pub/sub), the message notification service (Amazon SNS), and DevOps tools (Microsoft Azure DevOps).
\end{itemize}

\begin{figure}
  \centering
  \includegraphics[width=.8\linewidth]{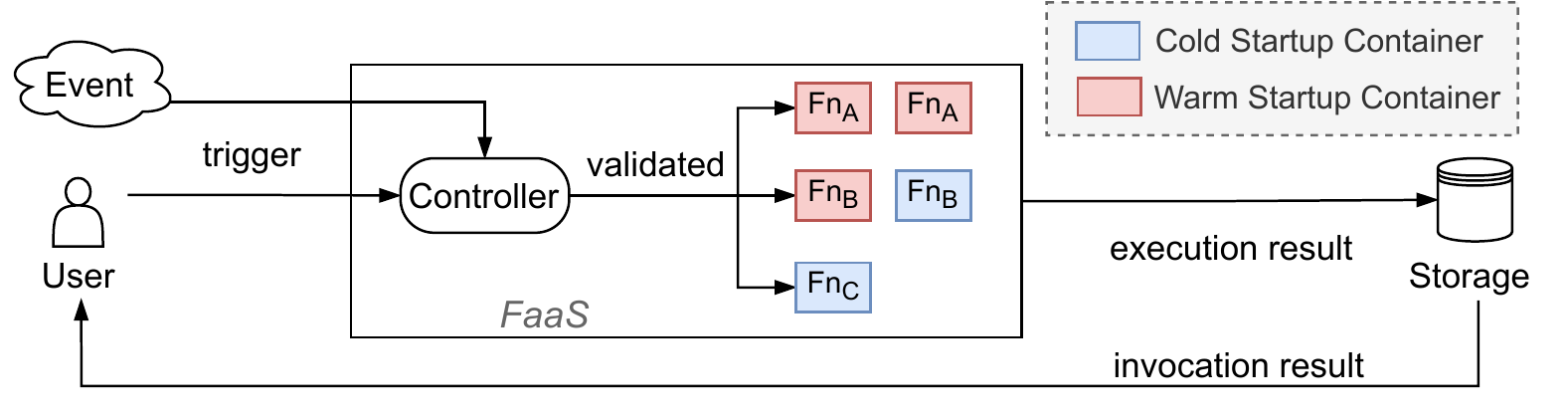}
  \vspace{-2mm}
  \caption{\label{fig:create_invoke}The example of an asynchronous invocation in serverless computing.}
\end{figure}

To depict the serverless computing model, we take the asynchronous invocation in Figure~\ref{fig:create_invoke} as an example. 
The serverless system receives triggered API queries from the users, validates them, and invokes the functions by creating new sandboxes (aka the \textbf{cold startup}~\cite{DBLP:journals/corr/abs-1902-03383,baldini2017serverless,DBLP:conf/eurosys/CaddenUADKA20}) or reusing running warm ones (aka the \textbf{warm startup}). The isolation ensures that each function invocation runs in an individual container or a virtual machine assigned from an access-control controller. Due to the event-driven and single-event processing nature, the serverless system can be triggered to provide on-demand isolated instances and scale them horizontally according to the actual application workload. Afterwards, each execution worker accesses a backend database to save execution results~\cite{DBLP:conf/usenix/BoucherKAK18}. By further configuring triggers and bridging interactions, users can customize the execution for complex applications (e.g., building internal event calls in a \{$Fn_A,Fn_B,Fn_C$\} pipeline). 

%Each function invocation runs in an individual runtime instance and is executed by calling through an access-control controller. By further configuring the environment settings and the triggers, users can achieve customized executing (e.g., external HTTP requests or internal event calls) for complex applications. 
%The serverless system receives triggered API queries from the users, validates them, and invokes the functions by creating new isolated instances (aka the cold statrup) or reusing running warm containers or virtual machines (aka the warm startup). Finally, it accesses a backend database to save execution results~\cite{DBLP:conf/usenix/BoucherKAK18}. Due to the event-driven and single-event processing nature, the serverless system can be triggered to provide on-demand isolated instances and scale them horizontally according to the actual application workload.

In the broader scenario, we think that the serverless computing model should be identified with the following features:
%Overall, the undeterminate formal definition and narrow perception of the FaaS-based serverless model may lack extensibility. In the broader scenario, serverless computing should be identified from the following features:
\begin{itemize}
  \item {\bf Auto-scaling.} Auto-scalability should not be only narrowed to the FaaS model (e.g., container black boxs as scheduling units in OpenWhisk~\cite{OpenWhiskruntime}). The indispensable factor in identifying a serverless system is performing horizontal and vertical scaling when accommodating workload dynamics. 
  %The most important factor to identify a serverless system is the ability to scale the instances to zero. 
  Allowing an application to scale the number of instances to zero also introduces a worrisome challenge - cold startup. When a function experiences the cold startup, instances need to start from scratch, initialize the software environment, and load application-specific code. These steps can significantly drag down the service response, leading to QoS (Quality-of-Service) violations.
  
  \item {\bf Flexible scheduling.} Since the application is no longer bound to a specific server, the serverless controller dynamically schedules applications according to the resource usage in the cluster, while ensuring load balancing and performance assurances. 
  %to guarantee the SLA (service-level agreements) of services and make the load balanced. 
  Moreover, the serverless platform also takes the multi-region collaboration into account~\cite{DBLP:conf/IEEEcloud/ZhengP19}. For a  more robust and available serverless system, flexible scheduling allows the workload queries to be distributed across a broader range of regions~\cite{DBLP:journals/corr/abs-1810-09679}. It avoids serious performance degradation or damage to the service continuity in case of unavailable or crash nodes.
  
  \item {\bf Event-driven.} The serverless application is triggered by events, such as the arrival of RESTful HTTP queries, the update of a message queue, or new data to a storage service. 
  By binding events to functions with triggers and rules, the controller and functions can use metadata encapsulated in context attributes. It makes relationships between events and the system detectable, enabling different collaboration responses to different events. 
  Cloud-Native Computing Foundation (CNCF) serverless group also published CloudEvents specifications for commonly describing event metadata to provide interoperability.
  
  \item {\bf Transparent development.} On the one hand, managing underlying host resources will no longer be a bother for application maintainers. It is because they are agnostic about the execution environment. Simultaneously, cloud vendors should ensure available physical nodes, isolated sandboxes, software runtimes, and computing power while making them transparent to maintainers. On the other hand, serverless computing should also integrate DevOps tools to help deploy and iterate more efficiently.
  
  \item {\bf Pay-as-you-go.} The serverless billing model shifts the cost of computing power from a capital expense to an operating expense. This model eliminates the requirement from users to buy exclusive servers based on the peak load. By sharing network, disk, CPU, memory, and other resources, the pay-as-you-go model only indicates the resources that applications actually used~\cite{DBLP:conf/sigsoft/AdzicC17,DBLP:journals/csur/BuyyaSCCSVGJVNT19,DBLP:journals/csur/AdhikariAS19}, no matter whether the instances are running or idle. 
\end{itemize}

We regard an elastic computing model with the above five features incorporated as the key to the definition of serverless computing. Along with the serverless emergence, application maintainers would find it more attractive that resource pricing is billed based on the actual processing events of an application rather than the pre-assigned resources~\cite{DBLP:conf/sigsoft/AdzicC17}.
Nowadays the serverless computing is commonly applied in backend scenarios for batch jobs, including data analytics (e.g., distributed computing model in PyWren~\cite{DBLP:conf/cloud/JonasPVSR17}), machine learning tasks (e.g., deep learning)~\cite{DBLP:journals/corr/abs-2004-03276,DBLP:journals/jss/LeitnerWSH19}, and event-driven web applications. 
%It makes the serverless computing model more attractive for application maintainers that resource pricing is billed based on the actual processing events of an application rather than the IaaS-based instances with pre-assigned resources~\cite{DBLP:conf/sigsoft/AdzicC17}. Nowadays the serverless computing is commonly applied in backend scenarios for batch jobs, including data analytics (e.g., distributed computing model in PyWren~\cite{DBLP:conf/cloud/JonasPVSR17}), machine learning tasks (e.g., deep learning)~\cite{DBLP:journals/corr/abs-2004-03276,DBLP:journals/jss/LeitnerWSH19}, and event-driven web applications. 

%However, the lack of general technical implementations cannot mentor inexperienced or context-specialized researchers to further dig into the nature and application of serverless computing. So it inspires us to propose a layered design for the general implementation of the serverless architecture.

%Although we have presented five features that define serverless computing, the abstract definitions cannot help researchers achieve a good understanding of serverless architecture. The lack of general technical implementations inspires us to further dig into the nature of serverless computing.

\subsection{Survey Method by the Layered Serverless Architecture}
\begin{figure}
  \centering
  \includegraphics[width=.87\linewidth]{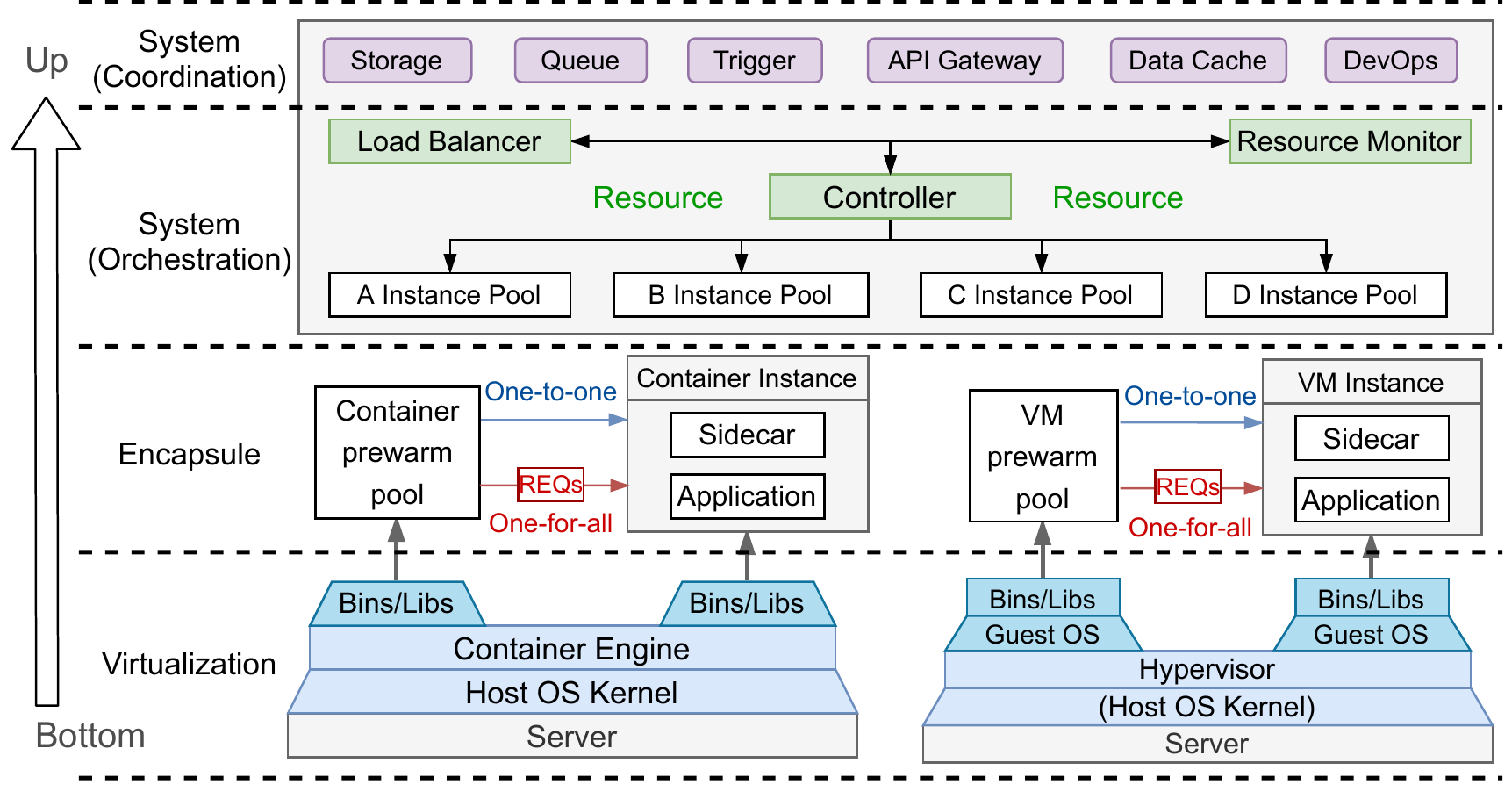}
  \vspace{-2mm}
  \caption{\label{fig:architecture}General implementation of the serverless architecture.}
\end{figure}

Several surveys in serverless computing have discussed the characteristics of serverless generalization~\cite{DBLP:journals/cacm/Schleier-SmithS21,shafiei2021serverless,9183650,baldini2017serverless,DBLP:journals/corr/abs-1902-03383,DBLP:journals/jcloudc/HassanBS21}. However, they only propose literature reviews from a high-level perspective while ignoring to provide enough architecture implications. As a result, researchers and serverless vendors may find it struggling to grasp and comprehend each issue in the real serverless architecture. In the lack of systematic knowledge, 
challenges and proposed solutions will lack high portability and compatibility for various serverless systems. To this end, this survey is inspired to propose a layered design and summarize the research domains from different views. It can help researchers and practitioners to further understand the nature of serverless computing.
As shown in Figure~\ref{fig:architecture}, we analyze its design architecture with a bottom-up logic and decouple the serverless computing architecture into four stack layers: Virtualization, Encapsule, System Orchestration, and System Coordination.

\textbf{Virtualization layer.} The Virtualization layer enables function isolation within a performance and functionality secured sandbox. The sandbox serves as the runtime for application service code, runtime environment, dependencies, and system libraries. To prevent access to resources in the multi-application or multi-tenant scenarios, cloud vendors usually adopt containers/virtual machines to achieve isolation. Currently, the popular sandbox technologies are Docker~\cite{Docker}, gVisor~\cite{gvisor}, Kata~\cite{kata}, Firecracker~\cite{DBLP:conf/nsdi/AgacheBILNPP20}, and Unikernel~\cite{DBLP:conf/asplos/MadhavapeddyMRSSGSHC13}. Section~\ref{sec:Virtualization} introduces these solutions to isolate functions and analyze their pros and cons.
  
\textbf{Encapsule layer.} Various middlewares in the Encapsule layer enable customized function triggers and executions, and they also provide data metrics collection for communicating and monitoring. We call all these additional middlewares the \emph{sidecar}. It separates the service's business logic and enables loose coupling between the functions and the underlying platform. 
Meanwhile, to speed up instance startup and initialization, the prewarm pool is commonly used in the Encapsule layer~\cite{azureprewarm,fissionprewarm,OpenWhiskprewarm,DBLP:conf/hotcloud/MohanSDENS19}. In addition, serverless systems may use prediction by analyzing the load pattern to prewarm each by one-to-one approach~\cite{DBLP:conf/icpads/XuZGWM19,DBLP:conf/usenix/ShahradFGCBCLTR20}, or build a template for all functions to dynamically install requirements (REQs) according to the runtime characteristics by a one-for-all approach. Those concepts are introduced and compared in Section~\ref{sec:Encapsule}.
  
\textbf{System Orchestration layer.} The System Orchestration layer allows users to configure triggers and bind rules, ensuring the high availability and stability of the user application by dynamically adjusting as load changes. 
Through the cloud orchestrator, the combination of online and offline scheduling can avoid resource contention, recycle idle resources and ease the performance degradation for co-located functions. The above implementations are also typically integrated into container orchestration services (e.g., Google Kubernetes and Docker Swarm). While in the serverless system, \textbf{resource monitor}, \textbf{controller}, and \textbf{load balancer} are consolidated to resolve scheduling challenges~\cite{DBLP:journals/tpds/HoseinyFarahabady18, DBLP:journals/icl/GuanWCSZ17, DBLP:conf/infocom/WangNL19,DBLP:journals/tpds/KimFLZ20,DBLP:conf/usenix/AkkusCRSSBAH18,DBLP:conf/globecom/ChangYYLJ17,DBLP:conf/cloud/KaffesYK19, DBLP:conf/cascon/MahmoudiLKL19}. They enable the serverless system to achieve scheduling optimizations in three different levels: resource-level, instance-level, and application-level, respectively. Section~\ref{sec:Orchestration} detailly analyzes the scheduling methodology from these three angles.
  
\textbf{System Coordination layer.} The System Coordination layer consists of a series of Backend-as-a-Service (BaaS) components that use unified APIs and SDKs to integrate backend services into functions. Distinctly, it differs from the traditional middlewares that use local physical services outside the cloud. These BaaS services provide the storage, queue service~\cite{DBLP:journals/jsa/NettoLCLS17, DBLP:conf/icdcsw/McGrathB17}, trigger binding~\cite{DBLP:conf/IEEEcloud/LeeSF18,k8s:cronjob}, API gateway, data cache~\cite{DAX,AWScache}, DevOps tools~\cite{istio,IOpipe,sonarqube,jenkins}, and other customized components for better meeting the System Orchestration layer's flexibility requirements. Section~\ref{sec:Coordination} discusses these essential BaaS components in a serverless system.

Each stack layer plays an essential role in the serverless architecture. Therefore, based on the above hierarchy, we conclude the contributions of this survey as follows:

\begin{enumerate}
  \item Introduce the serverless definition and summarize the features.
  \item Elaborate the architecture design based on a four-layer hierarchy, and review the significant and representative works in each layer.
  \item Analyze the current serverless performance and its limitations.
  \item Explore the challenges, limitations, and opportunities in serverless computing.
\end{enumerate}

The rest of the survey is organized as follows: Sections~\ref{sec:Virtualization}-\ref{sec:Coordination} introduce the four stack layers and elaborate current research domains in serverless computing. Section~\ref{sec:performance} analyzes several factors that degrade performance, and compares the current production serverless systems. Finally, the challenges, limitations, and opportunities of serverless computing are given in Section~\ref{sec:challenges}-\ref{sec:opportunities}. We conclude this paper in Section~\ref{sec:conclusion}.

%\section{State-of-the-art in Serverless Computing}
%\label{sec:2}
%In this section, we will elaborate on the distinct level of functionality in Virtualization, Encapsule, System Orchestration, and System Coordination layer step by step, and make suggestions for serverless architecture design in each layer.

\section{Virtualization Layer}
\label{sec:Virtualization}
Whenever a user function is invoked in serverless computing, it will be loaded and executed within a virtualized sandbox. 
A function can either reuse a warm sandbox or create a new one, but usually not co-run with different user functions. In this premise, most of the concerns in virtualization are isolation, flexibility, and low startup latency.
The isolation ensures that each application process runs in the demarcated resource space, and the running process can avoid interference by others. The flexibility is demonstrated by the ability of testing and debugging, and the additional supports for extending over the system. Low startup latency requires a fast response for the sandbox creation and initialization.
%The current research on the Virtualization layer mainly focuses on the above three features, and there are four representative sandboxing mechanisms: traditional VM (Virtual Machine), container, secure container and Unikernel. 
The current sandboxing mechanism on the Virtualization layer is broken into four representative categories: traditional VM (Virtual Machine), container, secure container, and Unikernel. Table~\ref{tab:virtualization} compares these mainstream approaches in several respects.
%We show them by detailing their representative works in Table~\ref{tab:virtualization}.

In the table, ``Startup latency'' represents the response latency of cold startup. ``Isolation level'' indicates the capacity of functions running without interference by others. ``OSkernel'' shows whether the kernel in GuestOS is shared. ``Hotplug'' allows the function instance to start with minimal resources (CPU, memory, virtio blocks) and add additional resources at runtime. ``OCI supported'' means whether it provides the Open Container Initiative (OCI), an open governance structure for expressing container formats and runtimes. Moreover, ``\Checkmark'' in all tables of this survey means this technique or strategy is used, and vice versa.

\begin{table*}
  \scriptsize
  \renewcommand\arraystretch{1.4}
  \caption{Techniques in the Virtualization layer.}
  \begin{tabular}{l|ccccccc}
    \toprule
    Virtualization & Startup latency & Isolation level & OSkernel & Hotplug & Hypervisor & OCI supported & Backed by\\ 
    \midrule
    Traditional VM & >1000ms & Strong  & unsharing &  & \Checkmark & \Checkmark & / \\
    Docker~\cite{Docker} & 50ms-500ms & Weak  & host-sharing & \Checkmark &  & \Checkmark & Docker \\
    SOCK~\cite{DBLP:conf/usenix/OakesYZHHAA18} & 10ms-50ms & Weak  & host-sharing & \Checkmark &  & \Checkmark & / \\
    Hyper-V~\cite{hyper} & >1000ms & Strong  & unsharing & \Checkmark & \Checkmark & \Checkmark & Microstft \\
    gVisor~\cite{gvisor} & 50ms-500ms & Strong  & unsharing &  & \Checkmark & \Checkmark & Google \\
    %Kata~\cite{kata} & Fast & Strong  & unsharing & \Checkmark & KVM/QEMU & \Checkmark & OpenStack \\
    Kata~\cite{kata} & 50ms-500ms & Strong  & unsharing & \Checkmark & \Checkmark & \Checkmark & OpenStack \\
    %FireCracker~\cite{Firecracker, DBLP:conf/nsdi/AgacheBILNPP20} & Fast & Strong  & unsharing &  & KVM & \Checkmark & Amazon \\
    FireCracker~\cite{DBLP:conf/nsdi/AgacheBILNPP20} & 50ms-500ms & Strong  & unsharing &  & \Checkmark & \Checkmark & Amazon \\
    %Hyper-V~\cite{hyper,hyperfaas} & Slow & Strong  & unsharing & \Checkmark & KVM/Xen & \Checkmark & Microstft \\
    Unikernel~\cite{DBLP:conf/asplos/MadhavapeddyMRSSGSHC13} & 10ms-50ms & Strong & Built-in &  & \Checkmark &  & Docker \\
    \bottomrule
  \end{tabular}
  \label{tab:virtualization}
\end{table*}

%The traditional VM-based isolation adopts a VMM (virtual machine manager, e.g., hypervisor) mediate access to all shared resources by provided interfaces. VMM provides virtualization capabilities to GuestOS and manages all physical resources (or using Qemu/KVM).
%Each VM instance flexible to dynamically modify the within the GuestOS. 
The traditional VM-based isolation adopts a VMM (virtual machine manager, e.g., hypervisor) which provides virtualization capabilities to GuestOS. It can also mediate access to all shared resources by provided interfaces (or using Qemu/KVM).
With snapshots, VM shows high flexibility in quick failsafe when patch performing on applications within each VM instance.
Though VM provides a strong isolation mechanism and flexibility, it lacks the benefits of lower startup latency for user applications (usually >1000ms). This tradeoff is fundamental in the serverless computing, where functions are small while the relative overhead of VMM and guest kernel is high.

\textbf{\textit{Container customization: provide high flexibility and performance.}}

Another common function isolation mechanism in serverless computing is using containers. The container engine leverages the Linux kernel to isolate resources, and creating containers as different processes in HostOS~\cite{DBLP:journals/ibmrd/BarlevBKPRS16,DBLP:conf/cns/MattettiSACDF15}. 
Each container shares the HostOS kernel with the read-only attribute, which typically includes binaries and libraries.
The high flexibility is also attached to the container with the UnionFS (Union File System), which enables the combination of the layered container image by read-only and read-write layers.
Essentially, a container achieves the isolation through namespace to enable processes sharing the same system kernel and Linux Cgroups to set resource limits. 
%By container images, an inert, immutable snapshot of a container, 
Without hardware isolation, container-based sandboxing shows lower startup latency than coarse-grained consolidation strategies~\cite{DBLP:journals/tpds/YeWWZSJZ15,DBLP:journals/fgcs/ArmantCBO18} in hypervisor-based VMs. 

The representative container engine is Docker~\cite{Docker}. 
Docker packages software into a standardized RunC container adapted to the environment requirements, including libraries, system tools, code, and runtime. 
Docker container has been widely applied to various serverless systems for its lightweight nature.
%Due to the lightweight nature of the container, 
%it has been widely applied to various serverless systems. 
Some works further optimize the container runtime for better adaption to the application requirements in the serverless system.
SOCK~\cite{DBLP:conf/usenix/OakesYZHHAA18} proposes an integration solution for serverless RunC containers, where
redundant mechanisms in Docker containers are discarded in this lean container. By only constructing a root file system, creating communication channels, and imposing isolation boundaries, SOCK container makes serverless systems running more efficiently in startup latency and throughput. The startup latency of SOCK container is reduced to 10ms-50ms, comparing to docker containers that usually takes 50ms-500ms.
Different from condensing redundance in lean containers, as additional tools (e.g., debuggers, editors, coreutils, shell) enriching the container and increasing the image size, CNTR~\cite{DBLP:conf/usenix/ThalheimBFK18} splits the container image into ``fat'' and ``slim'' parts. A user can independently deploy the ``slim'' image and expand it with additional tools by dynamically attaching the ``fat'' image to it. 
The evaluation of CNTR shows that the proposed mechanism can significantly improve the overall performance and effectively reduce the image size when extensively applied in the data center. 

\textbf{\textit{Secure Container: compromise security with high flexibility and performance.}}

At the same time, security concerns in Virtualization Layer arise for the relatively low isolation level of containers. Any process-based solution must include a relaxation of the security model for its insufficiency for mutually-untrusted functions. It requires containers to prevent code vulnerabilities in the case of shared kernel architecture. Side-channel attacks such as Meltdown~\cite{DBLP:journals/cacm/LippSGPHHMKGYHS20}, Zombieload~\cite{DBLP:conf/ccs/0001LMBS0G19}, Spectre~\cite{DBLP:journals/cacm/KocherHFGGHHLMP20} prompt mitigation approaches toward vulnerabilities, especially for multi-tenants in serverless context. In this case, container isolation should concern with preventing privilege escalation, information, and communication disclosure side channels~\cite{DBLP:conf/nsdi/AgacheBILNPP20}. The state-of-the-art solution to this issue is leveraging Secure Container. For example, Microsoft proposes their Hyper-V Container for Windows~\cite{hyper}. Hyper-V offers enhanced security and broader compatibility. Each instance runs inside a highly optimized MicroVM and does not share its kernel with others on the same host. However, it is still a heavy-weight virtualization that can introduce more than 1000ms of startup latency. In Google gVisor~\cite{gvisor}, the kernel in it acts as a non-privileged process to restrict \textit{syscalls} that called in userspace. However, the overhead introduced during interception and processing \textit{syscalls} in a sandbox is high. As a result, it is not well-suited for applications with heavy \textit{syscalls}. In order to isolate different tenants with affordable overhead, FireCracker~\cite{DBLP:conf/nsdi/AgacheBILNPP20} creates MicroVMs by customizing VMM for cloud-native applications. Each Firecracker sandbox runs in user space and is restricted by Seccomp, Cgroup, and Namespace policies.
With a container engine built-in MicroVMs, Kata~\cite{kata} adopts an agent to communicate with the kata-proxy located on the host through the hypervisor, thus achieve a secure environment in a lightweight manner. 
%Meanwhile, it clones a running Kata VM instance and shares it with other started Kata VMs. 
Both FireCracker and Kata containers can significantly reduce startup latency and memory consumption, and they all need only 50ms-500ms to start a sandbox.
With secure containers, it can provide complete and strong isolation from the HostOS and other tenants, at the cost of the limited flexibility in the condensed MicroVM. Meanwhile, it still results in instances' long startup latency due to the additional application initialization, e.g., JVM or Python interpreter setup.

\textbf{\textit{Specialized Unikernel: enhance flexibility with high security and performance.}} 

Another emerging virtualization technique is called Unikernel~\cite{DBLP:conf/asplos/MadhavapeddyMRSSGSHC13}, which leverages libraryOS, including series of essential dependent libraries
to construct a specialized, single-address-space machine image. Because the Unikernel runs as a built-in GuestOS, the compile-time invariance rules out runtime management, which significantly reduces the applicability and flexibility of Unikernel. However, unnecessary programs or tools such as $ls, cd, tar$ are not contained, so the image size of a Unikernel is smaller (e.g., 2MB by mirage-skeleton~\cite{mirage-skeleton} that compiled from Xen), the startup latency is much less (e.g., startup within 10ms), and the security is more substantial than containers. 
Based on it, LightVM~\cite{DBLP:conf/sosp/MancoLSMKSYRH17} replaces the time-consuming XenStore and implements the split tool stack, 
separating functionality that runs periodically from that which must be carried out, thus improving efficiency and reducing VM startup latency.
From the perspective of the software ecosystem, to solve the challenge that traditional applications are struggling to be transplanted to the Unikernel model~\cite{DBLP:conf/asplos/MadhavapeddyMRSSGSHC13,DBLP:conf/sigcomm/Schmidt17}, Olivier proposes HermitTux~\cite{DBLP:conf/vee/OlivierCLMR19}, a Unikernel model compatible with Linux binary. HermitTux makes the Unikernel model compatible with Linux Application Binary Interface while retaining the benefits of Unikernel.
However, Unikernel is not adaptable for developers once built, making it inherently inflexible for applications, let alone the terrible DevOps environment. Furthermore, in heterogeneous clusters, the heterogeneity of the underlying hardware forces Unikernel to update as drivers change, making it the antithesis of serverless philosophy.
%However, Unikernel is inherently inflexible for user applications, and the DevOps environment is terrible for maintainers, for that it is not modifiable for developers once built. And in heterogeneous clusters, the heterogeneity of the underlying hardware forces Unikernel to update as drivers change.
% Second, to ensure the performance of a hypervisor-based sandbox, 
% Cloud vendors tend to allocate continuous physical address space for mapping during virtualization. Thus hypervisor-based Unikernel will cause more memory fragmentation when hundreds of instances coexist. Finally, 

\begin{figure}
  \centering
  \includegraphics[width=.82\linewidth]{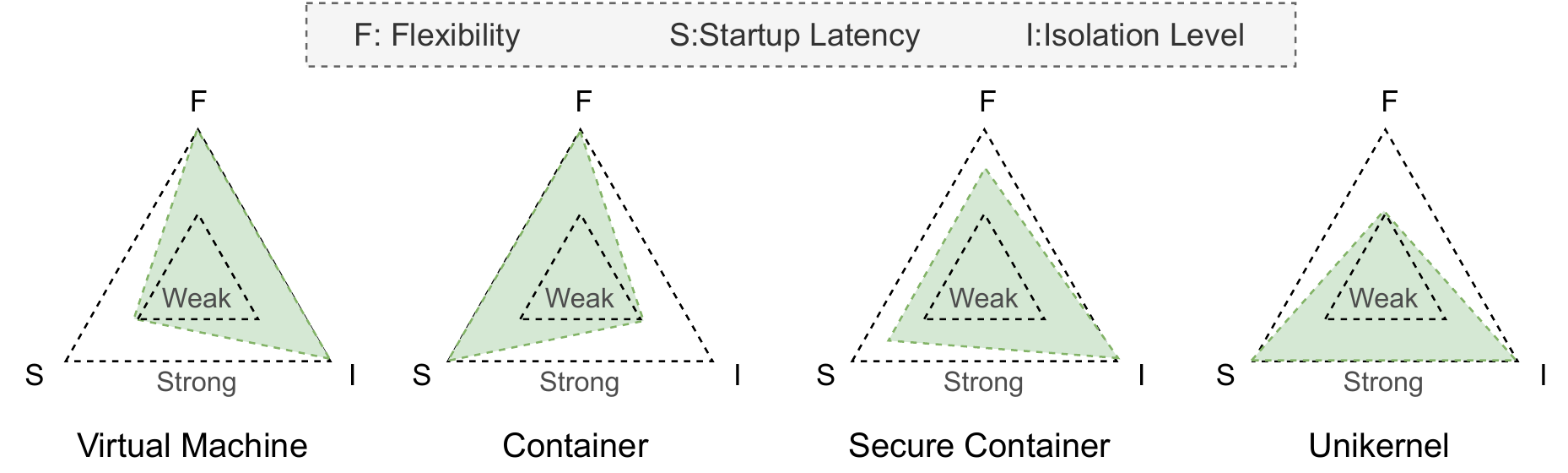}
  \caption{\label{fig:four}The flexibility, startup latency, and isolation level of four virtualization mechanisms.}
\end{figure}

\textbf{\textit{Tradeoffs among Security, performance, and flexibility.}} 

At last, we make the indicatrix diagram of these four technologies in Figure~\ref{fig:four} to show the tradeoffs among security, performance, and flexibility. %and summarize the \textbf{security concerns in Virtualization Layer} into the following aspects. 
% Firstly, when offering high flexibility with modifiable images by VM and container, it is critical to avoid that built ones are signed and originated from an unsafe pedigree, with the solutions~\cite{khan2017key,tak2017understanding} by keeping continuous vulnerability assessment and remediation program. 
%Secondly, relatively low isolation level requires containers to prevent the code vulnerabilities in the case of shared kernel architecture. Side-channel attacks such as Meltdown~\cite{DBLP:journals/cacm/LippSGPHHMKGYHS20}, Zombieload~\cite{DBLP:conf/ccs/0001LMBS0G19}, Spectre~\cite{DBLP:journals/cacm/KocherHFGGHHLMP20} prompt mitigation approaches toward vulnerabilities, especially for multi-tenants in serverless context. In this case, container isolation should concerns with preventing privilege escalation, information and communication disclosure side channels~\cite{DBLP:conf/nsdi/AgacheBILNPP20}.
To conclude, hypervisor-based VM shows better isolation and flexibility, while the container can make the instance start faster and flexible to customize the runtime environment. Secure Container offers both high security and relatively low startup latency with flexibility compromise. Unikernel demonstrates great potential in terms of performance and security, but it loses flexibility. 
% Given the ecology of existing technologies, we recommend using VM-based (for better security), container-based (for lower startup latency) or Secure Container-based (for security and performance tradeoff) implementation in the virtualization layer. 
When offering adaptable images in the production environment by either virtualization mechanism, it is also critical to avoid that built ones are signed and originated from an unsafe pedigree, with the solutions~\cite{khan2017key,tak2017understanding} by keeping continuous vulnerability assessment and remediation program.

\section{Encapsule Layer}
\label{sec:Encapsule}
A cold startup in serverless computing may occur when the function fails to capture a warm running container, or experiences a bursty load.
%In serverless computing, there are two main scenarios where a cold startup may occur. 
In the former, a function is invoked for the first time, or scheduled with a longer invocation interval than the instance lifetime. 
%instances are already recycled for a short lifetime and long invocation interval.
%its invocation interval is longer than the lifetime of the instances recycling. 
The typical characteristic is that instances (or pods) must get started from scratch. 
In the latter case of a bursty load, instances need to perform horizontal scaling during a surge in user workloads.
Function instances will autoscale as load changes to ensure adequate resource allocation. 
Besides taking less than one second to prepare a sandbox in the Virtualization layer, the initialization of software environment, e.g., load Python libraries;
and application-specific user code can dwarf the former~\cite{DBLP:journals/corr/abs-1902-03383,DBLP:journals/usenix-login/LionC0ZGY17,DBLP:conf/micro/ShahradBW19,DBLP:conf/asplos/DuYXZYQWC20,DBLP:conf/usenix/OakesYZHHAA18}. 
While we can provide a more lightweight sandboxing mechanism to reduce the cold startup latency in the virtualization layer, 
the state-of-the-art sandboxing mechanism may not demonstrate perfect compatibility for containers or VMs when migrated to the existing serverless architecture.
In response to the tradeoff between performance and compatibility, an efficient solution is to prewarm instances in the Encapsule layer. This approach is known as the prewarm startup, which has been widely researched. Representative work about instance prewarm is listed in Table~\ref{tab:encapsule}.
%In order to better optimize the cold startup and decrease the initialization cost while ensuring the compatibility of existing architecture,  
%the efficient way in the Encapsule layer is to prewarm instances, which is known as the prewarm startup. Representative work about instance prewarm is listed in Table~\ref{tab:encapsule}.

Before giving the detailed analysis and comparison, we first describe the taxonomy in each column. ``Template'' reflects whether the cold startup instance comes from a template. ``Static image'' shows whether the VM/container image for prewarm disables dynamically updating in each cold startup. ``Pool'' indicates whether there is a prewarm pool for function cold startups. ``Exclusive'' and ``Fixed-size'' represents whether the prewarmed instance and its prewarm pool is exclusive and size-fixed. ``Predict/Heuristic'' points out whether the prediction algorithm or heuristic-based method is used to prewarm instances. ``REQs'' reflects whether the runtime libraries and packages are dynamically loading and updating in the prewarm instance. ``C/R'' reflects whether it supports checkpoint and restore to accelerate the startup. ``Sidecar based'' represents whether the relevant technologies can be implemented or integrated into the sidecar. ``Imp'' indicates where it is implemented.

\newcommand{\tabincell}[2]{\begin{tabular}{@{}#1@{}}#2\end{tabular}}
\begin{table*}
  \scriptsize
  \renewcommand\arraystretch{1.4}
  \caption{Works in Encapsule layer.}
  \resizebox{\textwidth}{!}{
    \begin{tabular}{l|cccccccccc}
      \toprule
      Representative work & \tabincell{c}{Template} & \tabincell{c}{Static\\image} & \tabincell{c}{Pool} & \tabincell{c}{Exclusive} & \tabincell{c}{Fixed\\size} & \tabincell{c}{Predict\\/Heuristic} & \tabincell{c}{REQs} & \tabincell{c}{C/R} & \tabincell{c}{Sidecar\\based} & Imp\\ 
      \midrule
      Pause container~\cite{DBLP:journals/usenix-login/HendricksonSOHV16,DBLP:conf/icdcsw/McGrathB17} & & \Checkmark & & \Checkmark & & & & \Checkmark & & /\\
      Azure functions~\cite{azureprewarm} & & \Checkmark & \Checkmark & \Checkmark & \Checkmark & & &\Checkmark & & AWS\\
      Fission~\cite{fissionprewarm} & & \Checkmark & \Checkmark & \Checkmark & \Checkmark & & & & & Kubernetes\\
      Adaptive Warm-up~\cite{DBLP:conf/icpads/XuZGWM19} & & \Checkmark &  & \Checkmark & & \Checkmark & & & \Checkmark & Kubernetes\\
      Serverless in the Wild~\cite{DBLP:conf/usenix/ShahradFGCBCLTR20} & & \Checkmark & & \Checkmark & & \Checkmark & & & \Checkmark & OpenWhisk\\
      Replayable Execution~\cite{wang2019a} & & \Checkmark & & & & & & \Checkmark & \Checkmark & FaaS FW\\
      Catalyzer~\cite{DBLP:conf/asplos/DuYXZYQWC20} & \Checkmark & \Checkmark &  &  & & & & \Checkmark & \Checkmark & gVisor-based\\
      Mohan \emph{et al.}~\cite{DBLP:conf/hotcloud/MohanSDENS19} & \Checkmark & & \Checkmark & & & & \Checkmark & & & OpenWhisk\\
      Apache OpenWhisk~\cite{OpenWhiskprewarm} & \Checkmark & \Checkmark & \Checkmark & & \Checkmark & & & & & /\\
      SOCK~\cite{DBLP:conf/usenix/OakesYZHHAA18} & \Checkmark & & & & & \Checkmark & \Checkmark & & \Checkmark & OpenLambda\\
      \bottomrule
    \end{tabular}
  }
  \label{tab:encapsule}
\end{table*}

% \begin{comment}
% \begin{table*}[h]
%   \caption{Works in Encapsule layer}
%   \label{tab:commands}
%   \renewcommand\arraystretch{1.2}
%   \begin{tabular}{ll}
%     \toprule
%     Representative work & Prewarm approach and key idea \\ 
%     \midrule
%     Azure functions~\cite{azureprewarm} & Pool-prewarm (Set a fixed-size prewarm pool)\\
%     Fission~\cite{fissionprewarm} & Pool-prewarm (Set a fixed-size prewarm pool)\\
%     Adaptive Warm-up~\cite{DBLP:conf/icpads/XuZGWM19} & Predict-prewarm (Prewarm by LSTM-based prediction)\\
%     Serverless in the Wild~\cite{DBLP:conf/usenix/ShahradFGCBCLTR20} & Predict-prewarm (Prewarm by time series prediction)\\
%     Apache OpenWhisk~\cite{OpenWhiskprewarm} & Zygote-prewarm (Prewarm constraint Zygote containers)\\
%     Mohan \emph{et al.}~\cite{DBLP:conf/hotcloud/MohanSDENS19} & Zygote-prewarm (Prewarm self-evolving Zygote containers)\\
%     SOCK~\cite{DBLP:conf/usenix/OakesYZHHAA18} & Zygote-prewarm (Prewarm REQs-awared Zygote containers)\\
%     Catalyzer~\cite{DBLP:conf/asplos/DuYXZYQWC20} & Zygote-C/R (Combination of Zygote and C/R)\\
%     Pause container~\cite{DBLP:journals/usenix-login/HendricksonSOHV16,DBLP:conf/icdcsw/McGrathB17,UnderstandingAWS} & C/R (Checkpoint when idle and restore on demand)\\
%     Replayable Execution~\cite{wang2019a} & C/R (CRIU-based optimization for contaienrs)\\
%     \bottomrule
%   \end{tabular}
% \end{table*}
% \end{comment}

There are two common prewarm startup approaches: \textbf{one-to-one} prewarm startup and \textbf{one-for-all} prewarm startup. In the one-to-one prewarm startup,
each function instance is prewarmed from a size-fixed pool, or by dynamic prediction based on the historical workload traces.
While in the one-for-all prewarm startup, instances of all functions are prewarmed from cached sandboxes, which are pre-generated according to a common configuration file. 
When a cold startup occurs, the function only needs to specialize these pre-initialized sandboxes by importing function-specific code blob URI and settings.
For higher scalability and lower instance initialization latency, C/R (Checkpoint/Restore) is also combined with prewarmed instances in a serverless system. C/R is a technique that can freeze a running instance, make a checkpoint into a list of files, and then restore the running state of the instance at the frozen point.
A common pattern in serverless implementations is to pause the instance when idle to save resources, 
and then recovery it for reusing when invoked~\cite{DBLP:journals/usenix-login/HendricksonSOHV16,DBLP:conf/icdcsw/McGrathB17}. 
%Some approaches require some prior knowledge, which can be the historical invocation trace or an environment setting that needs to be achieved from the sidecar. 

\textbf{\textit{One-to-one prewarm by size-fixed pool: effective but resource-unfriendly.}}

The one-to-one strategy prewarms instances for each function, which means it is crucial to determine the warm-up time. Otherwise, a slow or quick warm-up cycle will either result in less efficiency at reducing cold startup or unnecessary prewarmed instances wasting a mass of resources. 
The current solution is to build an exclusive prewarm pool with a fixed size for each function to maximize the service stability. For example, Azure Functions~\cite{azureprewarm} warms up instances of each function by setting up a fixed-size prewarm pool. Once the always-ready instance is occupied, prewarmed instances will be active and continue to buffer until reaching the limit.
The open-sourced Fission~\cite{fissionprewarm} also prewarms like Azure Function. It introduces a component called $poolmgr$, 
which manages a pool of generic instances with a fixed pool size and injects function code into the idle instances to reduce the cold start latency.

% However, the approach of a size-fixed prewarm pool would produce massive idle instances in the background and lead to waste. 

\textbf{\textit{One-to-one prewarm by predictive warm-up: ways to make resource friendly.}}

One-to-one strategy prewarms exclusive instances in each size-fixed prewarm pool, and load codes whenever invocations arrive. In this pattern, it is a safe strategy without introducing other security concerns. However, it would produce massive idle instances in the background and make the serverless system resource unfriendly. Such a deficiency inspires researchers to propose more flexible prewarm strategies like using prediction-based and heuristic-based methods.
Xu~\emph{et al.}~\cite{DBLP:conf/icpads/XuZGWM19} design an AWU (Adaptive Warm-up) (AWU) strategy by leveraging the LSTM (Long Short-Term Memory)
networks to discover the dependence relationships based on the historical traces. It predicts the invoking time of each function to prewarm the instances, and initialize the prewarmed containers according to the ACPS (Adaptive Container Pool Scaling) strategy once AWU fails. 
Shahrad \emph{et al.}~\cite{DBLP:conf/usenix/ShahradFGCBCLTR20} propose a practical resource management policy for the one-to-one prewarm startup. By characterizing the FaaS workloads, 
they dynamically change the instance lifetime of the recycling and provisioning instances according to the time series prediction. 
CRIU (Checkpoint/Restore In Userspace)~\cite{CRIU} is a software tool on Linux to implement Checkpoint/Restore functions. 
Replayable Execution~\cite{wang2019a} makes improvements based on CRIU, using ``mmap'' to map checkpoint files to memory and leveraging the Copy-on-Write in OS to share cold data among multiple containers. By exploiting the intensive-deflated execution characteristics, 
it reduces the container's cold startup time and memory usage.

\textbf{\textit{One-for-all prewarm with caching-aware: try to make the prewarm generalized and resource friendly with privacy guaranteed.}}

One-for-all prewarm startup shares a similar mechanism with the \textbf{Template} method, which is hatched and already pre-imported most of the bins/libs after being informed by the socket. When a new invocation arrives and requires a new instance, it only needs to initialize or specialize from the templates. In the process above, Catalyzer~\cite{DBLP:conf/asplos/DuYXZYQWC20} optimizes the restore process in C/R by accelerating the recovery on the retrenched critical path. Meanwhile, it proposes a sandbox fork to leverage a template sandbox that already has pre-loaded the specific function for state resuing. To make the cold startup less initialization together with more flat startup latency, Mohan \emph{et al.}~\cite{DBLP:conf/hotcloud/MohanSDENS19} propose a self-evolving pause container pool by pre-allocating virtual network interfaces with lazy pause containers binding. As performance improves, so arises vulnerability. The security concerns in the one-for-all prewarm strategy are usually referred to as privacy concerns in Encapsule Layer. In other words, it is essential to make the private packages/libraries (REQs) inaccessible. For example, the famous open-sourced Apache OpenWhisk~\cite{Openwhisk} resolves it by allowing that users can assign private packages in a \textit{ZIP} or \textit{virtualenv} to dynamically specialize the prewarmed container~\cite{OpenWhiskprewarm}. Zygote mechanism is a cache-based design used in Android for applications instantiation.
SOCK~\cite{DBLP:conf/usenix/OakesYZHHAA18} leverages the generalized Zygote mechanism that when creating prewarmed containers by considering the internal characteristics of workloads.
Specifically, it designs a packages-aware caching model to dynamically adjust cached packages with the highest benefit-to-cost ratio. Because that it performs the metric allocating with runtime sampling in the system-level, malicious activities cannot reveal the sensitive packages/libraires for a specific function.

% In the one-to-one and one-for-all prewarm methods, both static (e.g., size-fixed prewarm pool and C/R) and dynamic (e.g., prediction-based and heuristic-based) prewarm methods are used.
% If a prewarm strategy is not based on the allocation of prewarmed instance pools, it will require the developers to achieve the optimization by implementing metric monitoring, analyzing, and the extended plugins into the sidecar. Otherwise, the system will adopt the Checkpoint/Restore techniques to cache runtime states. If the REQs required by the function are dynamically loading and changing in the prewarm instance, then the serverless system may take a higher priority to build a Zygote rather than checkpoint/restore a static image.

\textbf{\textit{One-to-one and one-for-all prewarm: the challenging points.}}

For one-to-one prewarm startup and one-for-all prewarm startup, both can be beneficial for optimizing the cold startup in the Encapsule Layer of serverless architecture. 
Their respective flaws are also apparent. The one-to-one prewarm startup focuses on significantly less initialization latency by exchanging the memory resource. It meets the challenge that a warm-up time in point is usually hard to measure or predict while ensuring the reasonable allocation of memory resources, according to the research~\cite{DBLP:conf/usenix/ShahradFGCBCLTR20}. On the one hand, prediction-based and heuristic-based methods are particularly effective when historical data is sufficient to build an accurate model, but degrades when the trace is scarce. On the other hand, the prediction and iteration operations can introduce high CPU overhead when massive applications and function chains co-exist.  

The template mechanism in the one-for-all prewarm startup is adopted to ease the high cost of functions cold startup from scratch. In addition, maintaining a global prewarm pool introduces less additional memory resource
consumption than the one-to-one prewarm startup. However, it still suffers from several challenges, including the huge template image size~\cite{DBLP:conf/fast/HarterSLAA16,DBLP:conf/fast/AnwarMTLRCZNWLH18} and confliction of various pre-imported libraries. It may also potentially reveal the vicinity in which applications with a similar portrait are widely deployed.
It is very important to ``suit the remedy to the case'' for cold startups in different scenarios. 
For example, it is much more efficient to generate a template by one-for-all prewarm startup when the function is invoked for the first time, or with poor predictions during the trace analysis. The one-to-one prewarm startup performs better for functions with general rules or diurnal patterns, and vice versa.

\section{Orchestration Layer}
\label{sec:Orchestration}
\begin{figure}
  \centering
  \subfigure[Provision adjustment at resource-level]{
  \includegraphics[width=.34\textwidth]{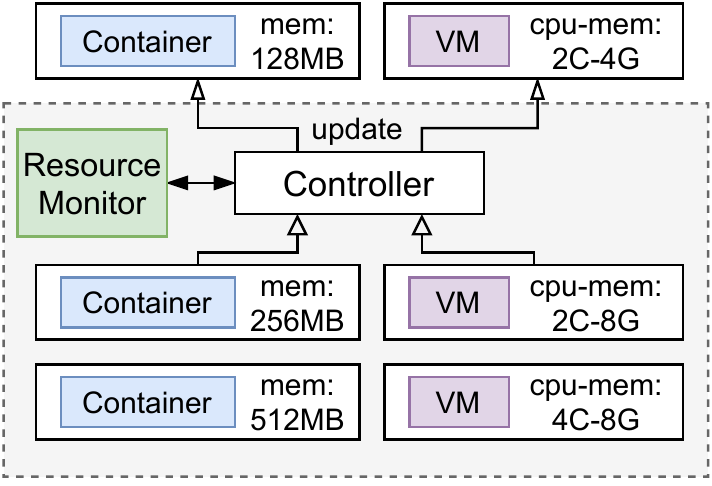}
  }
  \subfigure[Load balance at instance-level]{
  \includegraphics[width=.56\textwidth]{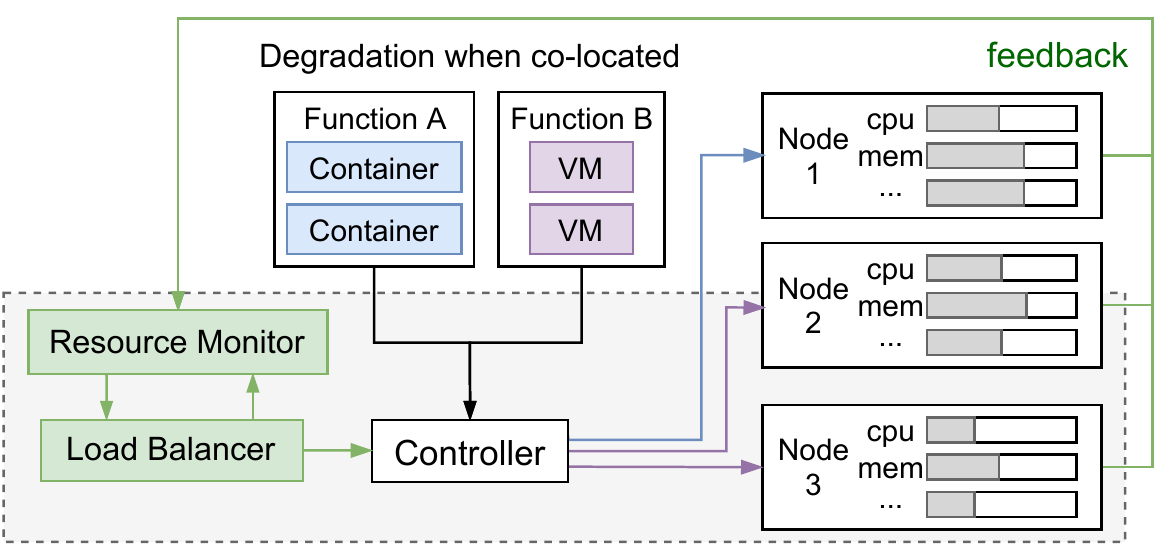}
  }
  \subfigure[Function partition at application-level]{
  \includegraphics[width=.84\textwidth]{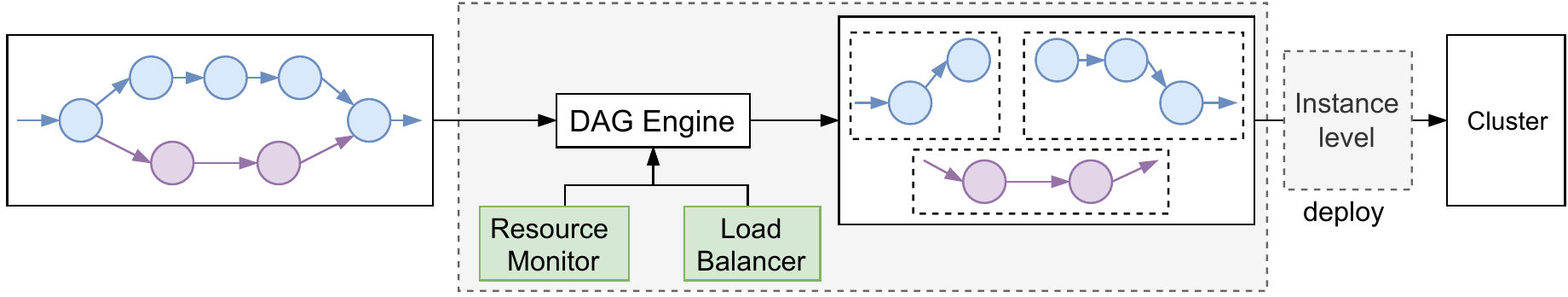}
  }
  \vspace{-2mm}
  \caption{\label{fig:resource}System logic and scheduling levels in Orchestration Layer.}
\end{figure}

The main challenge in the System Orchestration Layer is the friendly and elastic support for different services. 
Even though the current serverless orchestrators are implemented differently, the challenges they encounter in the system are much the same.
% According to the underlying architecture, we can divide most of the current serverless system into two categories, Kubernetes-based system and orchestration-integrated system. 
% Although their implementation logic is different, the challenges they encountered in the system are much the same.
As hundreds of functions co-exist on a serverless node, it raises challenges for scheduling massive functions with inextricable dependencies. Similar to the traditional solutions~\cite{DBLP:journals/ijcc/SriramaO18,DBLP:conf/ucc/ImaiCV13,DBLP:journals/csur/BuyyaSCCSVGJVNT19,DBLP:conf/icdcs/KwanWJM19,DBLP:conf/asplos/ChenDM19}, the serverless model also concerns the ability to predict the on-demand computing resources, and an efficient scheduling strategy for services. As shown in Figure~\ref{fig:resource}, researchers usually propose to introduce the load balancer and resource monitor components into the controller, to resolve provision and scheduling challenges. The load balancer aims to coordinate the resource usage to avoid overloading any single resource. Meanwhile, the resource monitor keeps watching the resource utilization of each node, and passes the updated information to the load balancer.
With the resource monitor and load balancer, a serverless controller can perform better scheduling strategies in three aspects: \textbf{resource-level}, \textbf{instance-level}, and \textbf{application-level}. We summarize the hierarchy in Table~\ref{tab:orchestration}.

% \begin{figure}
%   \centering
%   \includegraphics[width=.95\linewidth]{figures/resource.pdf}
%   \caption{\label{fig:resource}System logic and optimization in Orchestration Layer.}
% \end{figure}

Specifically, the ``focused hierarchy'' indicates an optimized method (aka resource adjusting) is designed besides an essential strategy for resource auto-provision, which can be classified into ``R'' (Resource-level),  ``I'' (Instance-level), or ``A'' (Application-level), respectively. ``Resource adjusting'' shows whether the scheduling provides an adjustment for resource provision. ``SLO'' reflects whether SLO constraints are considered. ``Intf'' represents whether the resource contention or interference is discussed. 
``Usage feedback'' reflects whether the feedback of resource usage in a physical node is considered. ``Dynamic strategy'' indicates whether it is a dynamic and runtime scheduling strategy. ``Trace driven'' indicates whether making choices depends on traces or collected data metrics. ``Predict/Heuristic'' reflects whether a prediction-based or heuristic-based method is used. ``Implement'' points out where it is implemented (``P'' represents it is a prototype). Finally, ``Insight'' summarizes its unique insight and key motivation.

\begin{table*}
  \scriptsize
  \renewcommand\arraystretch{1.3}
  \caption{Works by focused hierarchy in System Orchestration layer.}
  \resizebox{\textwidth}{!}{
    \begin{tabular}{l|ccccccccccc}
      \toprule
      \tabincell{c}{Representative\\work} & \tabincell{c}{Focused\\hierarchy} &\tabincell{c}{Resource\\adjusting} & SLO & Intf & \tabincell{c}{Usage\\feedback} & \tabincell{c}{Dynamic\\strategy} & \tabincell{c}{Trace\\driven} & \tabincell{c}{Predict\\/Heuristic} & Implement & Insight\\ 
      \midrule
      Pigeon~\cite{9071414} & R & \Checkmark &  &  & \Checkmark & \Checkmark &  &  & kubernetes & Static pool\\
      FlowCon~\cite{DBLP:conf/icpp/ZhengTGMCH19} & R\&(I) & \Checkmark &  &  & \Checkmark & \Checkmark & \Checkmark &  & P & DL tasks\\
      SIREN~\cite{DBLP:conf/infocom/WangNL19} & R\&(I) & \Checkmark &  &  &  & \Checkmark & \Checkmark & \Checkmark & AWS Lambda & ML tradeoff\\
      CherryPick~\cite{DBLP:conf/nsdi/AlipourfardLCVY17} & R & \Checkmark & \Checkmark &  & \Checkmark & \Checkmark & \Checkmark & \Checkmark & P & Bayesian Opt\\
      Lin~\emph{et al.}~\cite{DBLP:journals/tpds/LinK21} & R\&(A) & \Checkmark & \Checkmark &  & \Checkmark & \Checkmark & \Checkmark & \Checkmark & AWS Lambda & Profiling\\
      MPC~\cite{DBLP:journals/tpds/HoseinyFarahabady18}  & R & \Checkmark & \Checkmark & \Checkmark & \Checkmark & \Checkmark & \Checkmark & \Checkmark & OpenWhisk & /\\
      \midrule
      Chang~\emph{et al.}~\cite{DBLP:conf/globecom/ChangYYLJ17} & I\&(R) &\Checkmark & \Checkmark & & \Checkmark & \Checkmark & \Checkmark & & kubernetes & /\\
      Kaffes~\emph{et al.}~\cite{DBLP:conf/cloud/KaffesYK19} & I\&(R) &  &  &  & \Checkmark & \Checkmark &  &  & P & /\\
      FnSched~\cite{DBLP:conf/middleware/SureshG19} & I\&(R) & \Checkmark & \Checkmark &  & \Checkmark & \Checkmark & \Checkmark &  & OpenWhisk & /\\
      Guan~\emph{et al.}~\cite{DBLP:journals/icl/GuanWCSZ17} & I\&(A) & \Checkmark &  &  & \Checkmark & \Checkmark &  &  & P & Library\\
      McDaniel~\emph{et al.}~\cite{DBLP:conf/cluster/McDanielHT15} & I &  & \Checkmark & \Checkmark & \Checkmark & \Checkmark &  &  & Docker Swarm & Two-tiered\\
      Kim~\emph{et al.}~\cite{DBLP:journals/tpds/KimFLZ20} & I\&(R) & \Checkmark & \Checkmark & \Checkmark & \Checkmark & \Checkmark & \Checkmark & \Checkmark & P & CPU cap\\
      Smart spread~\cite{DBLP:conf/cascon/MahmoudiLKL19} & I &  & \Checkmark & \Checkmark & \Checkmark &  & \Checkmark & \Checkmark & AWS Lambda & Profiling\\
      \midrule
      Xanadu~\cite{DBLP:conf/middleware/DawBK20} & A\&(R) & \Checkmark &  &  & \Checkmark & \Checkmark & \Checkmark & \Checkmark & OpenWhisk & Profile\&predict \\
      Step Functions~\cite{StepFunctionstype} & A &  &  &  & \Checkmark &  &  &  & AWS Lambda & Health check\\
      WUKONG~\cite{DBLP:journals/corr/abs-1910-05896,10.1145/3419111.3421286} & A &  &  &  & \Checkmark & \Checkmark &  &  & AWS Lambda & Graph to seq\\
      Viil~\emph{et al.}~\cite{DBLP:journals/tjs/ViilS18} & A\&(R) & \Checkmark &  &  &  & \Checkmark &  & \Checkmark & Pegasus & Partition\\ 
      SAND~\cite{DBLP:conf/usenix/AkkusCRSSBAH18} & A & \Checkmark &  & \Checkmark &  &  &  &  & P & Colocation\\
      GlobalFlow~\cite{DBLP:conf/IEEEcloud/ZhengP19} & A\&(I) &  &  &  &  &  & \Checkmark &  & AWS Lambda & Cross-regions\\
      SONIC~\cite{DBLP:conf/cloud/TariqPNRL20} & A\&(R) & \Checkmark &  &  & \Checkmark & \Checkmark & \Checkmark & \Checkmark & AWS Lambda & Hybrid exchange\\    
      \bottomrule
    \end{tabular}
  }
  \label{tab:orchestration}
\end{table*}

\subsection{\textbf{Dynamic Adjustment of Resource Provision (Resource-level)}}
Resources including CPU, Memory serve as the basic scheduling objects in serverless computing. For isolation and stability, resources are configured by an orchestrator, and access to them is restricted. The serverless controller will allocate resources for a new instance and isolate the execution environment when a cold startup occurs.
Therefore, the key to building an efficient serverless controller is auto-scaling just the right amount of resources to satisfy the resilient workloads.
However, the controller component itself cannot adjust appropriately because the resource is highly dynamic in the cluster and the potential inaccurate resource specifications by default.

\textbf{\textit{Make resource provision of the container ``just the right amount''.}}

The common solution for avoiding resource over-provisioning is building feedbacks regarding historical traces. For example, works in~\cite{DBLP:conf/icnp/ChenSP16,DBLP:conf/infocom/ChenS17} optimize the original resource settings by varying the trace-driven patterns for VMs.
In serverless computing systems with more fine-grained functions, a real-time resource monitor can be employed to help the controller make dynamic resource adjustments, as shown in Figure~\ref{fig:resource}(a).
For example, Pigeon~\cite{9071414} builds a serverless framework, introducing a function-level resource scheduler and an oversubscribed static pool. The scheduler assigns containers with different resource configurations to the queries based on the node capacity and function requirement. However, their container pool is based on a static configuration, which may lead to resource segmentation and low utilization.
FlowCon~\cite{DBLP:conf/icpp/ZhengTGMCH19} facilitates dynamic resource allocation for container-based DL training tasks in the near future and resets resource configuration elastically. Though they design a dynamic auto-provision strategy based on the monitor feedback, the SLO constraints and resource interference are not considered further. It results in their lack of flexibility and practicality in a real production environment.
DRL (Deep Reinforcement Learning), evolving from Deep Q-learning, is a widely used combination algorithm by learning control strategies from higher-dimensional perceptual inputs~\cite{DBLP:journals/corr/MnihKSGAWR13}, which can be used to make resource provision decisions.
% Based on it, Wang~\emph{et al.}~\cite{DBLP:conf/infocom/WangNL19} propose a serverless scheduler for ML training jobs to learn the best solution in achieving higher model quality while minimizing the training time under a given cost.
For example, Wang~\emph{et al.}~\cite{DBLP:conf/infocom/WangNL19} propose a serverless scheduler based on DRL for ML training jobs. It can dynamically adjust the number of function instances needed and their memory size, to balance high model quality and the training cost.

\textbf{\textit{The keys to making resource provision robust to performance.}}

Recent works take the SLA into account to ensure stability when functions are invoked in a shared-resource cloud. 
CherryPick~\cite{DBLP:conf/nsdi/AlipourfardLCVY17} leverages the Bayesian optimization, which estimates a confidence interval of an application's running time and cost, to help search the optimal resource configurations. Unlike static searching solutions, it builds a performance model to distinguish the unnecessary iteration trials, thus accelerating the convergence. However, CherryPick's performance model targets big data applications specifically, not generalized to other applications.
Similarly, Lin~\emph{et al.}~\cite{DBLP:journals/tpds/LinK21} build an analytical model to help general serverless applications deployment. It can predict the application's end-to-end response time and the average cost under a given configuration. They also propose a Probability Refined Critical Path Greedy algorithm (PRCP) based on the transition probability, recursively searching the critical path of execution order. With PRCP, they can achieve the best performance with a specific configuration under budget constraints or less cost under QoS constraints.
Besides SLA, shared-resource contention should also be noticed in the multi-tenant environment.
HoseinyFarahabady~\emph{et al.}~\cite{DBLP:journals/tpds/HoseinyFarahabady18} discuss this topic.
Their proposed MPC optimizes the serverless controller for resource predictively allocation. By introducing a set of cost functions, it reduces the QoS violation, stabilizes the CPU utilization, and avoids serious resource contention.
However, these resource and workload estimations based on  ML (Machine Learning) or AI (Artificial intelligence) usually achieve a trade-off between an optimal global solution and robust performance to inaccurate workload information~\cite{DBLP:conf/ccgrid/ImaiPV18,DBLP:journals/csur/BuyyaSCCSVGJVNT19,DBLP:journals/tjs/CasasTRZ17,DBLP:conf/ccgrid/TokaDFS20}. Whether they can avoid fragile robustness and improve resource utilization in the production environment is unknown and remains a critical avenue to explore.
%Mean-field Games can also help achieve fast, resilient, and scalable responses by the potential chain and heterogeneous inference~\cite{DBLP:journals/cm/SemasingheMH17}.

% However, the traditional resource and workload estimation based on the ML (Machine Learning) or AI (Artificial intelligence) may be fragile due to the trade-off between global optimal solution and robustness to inaccurate workload information~\cite{DBLP:conf/ccgrid/ImaiPV18,DBLP:journals/csur/BuyyaSCCSVGJVNT19,DBLP:journals/tjs/CasasTRZ17,DBLP:conf/ccgrid/TokaDFS20}.
% With multiple tenants sharing the serverless infrastructure to run functions, distributed resources configuring across multiple dimensions (e.g., CPU, memory, I/O, bandwidth, and so on) in a fine-grained manner, as well as to achieve low-cost allocations that meet SLOs remains a critical avenue to explore.

\subsection{\textbf{Load Balancing for Instance Scheduling (Instance-level)}}
In addition to dynamic adjustment at the resource level, the most important part of a serverless system is instance-level scheduling.  
From the perspective of cloud vendors, they hope to achieve either higher throughput, higher resource utilization, or less energy consumption. At the same time, users prefer cheaper deployment costs and less end-to-end invocation latency. To this end, instances from multi-functions or multi-tenants should be carefully scheduled across the cluster to achieve the above targets.

The mainstream solution is leveraging load balancer, which is also shown in Figure~\ref{fig:resource}(b). It is designed as the queries router to help schedule functions and achieve the load balancing between nodes in the cluster.
The strategies can be classified into two categories: \textbf{Hash-based} and \textbf{Multi objective-based} methods. In the hash-based method, the controller uses a hash function to decide a home node (or executor) of a given function for default routing. Then it will set a stepsize to recursively filter out an alternative if the home is not available or under resource-constrained. They are usually done by a health check in each physical node. Until the cloud provider has a full understanding of the characteristics of workloads running in its serverless system and the cluster, we recommend using the hash-based method to implement a load balancer.
In the multi-objective-based balancing method, the load balancer aims at multiple optimizations, for example, throughput, response time, resource utilization, etc. Therefore, it should balance different factors to satisfy both the cloud vendors and users.

\textbf{\textit{Leverage resource monitor and load balancer to make scheduling decisions.}}

% Some researchers leverage game theory methods and market-oriented models for resource regulation within the cloud to solve such multi-optimization problems~\cite{DBLP:journals/csur/BuyyaSCCSVGJVNT19}. For example, 
% Kaffes~\emph{et al.}~\cite{DBLP:conf/cloud/KaffesYK19} tries to introduce a centralized and core-granularity scheduler to maintain a global view of cores in the cluster and route queries safely to workers.
Resource monitor can provide a global view of resource status in a cluster, which helps the load balancer make better scheduling decisions.
Chang~\emph{et al.}~\cite{DBLP:conf/globecom/ChangYYLJ17} design a comprehensive monitoring mechanism for the Kubernetes-based system. It can provide a variety of runtime information to the scheduler, including system resource utilization and the QoS performance of an application. The flaw of the study is that it does not provide a complicated resource scheduling algorithm.
Kaffes~\emph{et al.}~\cite{DBLP:conf/cloud/KaffesYK19} propose a centralized and core-granular scheduler. Centralization provides a global view of the cluster to the scheduler so that it can eliminate heavy-weight function migrations. Core-granularity binds cores with functions and therefore avoids core-sharing among functions and promises performance stability. However, they only consider the scheduling of CPU resource, but ignore other important resources like memory.
% FnSched~\cite{DBLP:conf/middleware/SureshG19} classify serverless functions into different categories,   takes into account the resource consumption and lifetime
% patterns of serverless functions by classifying them into different
% categories
FnSched~\cite{DBLP:conf/middleware/SureshG19} regulates CPU-shares to accommodate the incoming application invocations by checking the available resource. A key advantage of employing a greedy algorithm is that fewer invoker instances are scheduled by concentrating invocations in response to varying workloads. Though FnSched makes a tradeoff between scalable efficiency and acceptable response latencies, it is limited by the assumption that function execution times are not variable.
% In order to further supporting auto-scaling and minimizes the cost of deployment,
Guan~\emph{et al.}~\cite{DBLP:journals/icl/GuanWCSZ17} propose an AODC-based (Application Oriented Docker Container) resource allocation algorithm by considering both the available resources and the required libraries.
They model the container placement and task assignment as an optimization problem, and then take a Linear Programming Solver to find the feasible solution. The Pallet Container performs the AODC algorithm, serving as both a load balancer and resource monitor. The downside is that plenty of containers will occupy the memory space as the number of functions increases. 

\textbf{\textit{Take the performance interference and QoS constraints into consideration.}}

While improving utilization, load balancing strategies also bring the interference challenge that sharing resources between instances may result in performance degradation and QoS violation.
Different functions' sensitivities to different resources may vary, which means that we should avoid physical co-location of functions that are sensitive to the same resource (e.g., CPU-sensitive containers may cause serious CPU contention when co-located). The load balancer should notice and moderate the interference when scheduling containers.
McDaniel~\emph{et al.}~\cite{DBLP:conf/cluster/McDanielHT15} manage the I/O of containers at both the cluster and the node levels to reduce resource contention and eliminate performance degradation effectively. Based on a resource monitor in Docker Swarm, it refines the container I/O management by providing a client-side API, thus enforcing proportional shares among containers for I/O contention.
Kim~\emph{et al.}~\cite{DBLP:journals/tpds/KimFLZ20} present a fine-grained CPU cap control solution by automatically and distributedly adjusting the allocation of CPU capacity. Based on performance metrics, applications are grouped and allowed to make adjustment decisions, and application processes of the group consume only up to the quota of CPU time. Hence, it minimizes the response time skewness and improves the robustness of the controller to performance degradation. 
Smart spread~\cite{DBLP:conf/cascon/MahmoudiLKL19} proposes an ML-based function placement strategy by considering several resource utilization statistics. It can predictively find the best performing instance as well as incurs the least degradation in performance of the instance.

%Nguyen \emph{et al.}~\cite{DBLP:journals/cm/NguyenK20} analyze that some applications may run multiple replicas to improve performance and availability. So they propose a leader-based consistency maintenance mechanism to maintain consistency and coordinate tasks for container-based scheduling. 

\subsection{Data-driven Workflows for Application Deployment (Application-level)}

At the application level, load balancing strategies can be categorized into two kinds: the spread strategy, which distributes functions of an application across all the physical nodes, and the bin-pack strategy, which tries to schedule functions of an application to the same node first~\cite{DBLP:journals/tpds/HoseinyFarahabady18}.
% Essentially, the load balancing should make a tradeoff between the spread strategy, which distributes functions across all accessible physical nodes, and the bin-pack strategy, which occupies a single node first~\cite{DBLP:journals/tpds/HoseinyFarahabady18}.
Intuitively, the spread strategy seems to better balance the workloads on the nodes while avoiding serious resource contention.
However, it weakens the data locality, which means that the spread strategy will introduce more transmission overhead than bin-pack if functions are data-dependent.
So it is why we need to optimize the scheduling from the perspective of the application.

\textbf{\textit{Invocation patterns and workflow execution models.}}

\begin{figure}
  \centering
  \subfigure[Invocation patterns]{
  \includegraphics[width=.37\textwidth]{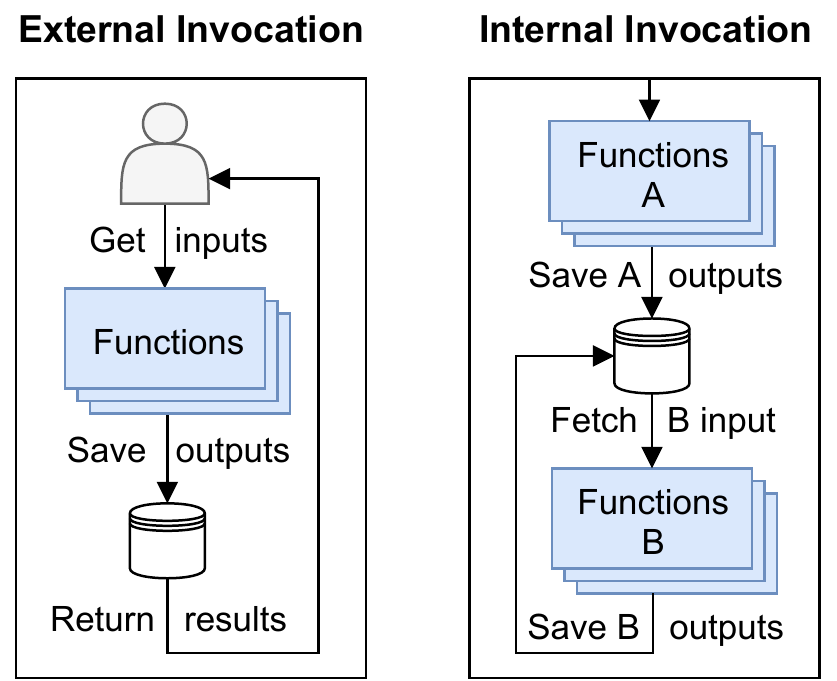}
  }
  \hspace{10mm}
  \subfigure[Workflow execution models]{
  \includegraphics[width=.37\textwidth]{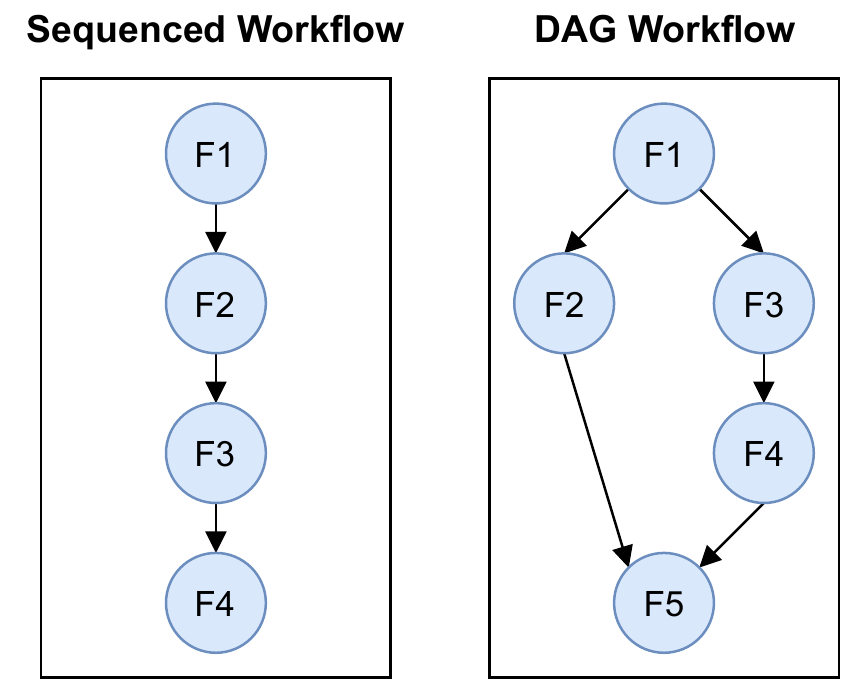}
  }
  \vspace{-2mm}
  \caption{Two Invocation patterns for funtions and two execution models of workflows.}
  \label{fig:workflow}
\end{figure}

As shown in Figure~\ref{fig:workflow}(a), if a function is invoked from user queries via the RESTful API or other triggers, it is called external invocation. The instance-level load balancing can perform well in external invocation scenarios. However, the emerging cloud applications may consist of several functions, and there are data dependencies between multiple functions. For example, the implementation of a real-world social network consists of around 170 functions~\cite{DBLP:conf/sigsoft/AdzicC17}. In this case, functions in such an application will get active by various triggers which may come from the user query or another function. If a function is initialized or assigned by other functions, it follows the internal invocation pattern. Currently, researchers raise their vision to the data-driven scheduling for internal invocations from the perspective of application-level topology.
% Different from the scenario of instance-level load balancing where multiple functions have no data dependencies, application-level scheduling handles more complex scenarios. In this case, 
% The emerging cloud applications may consist of several functions, for example, the implementation of a real-world social network consisting of around 170 functions~\cite{DBLP:conf/sigsoft/AdzicC17}.
% Because serverless computing is an event-driven system, functions in such an application will get active by various triggers which may come from the user query or another function. 
% As shown in Figure~\ref{fig:workflow}(a), if a function is invoked from user queries via the RESTful API or other triggers, it is known as the \textbf{External Invocation}. 
% Otherwise a function initialized or assigned by other functions follows the \textbf{Internal Invocation} pattern.

Workflow is the most common implementation of internal invocations, where functions are executed in a specified order to satisfy complex business logic.
The execution models of these data-driven workflows can be extracted into two approaches: sequence-based workflow and DAG-based workflow. As shown in Figure~\ref{fig:workflow}(b), functions are invoked in a pipeline through a registered dataflow in the sequence-based workflow.
The sequence-based workflow is the basic and the most common pattern in the serverless workflow, and most cloud vendors provide such execution mode for application definition. 
Obviously, there is more than one sequenced workflow in one complex application, and the same functions can be executed in various sequences. If we regard each function as a node and dataflow between nodes as a vector edge, 
such an application with multiple interlaced sequenced workflows can be defined by the DAG (hence the name ``DAG-based workflow'').
Nowadays, few Cloud vendors provide services for the application definition in the DAG (Direct Acyclic Graphs) form, aka serverless workflows~\cite{DBLP:journals/csur/AdhikariAS19,DBLP:conf/IEEEcloud/BessaiYOGN12,DBLP:journals/fgcs/BuyyaYVBB09}.

\textbf{\textit{The scheduling overhead introduced in serverless sequences.}}

With massive functions communicate with each other, scheduling of dataflow dependencies introduces more complexities. However, the existing serverless systems in the production environment commonly treat these workflows as simple recursion of internal invocations. 
It raises the challenge of reducing the overhead in the System Orchestration layer by scheduling function sequences~\cite{DBLP:conf/oopsla/BaldiniCFMMRST17}. Current policy to manage the function sequences is quite simple- functions are triggered following the first-come-first-served algorithm~\cite{DBLP:conf/cloud/TariqPNRL20}. However, as the length of the function sequence increases, cascading cold start overheads should be addressed to avoid seriously end-to-end latencies degradation of sequenced workflows~\cite{DBLP:conf/sac/BermbachKB20,DBLP:conf/middleware/DawBK20}. 
To this end, Xanadu~\cite{DBLP:conf/middleware/DawBK20} combines the prewarm strategy with a most-likely-path (MLP) estimation in the workflow execution. It prewarms instances by a speculative-based strategy and makes just-in-time resource provisioning. However, the prediction miss would introduce additional memory waste, especially in the scenario of multi-branch or DAGs.
Moreover, serverless workflow engines prefer the Master-Worker architecture where ready functions are identified by the state and invoked directly by the master without a queue~\cite{DBLP:journals/fgcs/MalawskiGZBF20,DBLP:conf/nsdi/FouladiWSBZBSPW17,DBLP:conf/cloud/AoIVP18,Hyperflow,DBLP:journals/corr/abs-1910-05896}, 
including AWS Step Functions~\cite{StepFunctionstype} and Fission Workflows~\cite{Fissionwf}. 
As shown in Figure~\ref{fig:workflow}(a), the deficiency is that the additional overhead is introduced in the function workflow through unnecessary middlewares (e.g., unnecessary storage in an internal invocation).

\textbf{\textit{Enhance the data locality for efficient serverless DAG executions.}}

To help function workflow avoid undesired middlewares, researchers usually co-locate the functions into subgraphs to enhance the data locality, as shown in Figure~\ref{fig:resource}(c).
For example, Viil~\emph{et al.}~\cite{DBLP:journals/tjs/ViilS18} use multilevel k-way graph partitioning to provision and configure scientific workflows automatically into multi-cloud environments. However, their partition algorithm may not match well with serverless applications, where each node in the graph can auto-scale multiple replicas in such as, foreach steps. In this case, the connections and edge weights become unpredictable. 
In serverless context, WUKONG~\cite{DBLP:journals/corr/abs-1910-05896,10.1145/3419111.3421286} implements a decentralized DAG engine based on AWS Lambda, which combines static and dynamic scheduling. It divides the workflow of an application into subgraphs, before and during execution, thus improving parallelism and data locality simultaneously. However, WUKONG's colocation of multi-functions within a Lambda executor may introduce additional security vulnerabilities due to its weakened isolation. 
SAND~\cite{DBLP:conf/usenix/AkkusCRSSBAH18} presents a new idea to group these workflow functions into the same instance so that libraries can be shared across forked to reduce initialization cost, and additional transmission can be eliminated in the workflow due to the data locality. SAND performs a better isolation mechanism than WUKONG by using process forking for function invocations, however it ignores the colocation interference resulting from the resource contention.
When exchanging intermediate data of DAGs, SONIC~\cite{DBLP:conf/usenix/MahgoubSMKCB21} proposes to use the VM-storage-based transmission strategy when functions are co-located on the same node. The optimal transferring depends on application-specific parameters, such as the input size and node parallelism. By predicting such runtime metrics of functions in the workflow, it dynamically perform the data passing selection with a communication-aware function placement.
GlobalFlow~\cite{DBLP:conf/IEEEcloud/ZhengP19} considers a geographically distributed scenario where functions reside in one region and data in another region. It groups the functions in the same region into subgraphs and connects them with lightweight functions, so that it improves data locality and reduces transmission latency. As the authors stated, the combination of local and cross-region strategies in a holistic manner can be further explored.

\textbf{\textit{Summary of the challenges in the scheduling of serverless workflows.}}

Workflow scheduling is an NP-hard problem, and researchers have been designing various strategies for it~\cite{DBLP:journals/csur/AdhikariAS19,DBLP:journals/jnca/MasdariVSA16}.
Such optimization in the workflow aims to minimize the makespan, reduce the execution cost, and improve resource utilization while satisfying single or multiple constraints. 
To resolve the above challenges, leveraging enhanced data locality is a focus in serverless computing. The challenge is that the end-to-end latency of a workflow query could increase significantly due to frequent interactions with the storage from different nodes.  
Resource volatility becomes another focus in the serverless system, which can be unpredictable as the number of functions increases in the production environment. It introduces more difficulty to find an efficient workflow placement and scheduling strategy in a concise decision time (e.g., 10ms for load balancing). In order to evaluate the efficiency and performance for future workflow-based research, DAG-based or DG-based serverless benchmarks also urgently need to be published. They are better adapted based on real applications rather than simple micro-benchmarks~\cite{DBLP:journals/jss/Scheuner020} or function self-loops~\cite{DBLP:conf/cloud/YuLDXZLYQ020,DBLP:journals/tpds/LinK21}. Keeping a guaranteed QoS performance is also significant for applications in serverless computing, whereas it has not been widely investigated.

\subsection{Security Concerns in Orchestration Layer}

In the Orchestration Layer, the most serious security concern is how to resolve unavailability. It usually refers more to performance security than functional security. The attackers may establish destructive behaviors from either resource-level, instance-level, or application-level, resulting in unmatched in-memory footprint, concurrency exhaustion, or workflow exceptions. The efficient solution to these concerns is leveraging BaaS components to restrict access.

\textbf{\textit{In-memory footprint by unrestricted read-in.}}

Contrary to the intuitive believing, serverless architecture can make the programming more complicated because the decoupled microservices have higher requirements on the normalized input data. The unrestricted memory read-in from the input data may result in the timeout or breakdown for its oversized memory footprint (e.g., 300MB memory read-in within a 256MB-limit container). Function developers may overlook this vulnerability in a public cloud where an attack can easily disguise as a legitimate invocation. On the premise that the user code is fragile against such kind of attack, a serverless system needs to bind filtering rules in event trigger to help avoid this security concern.
% memory exhaustion attacks

% By controlling botnets, broilers, or proxy servers, an attacker continuously sends a large number of legitimate requests to the target host, making the request processing of normal users extremely slow.  

\textbf{\textit{Concurrency exhaustion by DDoS (Distributed Denial of Service) attack.}}

When an application is decoupled into a mass of functions, its concurrency bottleneck depends on the maximum throughput of all function nodes. 
% This phenomenon indicates that any unavailable function node can seriously take down the entire application. 
% So the serverless architecture may introduce more security vulnerability than the traditional monolithic architecure for a wilder attack surface. 
In this case, the DDoS attack on serverless is a more significant threat. Any unavailable function node can seriously take down the entire application with a wilder attack surface. It can cause seriously degraded QoS by invocation exhaustion attacks in a single function node, or generate a large bill for the application account. The latter is known as the DoW (Denial of Wallet) attack in serverless computing. This security risk can be mitigated by setting an upper limit on invocations concurrency and instances quota on function creating.
% It could take down a single FaaS account, and generate a large bill in the process for the application account, which is also known as the DoW.

\textbf{\textit{Workflow exception by malicious dataflow.}}

When triggering invocations, function input parameters and preferences are also passed within the dataflow. Therefore, an internal invocation is more fragile than an external one if the function maintainer fails to provide essential verification of queries source. In addition, attackers can inject malicious data into the queries to generate the invocation exceptions, or make the workflow execution order out of control.
In the case of serverless architecture, all function nodes in the application need to authorize their access permission to identify whether the input data tamper with the current invocation.

\begin{figure}
  \centering
  \includegraphics[width=.84\linewidth]{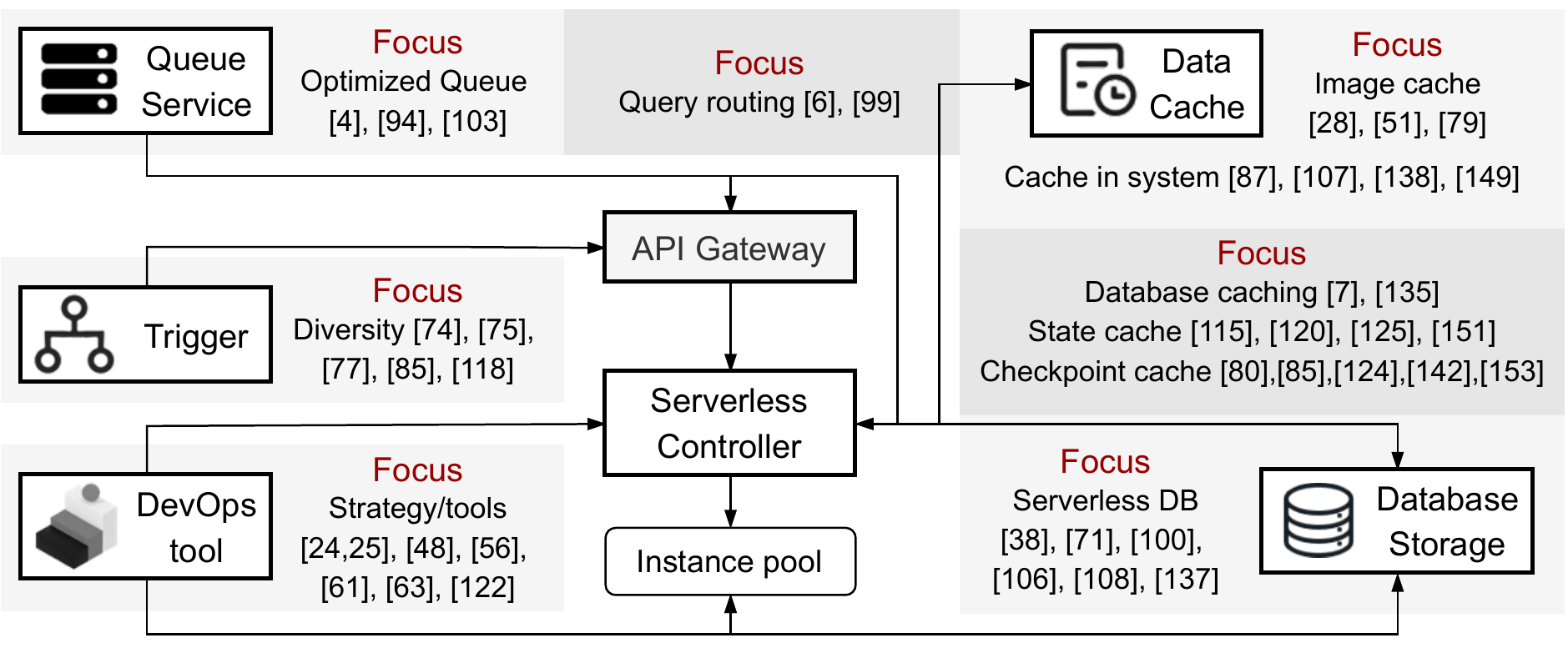}
  \caption{\label{fig:baas}Techniques and works about BaaS components in System Coordination layer.}
\end{figure}

\section{Essential BaaS Components in Coordination Layer}
\label{sec:Coordination}
We have examined the most critical implementations in Section~\ref{sec:Virtualization}, \ref{sec:Encapsule}, and \ref{sec:Orchestration}, and there are also some other components in the System Coordination layer introduced to support or enhance the serverless system. We also outline the relevant techniques and research in Figure~\ref{fig:baas}. In terms of implementation, a serverless system needs to integrate the six significant components or services: Storage, Queue, API gateway, Trigger, Data cache, and DevOps tools. Most of the literature focuses on Data cache, Queue service, and function storage from an academic perspective. In contrast, the cloud vendors mainly focus on productions about Trigger service and DevOps tools. We will discuss each of these components detailly in the following text.

\subsection{\textbf{Storage Service}}
One of the key requirements of a serverless workload is efficiently sharing ephemeral data between functions or saving results for asynchronous invocation. 
Therefore, a natural way to communicate between them is to exchange the data through a remote store. 

\textbf{\textit{Different phases of storage during the function execution.}}

During a serverless invocation, there are three phases where the database service is required: Authentication, In-Function, and Log.
Authentication is usually performed ahead of controller scheduling to avoid security issues, and it should get a fast response for access. 
Using an MMDB (Main Memory DataBase) to implement the Authentication phase is recommended in a serverless system, such as Redis, a high-performance key-value database.
The calls of storage APIs during the function execution make up the in-Function phase. Users can choose to use either a DRDB (Disk-Resident DataBase, e.g., MySQL) or an MMDB by different BaaS interfaces for ephemeral storage. 
The Log phase builds the bridge for users to return invocations results, especially for the functions invoked in an asynchronous manner. A detailed record in JSON format, including runtime, execution time, queue time, states, will be ephemerally or permanently stored and returned (e.g., CouchDB in OpenWhisk). It is recommended to be designed as serverless storage, following the invocation patterns to only pay for queries consumed by the storage operation and the storage space consumed when logging. However, the throughput of existing storage is a major bottleneck due to the frequent and vast functions interactions~\cite{DBLP:conf/cloud/JonasPVSR17,DBLP:journals/corr/abs-1902-03383}. Although current serverless systems support NAS (Network Attached Storage) to help reduce storage API calls, these shared access protocols are still network-based data communication essentially.

\textbf{\textit{IO bottleneck in storage: modeling in serverless context.}}

Traditional solutions use predictive methods~\cite{DBLP:conf/sosp/CortezBMRFB17,DBLP:conf/icac/NguyenSGSW13,DBLP:conf/green/WajahatGKK16} and active storage~\cite{DBLP:journals/fgcs/XieF0L16,DBLP:journals/internet/TinedoLASMRNNCO16,DBLP:conf/fast/WiresW17,DBLP:conf/fast/TinedoSZALMR17,DBLP:conf/IEEEcloud/SampeLA16,DBLP:conf/cloud/Zhang00S19} to automatically scale resources and optimize the data locality on demand. For serverless storage, researchers explore using a hybrid method to ease the I/O bottleneck. For example, Pocket~\cite{DBLP:conf/osdi/KlimovicWSTPK18} is strict with the separation of responsibilities across the control, metadata, and data planes. By using heuristics and combining several storage technologies, it dynamically rightsizes resources and achieves a balance between cost and performance efficiency. To alleviate the extremely inefficient execution for data analytics workloads in the serverless platform, Locus~\cite{DBLP:conf/nsdi/PuVS19} models a mixture of cheap but slow storage with expensive but fast storage. It makes a cost-performance trade-off to choose the most appropriate configuration variable and shuffle implementation. Middleware Zion~\cite{DBLP:conf/middleware/SampeALP17} enables a data-driven serverless computing model for stateless functions. It optimizes the elasticity and resource contention by injecting computations into data pipelines and running on dataflows in a scalable manner. 

Due to the data-shipping architecture of serverless applications, current works usually focus on designing a more elastic serverless storage, enhancing the data locality to ease the I/O contention of function communication on the DB-side. However, due to the potential heterogeneity of different functions, the uncertainty above still makes these technologies in practice challenging.

\subsection{\textbf{Specialized Queue}}
In various implementations of the serverless system, the Queue is acquiescently integrated into the System Orchestration layer, which passes messages between different system components. 
For instance, Apache Kafka, serves as a distributed message streaming platform that allows applications to write and subscribe to messages across different hosts. 
% The responsibility of the Queue depends on the execution order with the controller, which can be categorized into \textbf{Node queue} and \textbf{Function queue}.

\textbf{\textit{Interact with the controller by node queue and function queue.}}

The function queue can send messages between the controller and functions.
% If a queue is integrated ahead of the controller, it serves as the node queue for load balancing to schedule the functions to different nodes (e.g., a queue in the cluster manager).
In contrast, the node queue serves for load balancing to schedule the functions to different nodes (e.g., a queue in the cluster manager).
A representative of adopting function queue design is OpenWhisk~\cite{Openwhisk}.
% A topic has some producers that write messages to it and some consumers who subscribe to these messages. To increase parallelism, a topic is partitioned so that messages with the same consumer can be written to the same partition. 
When the OpenWhisk controller receives an invocation query, it decides which invoker should execute the instance and then sends it to the selected invoker via Kafka. To increase parallelism, it also leverages the topic partitioned so that messages with the same consumer can be written to the same partition. 
SAND \cite{DBLP:conf/usenix/AkkusCRSSBAH18} also follows this message queue design by introducing a two-level hierarchical message bus: a local message bus deployed on each host and a global message bus distributed across different hosts. 
A local message bus is partitioned into different message queues such that every function on the host subscribes to messages from this function queue. In this way, if a function and its successors are running on the same host, it can directly write its output into the local message queue subscribed by its successor. Otherwise, the output is written to the global message bus.
Other work dives into the shortcomings of the queue-based mechanism, which may lead to reduced performance and availability in the serverless context.
McGrath~\emph{et al.}~\cite{DBLP:conf/icdcsw/McGrathB17} propose to introduce the ``cold queue'' and the ``warm queue'' to assume different responsibility for function queries.
DORADO~\cite{DBLP:journals/jsa/NettoLCLS17} also uses shared memory to mediate communication and persist data. By such means, queries can be routed to any container that is replicated.

It is more convenient for developers to adopt scalable queues in serverless computing, as the scaling is delegated to cloud vendors. 
For example, Amazon Simple Queue Service provides scalability by processing each buffered invocation independently, scaling transparently to handle bursty loads without provisioning instructions.
%Nowadays, more and more cloud platforms are beginning to provide queue-less service, which means users can only pay for the messages queries write and subscribe to like they do in other serverless services. 
% For example, Amazon Simple Queue Service (SQS) \cite{aws:sqs} provides an API over which applications running in the cloud can write messages into and subscribe to messages from a queue.
% With the help of queue service, it is more convenient for developers to deploy applications in the cloud since they do not need to implement the message queue by themselves. 
The idea of the scalable queue also meets the requirements for serverless computing, such as pay-per-use, dependability, convenience, and flexibility.

\subsection{\textbf{API Gateway and Various Triggers}}
In serverless computing, instances are created on-demand, and invocations are not bound with a static address. When deploying the containers or VMs to the cluster, the system will dynamically assign addresses to services. In this case, containers on the same node in the default network can communicate by IP addresses, while containers across nodes need to allocate ports for forwarding. The dynamic port allocation raises the challenge to manage as it intensifies at scale. The API Gateway component can provide a unique entry point to ensure accurate services' addresses. When the queries join the API Gateway, the service registry is inquired, and the queries are forwarded to the available service instances according to the IP route. It should also consider the availability and reliability of the serverless system, for example, the lazy reaction for services incompatible due to the hardware heterogeneity.
% Meanwhile, a general API Gateway also maps HTTP parameters into input parameters required for function services. 

Meanwhile, serverless systems design various triggers to invoke functions in response to queries. 
A trigger defines how a function is invoked, and the binding rule represents a mapping between them. The trigger and binding rule together make up a probe for the detectable event and help avoid hardcoding access to other services~\cite{DBLP:conf/IEEEcloud/LeeSF18}.
Besides invoking an event-driven function, triggers can also provide a declarative way to connect data to the code (e.g., storage services).

There are four most popular triggers: HTTP, Queue, Timer, and Event.
HTTP trigger is widely used~\cite{DBLP:conf/usenix/ShahradFGCBCLTR20} to handle external invocations, by which a function can be easily invoked once an HTTP query arrives. 
% One example is Kubeless \cite{kubeless}, a framework simplifying the deployment of functions on Kubernetes, which enables users to create an HTTP trigger for a function invocation in Kubernetes. 
While HTTP trigger simplifies the external invocation for functions, it shows less efficiency in the case of internal invocations.
An alternative way to handle internal events is using Queue trigger, by which functions get triggered whenever an invocation enqueues.
For instance, Kubeless~\cite{kubeless} provides a Kafka-based queue trigger bond with a Kafka topic so that users can invoke the function by writing messages into the topic.   
Specific purposes also require more extensive triggers. For example, a timer trigger in Kubernetes can invoke a function periodically. 
It creates a CronJob \cite{k8s:cronjob} object, which is written in a Cron expression representing the set of invocation time, to schedule a job accordingly. 
An Event trigger invokes a function in response to an event, which is the atomic piece of information that describes something that happened in the system. A convincing example of such implementation is Triggerflow~\cite{triggerflow}, which maps a workflow by setting an event trigger in each edge.

\subsection{\textbf{Data Cache}}
\label{cache}
To ensure graceful performance in case that the workload bursts and reaches a hard limit in concurrency, the common practice among cloud-based applications are utilizing multiple levels of caches~\cite{DBLP:conf/fast/ArteagaCXS016}. 
% Even if the serverless system can auto-scale instances as workload changes, we still need to ensure caching service in case that the workload bursts and reaches a hard limit in concurrency.
Data caching can cut out unnecessary roundtrips for less response time when queries experience a full-end invocation path. 
One common idea is caching at API Gateway (e.g., caching solution for GET method~\cite{AWScache}), or caching the pages and only result in a storage I/O query (e.g., Amazon DAX~\cite{DAX} and Aurora~\cite{DBLP:conf/sigmod/VerbitskiGSBGMK17} for database caching).
Another common solution is to cache in the system~\cite{DBLP:conf/fast/WangZMARST0C20,DBLP:conf/osdi/RashmiCKSR16,DBLP:conf/sc/YuHWZL18,DBLP:conf/usenix/MahgoubSMKCB21}. It freezes maintainers from declaring inside the function by enabling the caching of static assets or large objects.
However, the cached data is only available in the ephemeral container, which makes sharing across all short-term instances of a function challenging. And this approach may not be as effective as it seems- the first invocation in every container will result in cache misses. 

\textbf{\textit{Image cache: on-demand loading and page sharing.}}

The simplest and popular method is to provide the image cache for accelerating. 
In a container-based serverless system, an image is composed of multiple layers and shared by numerous containers as needed.
When functions are invoked on a host for the first time, images need to be downloaded and cached locally. For example, 
Slacker~\cite{DBLP:conf/fast/HarterSLAA16} builds a Docker storage driver and uses block-level COW to implement snapshots in a VMstore. Docker images
are represented by VMstore’s read-only snapshots, and the pull and push operations only involve sharing snapshot IDs rather than large network transfers. It makes it possible for docker workers to fetch data lazily from shared storage as needed.
%lazily pulling images in stages, which makes it possible to share the same data in layers during different runs of the same image.
% The Unikernel snapshots in SEUSS~\cite{DBLP:conf/eurosys/CaddenUADKA20} caches the images by a snapshot stack to enable fast deployment. It applies the page sharing and significantly decreases the memory footprint of functions startup.
DADI~\cite{DBLP:conf/usenix/LiYDMLH20} also implements an image service merging a sequence of block-based layers, and it caches recently used data blocks adopting the overlay with a tree-structured design.
SEUSS~\cite{DBLP:conf/eurosys/CaddenUADKA20} factoring out common execution state shared in a snapshot stack, which expresses a lineage between snapshots. It uses CoW to capture into a snapshot only the pages that were modified for fast deployment. Furthermore, SEUSS proposes to use anticipatory optimization to reduce the number of written pages captured in each snapshot. 
% Several Cloud vendors nowadays have released such image services that improve over the current OCI image specification for faster image distribution and on-demanded image data loading, such as Alibaba Cloud Nydus~\cite{alitalk} and Microsoft Azure Teleport~\cite{teleport}.

\textbf{\textit{State cache: make the serverless applications stateful.}}

State cache can be further combined with an active database to enable function execution stateful. Bledi~\cite{DBLP:conf/osdi/ZhangCCAL20} extended from Olive~\cite{DBLP:conf/osdi/SettySLZCPR16} adopts the refined SSF (Stateful Serverless Function) instance with a built-in database. By saving a set of intent tables recording the SSF's state and function information, it provides a fault-tolerant workflow execution manner. SNF~\cite{DBLP:conf/cloud/SinghviKAB20} decouples the functions into computing units and state units, and relaxes the constraint of communication between cooperating units. When processing a subsequent flowlet in the same flow, the function's internal state is cached in the local memory. Then SNF can proactively replicate ephemeral state among compute units. Cloudburst~\cite{DBLP:journals/pvldb/SreekantiWLSGHT20} packs the local cache with a function executor in each instance, and periodically publishes a snapshot of cached keys to the key-value store. By such means, Cloudburst can enhance the data locality via physical colocation of mutable caches and enable state management with the remote auto-scaling key-value database.

\textbf{\textit{Checkpoint cache: enable functions with fault tolerance.}}

The demand for fault tolerance also inspires researchers to make relevant techniques applicable in serverless context, such as C/R-based~\cite{DBLP:conf/ic2e/LiK15,DBLP:conf/cloud/ZhangFPS20} and log-based~\cite{DBLP:conf/sosp/WangLNMMTS19,triggerflow}. One example of such implementation in serverless computing is AFT~\cite{DBLP:conf/eurosys/SreekantiWCGHF20}, which builds an interposition between a storage engine and a common serverless platform by providing atomic fault tolerance shim. It leverages the data cache and the shared storage to guarantee the isolation of atomic read, avoid storage lookups for frequent access, and prevent significant consistency anomalies. 

In addition to the implementations we discussed above, other caching mechanisms can be explored and integrated into any layer of our proposed serverless architecture. 
Of course, they are all recommended to follow the pay-as-go mode based on the used resource. 
In summary, data caching is still an essential and important component for higher flexibility and better performance. 

\subsection{\textbf{DevOps Tools}}
DevOps is a compound of development and operations, which improves collaboration and productivity. Since the responsibility of managing underlying resources and runtime environment are transferred to the cloud vendors in the serverless concept, developers only need to focus on the code logic. 
Operations teams are actually liberated from this process, where they are required to check, compile, pack images, and test deployment after developers submitting the code. There is a necessity for providing DevOps tools in the serverless system.
%This is the reason why we discuss the necessity for providing DevOps tools in the serverless system.
We agree with the pipeline implementation~\cite{DBLP:conf/profes/IvanovS18} to group the DevOps into CI (Continuous Integration), CD (Continuous Delivery), and CM (Continuous Monitoring) categories. 

CI refers to the continuous merging of developed code with others during software development while ensuring automatical validation and building. Not only should it make the operations within the function (e.g., Jenkins~\cite{jenkins} for integration test, Honeycomb~\cite{honeycomb} and SonarQube~\cite{sonarqube} for inspection test),
but also focus on availability between functions in an application (e.g., monitoring and debugging tool IOpipe for workflows~\cite{IOpipe}).

CD requires that the serverless system can automatically update the application instances with the old version while keeping the services still available. Nowadays the common solutions are \textbf{Rolling-update}, \textbf{RED-Black} (aka Blue-Green), and \textbf{Canary} deployment, which are adopted by Kubernetes~\cite{cncfk8s,istio}.
Essentially, these deployment strategies share the same mechanism to keep part of instances with old version serving and then gradually replace them with new ones. The difference between them is the complexity of rollback to the last available instances.

CM enables the DevOps team to receive feedback on the problems and errors during the above steps in time. It also provides a visualized interface for monitoring applications' runtime behavior and resource utilization.
Actually, the feedback tool is already implemented in most CI/CD tools and runs through the whole DevOps lifecycle of the application. By enriching the CM component, visualizing activities, and resource monitoring~\cite{khan2017key}, users can better understand their services running in the cloud.

The concept of DevOps in serverless typically appears more frequently in production scenarios, and most platforms provide various such tools to ensure good compatibility and flexibility. However, the introduction of DevOps may bring about new vulnerabilities threatening the security of instances, and serverless research should also provide more substantial support for detecting vulnerable containers~\cite{DBLP:conf/icdcsw/BilaDKWY17,khan2017key,tak2017understanding,VulnerabilityAdvisor}.

\section{Performance and Comparison}
\label{sec:performance}
This section first summarizes the performance with different VMs/containers, language runtime, and resource limits in serverless computing. Then, we analyze the current production serverless systems to show the preferences.

\subsection{Performance Analysis}
\begin{figure}
  \centering
  \includegraphics[width=.82\linewidth]{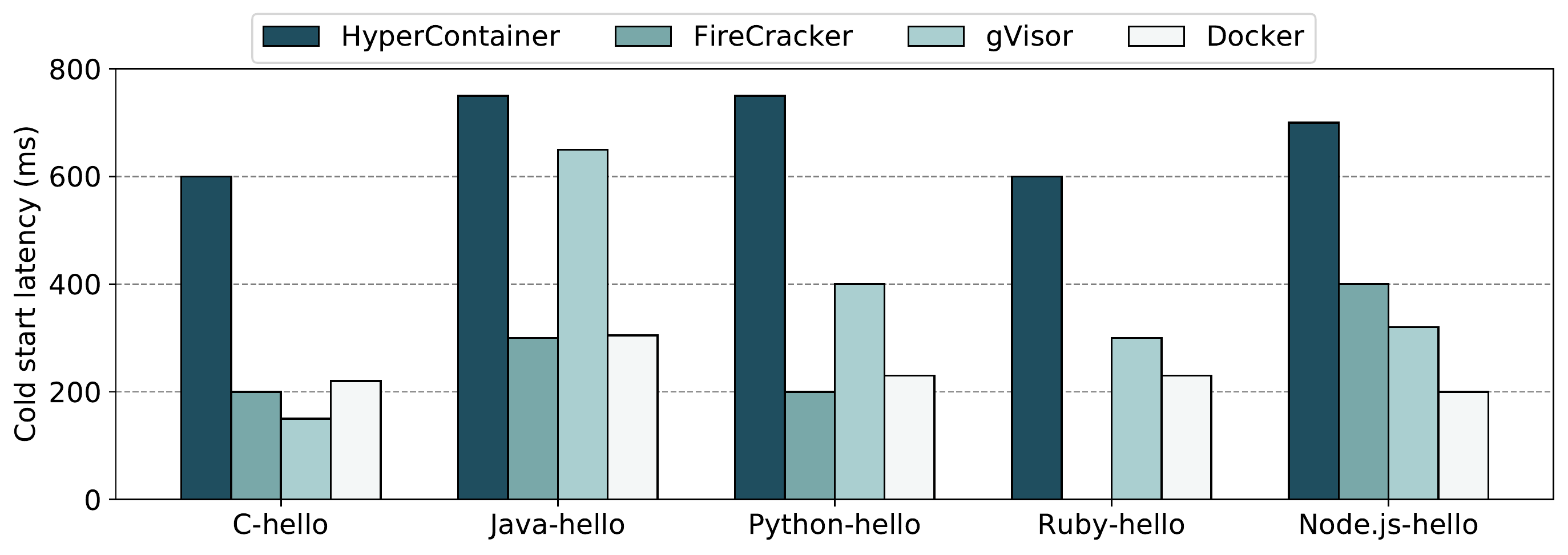}
  \vspace{-2mm}
  \caption{\label{fig:coldstart:runtime} Cold startup latency under different runtimes~\cite{DBLP:conf/asplos/DuYXZYQWC20}.}
\end{figure}

The runtime within the instance built from different virtualization technologies can exhibit different cold startup performances. 
%At present, the most popular runtimes are HyperContainer, FireCracker, gVisor, and Docker. There are also some other optimized virtualization startup techniques based on the above VM/container runtimes. 
Besides, the language runtime is another factor that can seriously affect the cold startup latency. For example, evaluations in Catalyzer~\cite{DBLP:conf/asplos/DuYXZYQWC20} show the cold startup latencies with different VM/container and language runtimes.
As shown in Figure~\ref{fig:coldstart:runtime}, HyperContainer introduces the highest cold startup latency with various language runtimes.  Process-based Docker runtime certainly performs significantly better than others. Generally speaking, the interpreted languages (e.g., Python) incur a higher initial cost and make startup times up to 10$\times$ slower~\cite{DBLP:conf/usenix/AkkusCRSSBAH18,DBLP:conf/usenix/OakesYZHHAA18} than the compiled languages (e.g., C) when cold startups.

However, according to the performance tests from Jackson~\cite{DBLP:conf/ucc/JacksonC18}, which measure the startup and execution latency of different language runtimes, the performance of compiled and interpreted language runtime also depends on the platform. For example, the cold startup latency of .NET C$\#$ on AWS Lambda is higher than that of Node.js, while it is the opposite on Azure Functions. It is because that Azure Functions would provide better support for C$\#$ based on its core technology .NET for Microsoft, and implements by running on windows containers rather than the open-source .NET CLR (Common Language Runtime) based on Linux containers.

\begin{figure}
  \centering
  \includegraphics[width=.84\linewidth]{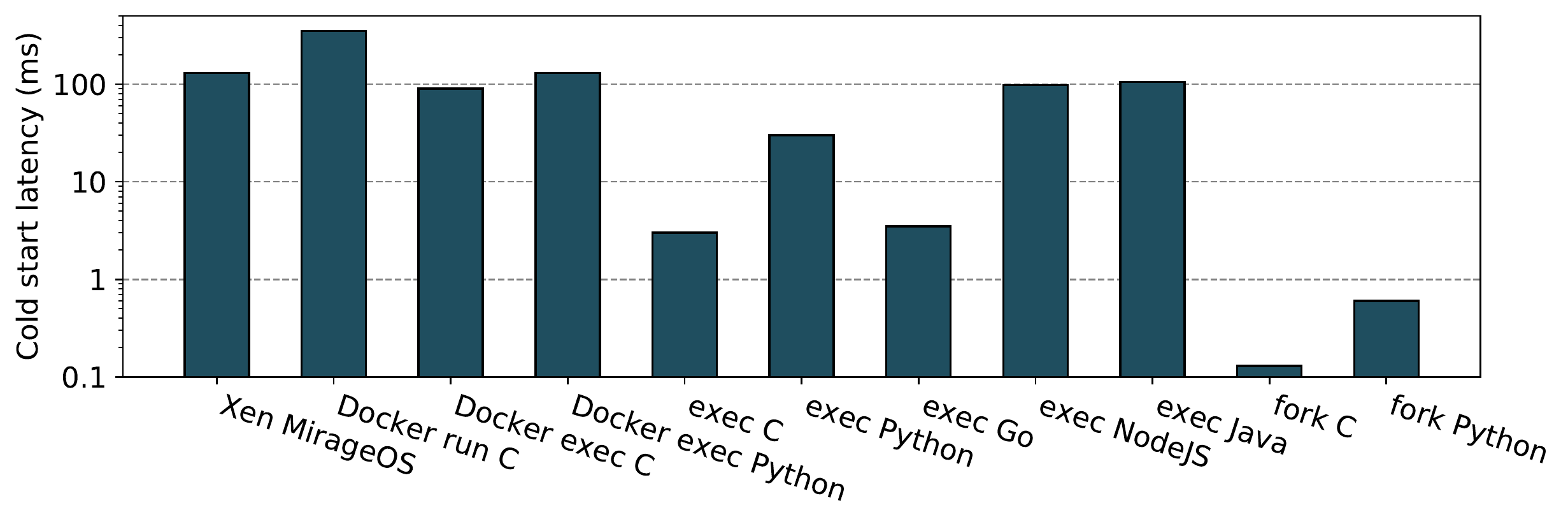}
  \vspace{-3mm}
  \caption{\label{fig:coldstart:method}Cold startup latency under different isolation mechanisms~\cite{DBLP:conf/usenix/AkkusCRSSBAH18}.}
\end{figure}

Besides the cold startup analysis of different language runtimes, SAND~\cite{DBLP:conf/usenix/AkkusCRSSBAH18} also measures several sandbox isolation mechanisms for function executions, and we show their results in Figure~\ref{fig:coldstart:method}. Native executions (exec and fork) are the fastest methods, while Unikernel (Xen MirageOS) performs similar to using a Docker container. Regardless of the recycled user code in memory in the paused container, using the Docker client interface to start a warm function (Docker exec C) is much faster than a cold startup (Docker run C). 

\begin{figure}
  \centering
  \includegraphics[width=.82\linewidth]{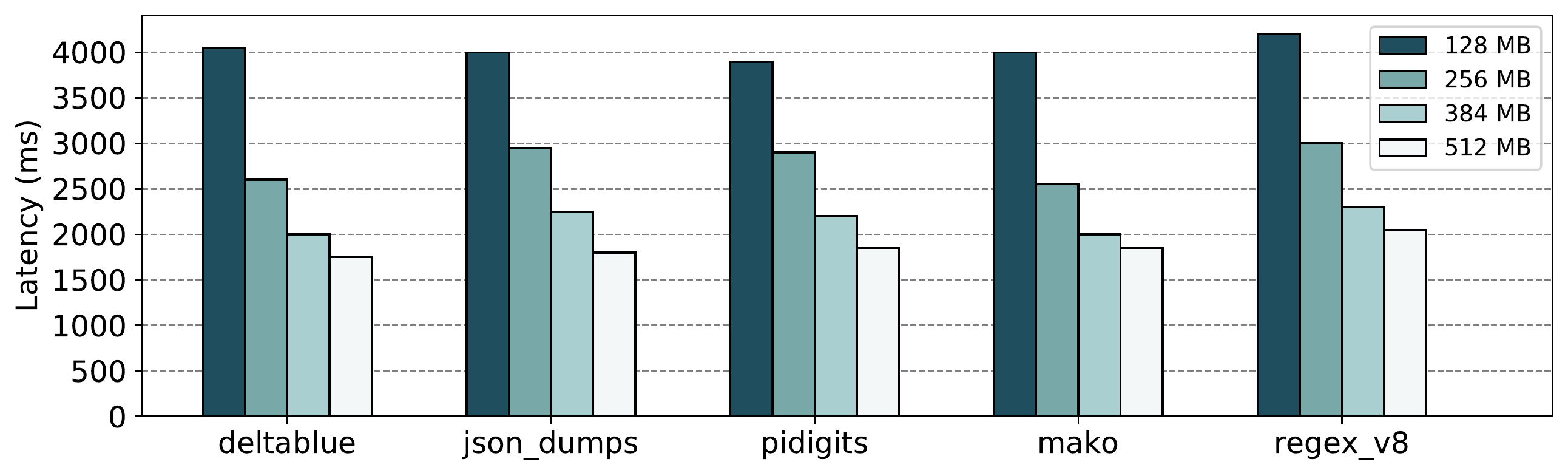}
  \vspace{-2mm}
  \caption{\label{fig:coldstart:memory}Cold startup latency with different container memory limits~\cite{DBLP:conf/micro/ShahradBW19}.}
\end{figure}

Another significant factor that slows down the cold startups for the container-based serverless system is the memory limit. The performance evaluation about memory allocation~\cite{DBLP:conf/micro/ShahradBW19} is shown in Figure~\ref{fig:coldstart:memory}. The cold startup latency of each microbenchmark function increases as stepping to smaller memory limits. We can also see that there is a significant decrease in container startup latency when stepping from 128MB to 256MB. And larger memory limit results in less obvious optimizations without the reasonable regime of marginal increases. It also explains why most serverless systems set 256MB as the default memory limit of the function container.

As the supplement of the above factors that affect serverless cold startup performance, Shahrad~\emph{et al.}~\cite{DBLP:conf/micro/ShahradBW19} explore other factors that may affect the function cold startup and execution time, such as MKPI (mispredictions per kilo-instruction), LLC (Last-level Cache) size, and memory bandwidth. Firstly, they find that a longer execution time usually appears with noticeably lower branch MKPI within a function. It is easy for us to understand that functions with short execution time spend most of the time on language runtime startup, and thus the branch predictor outputs more miss when stay trained. Secondly, the LLC size is not a significant factor that affects the cold startup latency and execution time. Higher LLC size cannot bring in better performance for serverless function execution because of the insensitivity. Only when the LLC size is very small (e.g., less than 2M) will it become a bottleneck for the function execution and cold startup. Cloud vendors usually set a default LLC size and pre-profile in the serverless system to avoid serious performance degradation. 
%Thirdly, container image loading will heavily occupy the memory bandwidth during the function cold startup, and it will decrease with function workload increases. It is because frequently invoked functions spend more on code execution instead of images keeping loaded into memory. 
BabelFish~\cite{DBLP:conf/isca/SkarlatosDGKT20} also finds that lazy page table management can result in heavy TLB stress in a containerized environment. To avoid redundant kernel jobs produced in the process of page table management, they try to share translations across containers in the TLB and page tables.

%\begin{figure}[h]
%  \centering
%  \includegraphics[width=.9\linewidth]{figures/coldstart_latency_manufacture.pdf}
%  \caption{\label{fig:coldstart:manufacture}Overall latency of serverless function under different manufacture's platforms.}
%\end{figure}

\subsection{Production Comparison}

With more attempts to enable the rapid development of cloud-native applications, 
Wang~\emph{et al.}~\cite{DBLP:conf/usenix/WangLZRS18} evaluate the performance of three commercial serverless platforms by invoking measurement functions with stepwise memory limits to collect various system-level metrics. 
Lee~\emph{et al.}~\cite{DBLP:conf/IEEEcloud/LeeSF18} also gives a detailed comparison between Amazon Lambda, Google Functions, Microsoft Azure Functions, and IBM OpenWhisk. They demonstrate the differences in terms of throughput, network bandwidth, I/O capacity, and computing performance. 
% All of Lee's experiments are evaluated using 1.5GB memory allocation (IBM OpenWhisk with 512MB), which is more instructional for deploying serverless applications. 
Based on these experiments, we summarize the metrics in Table~\ref{tab:diff}.

\begin{table*}
  \centering
  \scriptsize
  \caption{Comparing metrics of four serverless vendors~\cite{DBLP:conf/IEEEcloud/LeeSF18,DBLP:conf/icdcsw/McGrathB17} (``CCI'' means the concurrent invocations).}
  \label{tab:diff}
  \renewcommand\arraystretch{1.3}
  \resizebox{\textwidth}{!}{
    \begin{tabular}{c|cccc}
      \toprule
      Item & Amazon Lambda & Google Functions & Microsoft Azure Functions & IBM OpenWhisk\\ 
      \midrule
      GFLOPS per function & 19.63 & 4.35 & 2.15 & 3.19 \\
      TFLOPS in 3000 & 66.30 & 13.04 & 7.94 & 12.30 \\
      Throughput of 1-5 CCI & 20-55TPS & 1-25TPS & 60-150TPS & 1TPS \\
      Throughput of 2000 CCI & 400TPS & 40TPS & 120TPS & 210TPS \\
      CCI Tail latency & best & superior & worst & inferior \\
      CI/CD performance & best & fail frequently & long latency & balanced \\
      Read/Write (1-100 CCI) & 153/83 MB/s - 93/39.5 MB/s & 56/9.5 MB/s - 54/3.5 MB/s & 424/44 MB/s - NA & 68/8 MB/s- 34/0.5 MB/s\\
      File I/O (1-100 CCI) & 2-3.5 second & 10-30 & 3.5-NA & 15-60 \\
      Object I/O (1-100 CCI) & 1.3-2.4 second & 5-8 & 12-NA & 1-30 \\
      Trigger Throughput & 55-25-860 (HTTP-Object-DB) & 20-25-NA & 145-250-NA  & 50-NA-40 \\
      Language Runtime overhead & balanced 0.05s avg & (-0.06) 0.22s (+0.1) & (-0.02) 0.22s (+0.03) & (-0.02) 0.17s (+0.02) \\
      Dependencies overhead & (-0.5) 1.1s (+0.2) avg & (-0.5) 1.9s (+0.4) & (-1.3) 3.4s (NA) & NA \\
      Maximum Memory & 3008MB & 2048MB & 1536MB & 512MB \\
      Execution Timeout & 5 minutes & 9 minutes & 10 minutes & 5 minutes \\
      Price per Memory & \$0.0000166/GB-s & \$0.0000165/GB-s & \$0.0000016/GB-s &  \$0.000017/GB-s \\
      Price per Execution & \$0.2 per 1M & \$0.4 per 1M & \$0.2 per 1M & NA \\
      Free Tier & First 1 M Exec & First 2 M Exec & First 1 M Exec & Free Exec/40,000GB-s \\
      Idle instance lifetime & 5-7 min & 15 min & Mostly 20-30 minutes & default 15 min\\
      \bottomrule
    \end{tabular}
  }
\end{table*}

From this table, we can get a glimpse of their respective strengths and weaknesses. For example, AWS Lambda shows higher capacity and throughput of concurrent function invocations, however, performing poorly in trigger throughput. Microsoft Azure Functions enable fast read and write speed when queries are invoked in sequence, and show relatively higher function cold startup latency. Undoubtedly, all cloud vendors are aware of the challenges in serverless architecture and are actively optimizing the function invocation performance and relevant BaaS bottlenecks.

\section{Other Key Limitations and Challenges}
\label{sec:challenges}
% several research challenges of the intra-layer literature. In this section, we list 
% Sandbox isolation, serving as the basic technology in the Virtualization layer to support serverless, is already discussed and surveyed~\cite{DBLP:conf/sqamia/BargmannT19,DBLP:conf/ic2e/BachiegaSBS18}.
The limitations of the current works in each layer and challenges are already discussed in the corresponding sections. This section will highlight other key limitations and challenges in the Encapsule, Orchestration, and Coordination layer, respectively, as an orthogonal supplement. We refer readers for more detailed and focused discussions on the Virtualization Layer in other survey~\cite{DBLP:conf/sqamia/BargmannT19,DBLP:conf/ic2e/BachiegaSBS18}.

\subsection{Stateless within Encapsule Layer}
An essential feature of serverless is that the service is loaded and executed on-demand rather than deployed in a long-term running instance. To prevent a large number of instances from occupying memory resources, the serverless controller sets an instance lifetime to recycle them automatically. Because short life-span functions within the application are no longer associated with a particular instance or server, each query processed cannot be guaranteed to be invoked by the same function instance. In other words, the application's state cannot and will not be kept on the resumed instance~\cite{DBLP:journals/fgcs/KeshavarzianSS19}.
% To simplify function management, develo pers need to follow several technical and several architectural constraints in serverless-based functions.
%To prevent a large number of instances in the system from occupying memory resources, the serverless controller sets a execution timeout to recycle instances automatically. 
%Currently, most serverless systems limit the execution time of the function to reduce resource usage in the cluster. 
% Therefore, the serverless application is more suitable for services with a short execution time. 
% Thus serverless architecture developers have no choice but to follow the strangler pattern by decomposing a large service into more different microservices and then build fine-grained node interconnections. 
The stateless nature weakens the generality of the serverless architecture, limiting its scope to stateless applications, such as Web applications, IoT (Internet of Things), media processing, etc. 
% It should be noticed that we deny serverless architecture cannot be adopted in stateful scenarios, as the relevant works have been discussed in Section~\ref{cache} (see~\cite{DBLP:conf/osdi/ZhangCCAL20,DBLP:journals/pvldb/SreekantiWLSGHT20,DBLP:conf/cloud/SinghviKAB20}). 
Undoubtedly, the extension of stateful serverless architecture (see~\cite{DBLP:conf/osdi/ZhangCCAL20,DBLP:journals/pvldb/SreekantiWLSGHT20,DBLP:conf/cloud/SinghviKAB20} in Section~\ref{cache}) by saving state in object storage or key-value stores fails to provide low latency and high throughput simultaneously, makes it inferior to regular sticky sessions as IaaS or PaaS does. 

\subsection{Memory Fragmentation within Orchestration Layer}

\begin{figure}
  \centering
  \subfigure[Allocation fragementation]{
  \includegraphics[width=.32\textwidth]{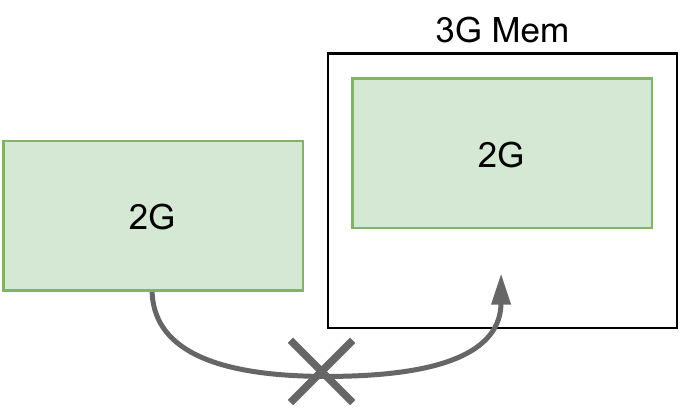}
  }
  \hspace{6mm}
  \subfigure[Scheduling fragementation]{
  \includegraphics[width=.46\textwidth]{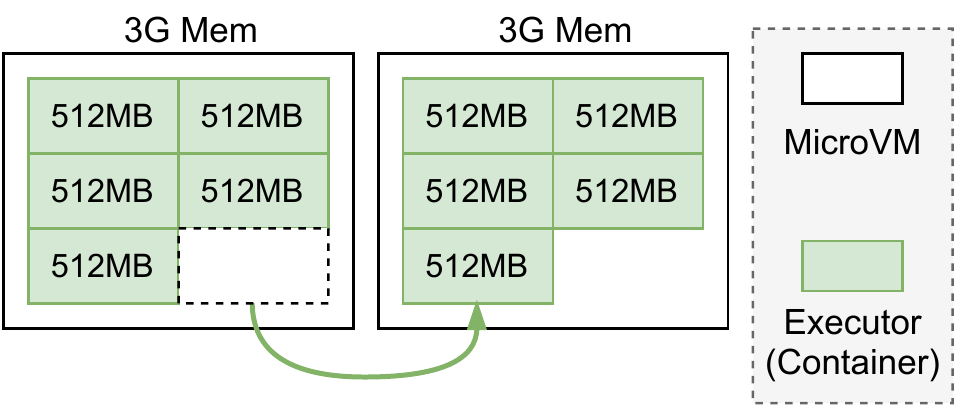}
  }
  \vspace{-2mm}
  \caption{\label{fig:fragmentation}Two scenarios where the memory fragmentation arises.}
\end{figure}

In the serverless architecture where multiple tenants co-exist, concurrent invocations are either processed in multiple containers and experience undesired cold startups in each one, or executed concurrently in one single container (e.g., OpenFaaS and OpenLambda). In the former, a container is allowed to execute only one invocation at a time for performance isolation. In this case, the memory footprint of massive sidecars prevents serverless containers from achieving high-density deployment and improved resource utilization~\cite{DBLP:conf/nsdi/AgacheBILNPP20}. The key to this challenge is slimming and condensing the container runtime by deduplication within the VMM and guest kernel, such as sharing the page cache across different instances on the host. 
% With the rising of secure containers, memory fragmentation becomes another significant challenges for researchers. Because of the co-location of multiple tenants in the cluster, the key to improve the resource utilization is to achieve a high density of containers deployment. 
In the latter, memory fragmentation becomes a top priority. 
%For example, containers are scheduled in microVMs as shown in Figure~\ref{fig:fragmentation}. 
The figure~\ref{fig:fragmentation} depicts two common scenarios where memory fragmentation may arise. 
% Take the containers scaling in microVMs as an example, there are two common scenarios that may cause memory fragmentation, called allocation fragementation and scheduling fragementation. 
% As shown in Figure~\ref{fig:fragmentation}, 
Allocation fragmentation is usually due to the improper provision of a microVM. Function executors can not fully utilize the memory allocated. Scheduling fragmentation is inevitable and usually caused by instance-level load balancing strategy when auto-scaling with workload changes. 
%An efficient measure of container deployment density is allocated memory-per-container.
% By measuring of Allocated Memory-Per-Container, the concerned challenge calls for an efficient methodology to instruct how to search for the global optimal solution of memory fragmentation. 
Since the serverless emergence, challenges remain in further achieving an efficient methodology in instructing how to search for a high-density container deployment solution. 
% The lack of research on memory fragmentation in secure containers prevents serverless platforms from high-density container deployment. 
% An efficient measure of container deployment density is Allocated Memory-Per-Container. Given the diversity of each function workload, there is a conflict between increasing the deployment density of a single function and the cluster. 

% \subsection{\textbf{Tradeoff between Performance and Security}}
% Under the serverless architecture, users cannot access the host on which the application is deployed. Applications of different users, and different applications under the same user, may share underlying resources at runtime. Although most serverless platforms provide timing for the function execution, it is still challenging to dive into the details with the profiling or APM (Application Performance Management) tools due to security concerns. Especially for some applications with high-security requirements, the platform has to use stronger virtualization mechanisms to avoid security risks.  There is no doubt that the loading and reloading of function runtimes during the cold startup will introduce longer latency and slow down the execution performance even more. The current compromise is to reuse the prewarmed instances for functions that are sensitive to latency or invoked frequently.

% Meanwhile, due to the closed-source of the containers/VMs in the production serverless platform, developers lack equipped testing and integration tools for the local execution environment. Only a few of the available tools are actually used. The access latency will increase as more components are added to the serverless system. Even though it can be optimized by using proprietary network protocols, RPC invocations, remaking data formats, or by scheduling instances on the same host to reduce latency, these approaches undoubtedly require a tradeoff between performance and security by each Cloud vendor.

\subsection{API and Benchmark Lock-in within Coordination Layer}
% The different VM/container adopted by cloud vendors also make the official language runtimes supported by each serverless platform different.
% We investigate the popular serverless platforms and list the language runtimes they support in Table~\ref{tab:LR} below.
% OpenWhisk, Fission and Kubeless are known as popular open-sourced serverless system for researchers, and the others are all well-known production serverless platforms. 
% As shown in the table, the most popular language runtimes are already supported by them. However, there are still some language runtimes that cannot provide for developers. 
When people talk about serverless vendor lock-in, they are concerned about the portability of functions. However, the real point of this problem depends on the API from other services rather than the function itself. Though some efforts such as Apex~\cite{Apex} and Sparta~\cite{Sparta} allow users to deploy functions to serverless platforms in languages that are not supported natively, the BaaS services from different platforms and their API definitions are still different. The challenge with API lock-in is derived from the tight coupling between the user functions and other BaaS components, which can add difficulty to the code migration between different FaaS platforms.
% For example, an application workflow contains two nodes: a function A written in PHP and a function B written in Go. It has only  options to deploy due to the language runtimes limitation.
% Worse still, applications with PHP and Ruby runtime functions can only be deployed in IBM Functions. 
% Even though there are some efforts such as Apex~\cite{Apex} and Sparta~\cite{Sparta} allow users to deploy functions to serverless platforms in languages that are not supported natively, the BaaS services from different platforms and their API definitions are also different. It is still the key challenge in serverless API lock-in, which can add difficulty to the code migration between different FaaS vendors.

The over-simplified benchmark is another problem with API lock-in. Easy-to-build micro-benchmarks are over-emphasized and used in 75\% of the current works~\cite{DBLP:journals/jss/Scheuner020}. We call for the establishment and open-source of cross-platform real-world application benchmarks besides scientific workflows~\cite{DBLP:journals/corr/abs-1810-09679,DBLP:journals/fgcs/MalawskiGZBF20,DBLP:conf/cloud/JonasPVSR17}. However, when decomposing a large service into different functions and then build fine-grained node interconnections, the complexity of the application architecture makes the grading of the function challenging to guide and determine.

% \begin{table*}
%   \centering
%   \footnotesize
%   \caption{The official supported language runtimes of different serverless platforms.}
%   \label{tab:LR}
%   \begin{tabular}{cccccccccc}
%     \toprule
%     Serverless Platforms & Node.js & Java & Python & Ruby & Go & .NET & PowerShell & JavaScript & PHP \\ 
%     \midrule
%     OpenWhisk  & \Checkmark & \Checkmark & \Checkmark & \Checkmark & \Checkmark & \Checkmark  & \Checkmark & \Checkmark & \Checkmark \\ 
%     Fission  & \Checkmark & \Checkmark & \Checkmark & \Checkmark & \Checkmark & \Checkmark  & \Checkmark &  & \Checkmark \\ 
%     Kubeless  & \Checkmark & \Checkmark & \Checkmark & \Checkmark & \Checkmark & \Checkmark  &  &  & \Checkmark \\ 
%     Azure Functions  & \Checkmark & \Checkmark & \Checkmark &  &  & \Checkmark  & \Checkmark & \Checkmark & \\ 
%     Google Functions  & \Checkmark & \Checkmark & \Checkmark & \Checkmark & \Checkmark & \Checkmark &  &  & \\ 
%     Function Compute  & \Checkmark & \Checkmark & \Checkmark &  & \Checkmark & \Checkmark &  &  & \Checkmark \\ 
%     AWS Lambda  & \Checkmark & \Checkmark & \Checkmark & \Checkmark & \Checkmark & \Checkmark & \Checkmark &  &  \\ 
%     IBM Functions  & \Checkmark & \Checkmark & \Checkmark & \Checkmark & \Checkmark & \Checkmark &  &  & \Checkmark \\
%     FunctionGraph  & \Checkmark & \Checkmark & \Checkmark &  & \Checkmark & \Checkmark &  &  & \Checkmark \\
%     \bottomrule
%   \end{tabular}
% \end{table*}

\section{Opportunities in Serverless Computing}
\label{sec:opportunities}
At last, we discuss some future opportunities that serverless computing faces and give some preliminary, constructive explorations to solutions.
%% TODO: Intro
\subsection{\textbf{Application-Level Optimization}}
\label{sec:opportunities:1}

Application-level optimization requires coordinating between different functions within the application instead of focusing on each general function. Complex interconnections like data dependence and caller-callee relation may conceal between functions. Future works could achieve application-level optimization in two ways: workflow support and workflow scheduling.
% workflow support
Workflow support means general support for the inter-connection among functions. We think the following supports are necessary:
\begin{itemize}
  \item \textbf{\textit{Better storage.}} In some cases, functions need to exchange large ephemeral files with each other. If we register intermediate storage, transferring between storage and functions will take up most of the I/O resource and significantly slow down the response. This consequence is exacerbated in the serverless workflow scenario. Therefore, better storage demands higher priority for metadata exchange between functions within an application.
  \item \textbf{\textit{Higher parallelism capacity.}} In an example of video processing, multiple recoding instances can be invoked simultaneously in a MapReduce way to speed up the transcoding. Distinctly, there is great potential in parallelism to optimize end-to-end latency. However, higher parallelism is hard to implement due to the considerations on resource utilization management in physical nodes. If a serverless system could provide superior parallelism with sustainable resource overhead, it can further empower users. For example, a serverless system allows multiple queries to be invoked concurrently within an instance with a guaranteed QoS, or optimizes the guest kernel, container runtime, and cgroups to achieve lighter virtualization in high concurrency and density scenarios.
  % \item \textit{Error Handling.} Errors happen from time to time, and it's important to handle errors in time. In our ML example, errors may happen when user upload an invalid image, and application should tell the user to upload a new image. However, due to isolation requirement in serverless, error message cannot be noticed by another function. It makes error handling code distubuted over every function, not easy to manage. Therefore, serverless system should provide better mechanism to handle errors in function.
\end{itemize}
% workflow scheduling
Workflow scheduling drives a scheduling strategy that takes functions' interconnection into account. We think the following considerations are missing in current works:
\begin{itemize}
  \item \textbf{\textit{Caller-callee relation.}} Caller-callee relation is common in a complex application. Usually, the callee will be invoked after the caller finishes, as Figure~\ref{fig:dataflow}(a) shows. It gives researchers an opportunity to explore: the system can prewarm function instances and execute them in advance with partial data by the dataflow architecture. As shown in Figure~\ref{fig:dataflow}(b), Function B, C can start execution earlier while Function A is not complete, thanks to the data dependency rather than function state dependency. In the case of providing an optimized interface with dataflow canonical patterns and applying directly to functions, cloud vendors could enable an application to achieve higher parallelism and lower response latency via a data pipeline. 

  \begin{figure}
    \centering
    \subfigure[Controlflow]{
    \includegraphics[width=.44\textwidth]{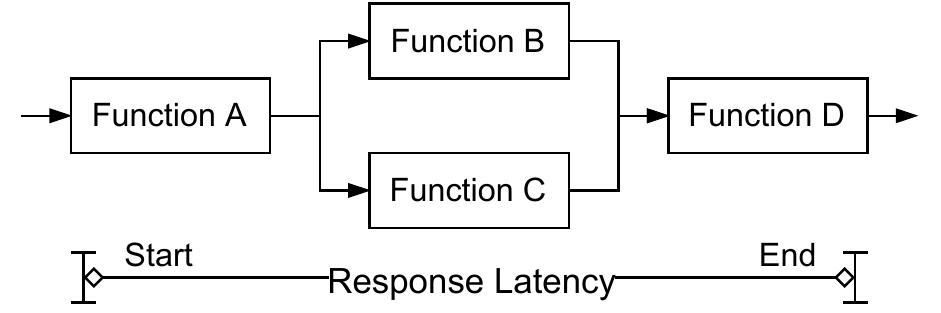}
    }
    \hspace{4mm}
    \subfigure[Dataflow]{
    \includegraphics[width=.42\textwidth]{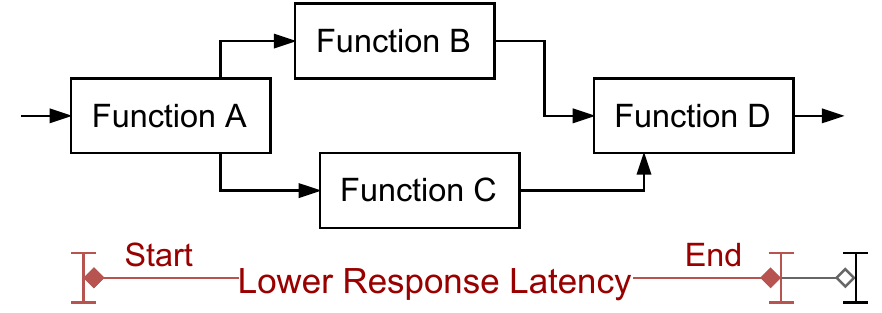}
    }
    \vspace{-2mm}
    \caption{\label{fig:dataflow}The dataflow architecture for serverless workflow.}
  \end{figure}

  \item \textbf{\textit{Data locality.}} We have mentioned that metadata exchange may continually happen between functions in an application. If two functions with data dependency are scheduled on the same physical node, the data transmission can be significantly reduced by middleware. However, the current serverless system is a data-shipping architecture, which sends data to the code node to parse instead of sending code to the data node. Thus, on the one hand, a serverless system cannot guarantee that the data stored and the workers scheduled are just in the same physical node.
  On the other hand, frequent code transferring should also be avoided due to security and privacy concerns. Improving data locality can effectively reform the application design from a data-shipping architecture into a code-shipping one~\cite{DBLP:conf/cidr/HellersteinFGSS19}.
\end{itemize}

\subsection{\textbf{Robust Performance of Cold Startup Alleviation}}
% eliminating coldstart latency is important
% coldstart may happen in vary situations, so robustness in coldstart latency elimination is important
% current work rely on coldstart prediction, which need history data
% some functions have few history data and are sensitive to coldstart latency

% the amount of resource greatly affect coldstart latency
% coldstart with limited resource happens
% coldstart with more resource, and execute with less?

Current works usually use predictive methods to reduce cold startups, while they all require functions' historical traces or system-level metrics. By predicting in the near future, the system will enlarge the container pool or prewarm template containers. Nonetheless, it is impractical for each function to collect enough data and build an accurate prediction model. Like Shahrad~\cite{DBLP:conf/usenix/ShahradFGCBCLTR20} shows in the Azure trace, about 40\% and 30\% of registered functions and applications, respectively, are invoked less than ten times daily. This fact also makes it more challenging to collect system-level information periodically for such kind of services.  
%For example, by regularly checking the invocation frequency for each function, a serverless system uses an LRU-based method to update the template containers with the cached images or packages. 
For example, an LRU-based template can maximize the cache hits for hotspot functions startup, whereas cold startups of non-hot functions can not benefit from the cache updating at the system level. 
The current compromise to this discrepancy is to use a reserved container pool for functions, in spite of a massive waste of resources.
%The current solution to deal with this challenge adopted by production serverless platforms is to use a reserved container pool for each function, although it leads to a massive waste of resources.

It is crucial to explore the warm-up strategy with strong robustness to performance, especially for functions that are sporadically triggered or sensitive to latency. It requires the serverless controller and load balancer to be more general enough to alleviate cold startups or reduce the performance degradation. They may make decisions based on the information inside the functions, such as the service category, the environment libraries used, and the context diagram, for cold startup prediction and alleviation. For example, a serverless system can build shared images and template containers for functions within the same category, or pack the functions with similar environment configurations and implement more fine-grained inside isolation mechanisms.

\subsection{\textbf{Accelerators in Serverless}}

Accelerators like GPUs and FPGAs are widely used in many applications such as databases~\cite{10.1145/1376616.1376670,DBLP:conf/cidr/ChenCBHHWC20} and graph processing~\cite{10.1109/TPDS.2013.111,10.1145/3431920.3439290}. They can significantly speed up the processing of specific tasks, like image processing and machine learning applications. To satisfy the demand for accelerators, cloud vendors furnish accelerators in IaaS (e.g., AWS EC2 P4 and F1 instance) and SaaS (e.g., AWS SageMaker) manner. However, the inflexibility of such accelerators impedes the instantiation in serverless computing. This circumstance leads to two obstacles: (1) it makes the usage of accelerators less convenient and flexible in the cloud; (2) it limits the range of applications that serverless can support. We think a multiplexing accelerator in serverless is the key to solving these obstacles. For example, some works~\cite{mark, NARANJO202032} integrate GPUs into serverless systems, and BlastFunction~\cite{blastfunction} makes FPGAs available in serverless. However, the current works are still insufficient. We think future research can focus on the following points:
\begin{itemize}
  \item \textbf{\textit{Accelerator-aware scheduling}}. Accelerators can also be considered a resource in serverless systems, except they have more irreplaceable features than others. Latency-aware scheduling and on-demanding scaling is more expensive on accelerators, stimulating the serverless controller to treat accelerators distinctively. In such a situation, the scheduling strategy should be more conservative when scheduling multiple tasks on one accelerator. 
  \item \textbf{\textit{Accelerator virtualization}}. Virtualization is an essential technology applied in a serverless system. It is used to fulfill runtime environment management, resource isolation, and high security. However, serverless accelerator schemes are not explored insofar as CPU virtualizations. It makes accelerators embarrassing to be integrated into the serverless system. Therefore, to better support accelerators in serverless, accelerator virtualization should be further explored.
  \item \textbf{\textit{Automatic batching}}. Accelerators usually have strong I/O bandwidth restrictions. Batching queries is a common operation to conquer these restrictions and make full use of accelerators' computation ability. However, the batching operation will introduce redundancy into end-to-end latency. Therefore, a serverless batching strategy that balances utilization and latency should be investigated in future research.
\end{itemize}

\section{Conclusion}
\label{sec:conclusion}
The rapid development of the cloud-native concept inspires developers to reorganize cloud applications into microservices. Elastic serverless computing becomes the best practice for these microservices. This survey explicates and reviews the fundamental aspects of serverless computing, and provides a comprehensive depiction of four-layered design architecture: Virtualization, Encapsule, System Orchestration, and
System Coordination layers. We elaborate on the responsibility and significance of each layer, enumerate relevant works, and give practical implications when adopting these state-of-the-art techniques. Serverless computing is still in its infancy, and the potential remains sealed in forthcoming years. 

\bibliographystyle{ACM-Reference-Format}
\bibliography{sample-base}


\begin{thebibliography}{156}

%%% the Web address.

\ifx \showCODEN    \undefined \def \showCODEN     #1{\unskip}     \fi
\ifx \showDOI      \undefined \def \showDOI       #1{#1}\fi
\ifx \showISBNx    \undefined \def \showISBNx     #1{\unskip}     \fi
\ifx \showISBNxiii \undefined \def \showISBNxiii  #1{\unskip}     \fi
\ifx \showISSN     \undefined \def \showISSN      #1{\unskip}     \fi
\ifx \showLCCN     \undefined \def \showLCCN      #1{\unskip}     \fi
\ifx \shownote     \undefined \def \shownote      #1{#1}          \fi
\ifx \showarticletitle \undefined \def \showarticletitle #1{#1}   \fi
\ifx \showURL      \undefined \def \showURL       {\relax}        \fi
% invisible to TeX
\providecommand\bibfield[2]{#2}
\providecommand\bibinfo[2]{#2}
\providecommand\natexlab[1]{#1}
\providecommand\showeprint[2][]{arXiv:#2}

\bibitem[\protect\citeauthoryear{Adhikari, Amgoth, and Srirama}{Adhikari
  et~al\mbox{.}}{2019}]%
        {DBLP:journals/csur/AdhikariAS19}
\bibfield{author}{\bibinfo{person}{Mainak Adhikari}, \bibinfo{person}{Tarachand
  Amgoth}, {and} \bibinfo{person}{Satish~Narayana Srirama}.}
  \bibinfo{year}{2019}\natexlab{}.
\newblock \showarticletitle{A Survey on Scheduling Strategies for Workflows in
  Cloud Environment and Emerging Trends}.
\newblock \bibinfo{journal}{\emph{{ACM} Comput. Surv.}} \bibinfo{volume}{52},
  \bibinfo{number}{4} (\bibinfo{year}{2019}), \bibinfo{pages}{68:1--68:36}.
\newblock
\urldef\tempurl%
\url{https://doi.org/10.1145/3325097}
\showDOI{\tempurl}


\bibitem[\protect\citeauthoryear{Adzic and Chatley}{Adzic and Chatley}{2017}]%
        {DBLP:conf/sigsoft/AdzicC17}
\bibfield{author}{\bibinfo{person}{Gojko Adzic} {and} \bibinfo{person}{Robert
  Chatley}.} \bibinfo{year}{2017}\natexlab{}.
\newblock \showarticletitle{Serverless computing: economic and architectural
  impact}. In \bibinfo{booktitle}{\emph{Proceedings of the 2017 11th Joint
  Meeting on Foundations of Software Engineering, {ESEC/FSE} 2017, Paderborn,
  Germany, September 4-8, 2017}}, \bibfield{editor}{\bibinfo{person}{Eric
  Bodden}, \bibinfo{person}{Wilhelm Sch{\"{a}}fer}, \bibinfo{person}{Arie van
  Deursen}, {and} \bibinfo{person}{Andrea Zisman}} (Eds.).
  \bibinfo{publisher}{{ACM}}, \bibinfo{pages}{884--889}.
\newblock
\urldef\tempurl%
\url{https://doi.org/10.1145/3106237.3117767}
\showDOI{\tempurl}


\bibitem[\protect\citeauthoryear{Agache, Brooker, Iordache, and Liguori}{Agache
  et~al\mbox{.}}{2020}]%
        {DBLP:conf/nsdi/AgacheBILNPP20}
\bibfield{author}{\bibinfo{person}{Alexandru Agache}, \bibinfo{person}{Marc
  Brooker}, \bibinfo{person}{Alexandra Iordache}, {and}
  \bibinfo{person}{Anthony Liguori}.} \bibinfo{year}{2020}\natexlab{}.
\newblock \showarticletitle{Firecracker: Lightweight Virtualization for
  Serverless Applications}. In \bibinfo{booktitle}{\emph{17th {USENIX}
  Symposium on Networked Systems Design and Implementation, {NSDI} 2020, Santa
  Clara, CA, USA, February 25-27, 2020}},
  \bibfield{editor}{\bibinfo{person}{Ranjita Bhagwan} {and}
  \bibinfo{person}{George Porter}} (Eds.). \bibinfo{publisher}{{USENIX}
  Association}, \bibinfo{pages}{419--434}.
\newblock
\urldef\tempurl%
\url{https://www.usenix.org/conference/nsdi20/presentation/agache}
\showURL{%
\tempurl}


\bibitem[\protect\citeauthoryear{Akkus, Chen, Rimac, Stein, Satzke, Beck,
  Aditya, and Hilt}{Akkus et~al\mbox{.}}{2018}]%
        {DBLP:conf/usenix/AkkusCRSSBAH18}
\bibfield{author}{\bibinfo{person}{Istemi~Ekin Akkus},
  \bibinfo{person}{Ruichuan Chen}, \bibinfo{person}{Ivica Rimac},
  \bibinfo{person}{Manuel Stein}, \bibinfo{person}{Klaus Satzke},
  \bibinfo{person}{Andre Beck}, \bibinfo{person}{Paarijaat Aditya}, {and}
  \bibinfo{person}{Volker Hilt}.} \bibinfo{year}{2018}\natexlab{}.
\newblock \showarticletitle{{SAND:} Towards High-Performance Serverless
  Computing}. In \bibinfo{booktitle}{\emph{2018 {USENIX} Annual Technical
  Conference, {USENIX} {ATC} 2018, Boston, MA, USA, July 11-13, 2018}},
  \bibfield{editor}{\bibinfo{person}{Haryadi~S. Gunawi} {and}
  \bibinfo{person}{Benjamin Reed}} (Eds.). \bibinfo{publisher}{{USENIX}
  Association}, \bibinfo{pages}{923--935}.
\newblock
\urldef\tempurl%
\url{https://www.usenix.org/conference/atc18/presentation/akkus}
\showURL{%
\tempurl}


\bibitem[\protect\citeauthoryear{Alipourfard, Liu, and Chen}{Alipourfard
  et~al\mbox{.}}{2017}]%
        {DBLP:conf/nsdi/AlipourfardLCVY17}
\bibfield{author}{\bibinfo{person}{Omid Alipourfard},
  \bibinfo{person}{Hongqiang~Harry Liu}, {and} \bibinfo{person}{Jianshu Chen}.}
  \bibinfo{year}{2017}\natexlab{}.
\newblock \showarticletitle{CherryPick: Adaptively Unearthing the Best Cloud
  Configurations for Big Data Analytics}. In \bibinfo{booktitle}{\emph{14th
  {USENIX} Symposium on Networked Systems Design and Implementation, {NSDI}
  2017, Boston, MA, USA, March 27-29, 2017}},
  \bibfield{editor}{\bibinfo{person}{Aditya Akella} {and} \bibinfo{person}{Jon
  Howell}} (Eds.). \bibinfo{publisher}{{USENIX} Association},
  \bibinfo{pages}{469--482}.
\newblock
\urldef\tempurl%
\url{https://www.usenix.org/conference/nsdi17/technical-sessions/presentation/alipourfard}
\showURL{%
\tempurl}


\bibitem[\protect\citeauthoryear{Amazon}{Amazon}{2021}]%
        {AWScache}
\bibfield{author}{\bibinfo{person}{Amazon}.} \bibinfo{year}{2021}\natexlab{}.
\newblock \bibinfo{title}{Enabling API caching to enhance responsiveness in
  AWS}.
\newblock
\newblock
\urldef\tempurl%
\url{https://docs.aws.amazon.com/apigateway/latest/developerguide/api-gateway-caching.html}
\showURL{%
\tempurl}


\bibitem[\protect\citeauthoryear{Amazon DynamoDB Accelerator (DAX): A fully
  managed, highly available, in-memory cache service}{Amazon DynamoDB
  Accelerator (DAX): A fully managed, highly available, in-memory cache
  service}{2021}]%
        {DAX}
Amazon DynamoDB Accelerator (DAX): A fully managed, highly available, in-memory
  cache service \bibinfo{year}{2021}\natexlab{}.
\newblock
\newblock
\urldef\tempurl%
\url{https://aws.amazon.com/dynamodb/dax/}
\showURL{%
\tempurl}


\bibitem[\protect\citeauthoryear{Anwar, Mohamed, Tarasov, Littley, and
  Rupprecht}{Anwar et~al\mbox{.}}{2018}]%
        {DBLP:conf/fast/AnwarMTLRCZNWLH18}
\bibfield{author}{\bibinfo{person}{Ali Anwar}, \bibinfo{person}{Mohamed
  Mohamed}, \bibinfo{person}{Vasily Tarasov}, \bibinfo{person}{Michael
  Littley}, {and} \bibinfo{person}{Lukas Rupprecht}.}
  \bibinfo{year}{2018}\natexlab{}.
\newblock \showarticletitle{Improving Docker Registry Design Based on
  Production Workload Analysis}. In \bibinfo{booktitle}{\emph{16th {USENIX}
  Conference on File and Storage Technologies, {FAST} 2018, Oakland, CA, USA,
  February 12-15, 2018}}, \bibfield{editor}{\bibinfo{person}{Nitin Agrawal}
  {and} \bibinfo{person}{Raju Rangaswami}} (Eds.). \bibinfo{publisher}{{USENIX}
  Association}, \bibinfo{pages}{265--278}.
\newblock
\urldef\tempurl%
\url{https://www.usenix.org/conference/fast18/presentation/anwar}
\showURL{%
\tempurl}


\bibitem[\protect\citeauthoryear{Ao, Izhikevich, Voelker, and Porter}{Ao
  et~al\mbox{.}}{2018}]%
        {DBLP:conf/cloud/AoIVP18}
\bibfield{author}{\bibinfo{person}{Lixiang Ao}, \bibinfo{person}{Liz
  Izhikevich}, \bibinfo{person}{Geoffrey~M. Voelker}, {and}
  \bibinfo{person}{George Porter}.} \bibinfo{year}{2018}\natexlab{}.
\newblock \showarticletitle{Sprocket: {A} Serverless Video Processing
  Framework}. In \bibinfo{booktitle}{\emph{Proceedings of the {ACM} Symposium
  on Cloud Computing, SoCC 2018, Carlsbad, CA, USA, October 11-13, 2018}}.
  \bibinfo{publisher}{{ACM}}, \bibinfo{pages}{263--274}.
\newblock
\urldef\tempurl%
\url{https://doi.org/10.1145/3267809.3267815}
\showDOI{\tempurl}


\bibitem[\protect\citeauthoryear{Apex: Serverless Architecture}{Apex:
  Serverless Architecture}{2021}]%
        {Apex}
Apex: Serverless Architecture \bibinfo{year}{2021}\natexlab{}.
\newblock
\newblock
\urldef\tempurl%
\url{https://apex.sh/}
\showURL{%
\tempurl}


\bibitem[\protect\citeauthoryear{Armant, Cauwer, Brown, and O'Sullivan}{Armant
  et~al\mbox{.}}{2018}]%
        {DBLP:journals/fgcs/ArmantCBO18}
\bibfield{author}{\bibinfo{person}{Vincent Armant}, \bibinfo{person}{Milan~De
  Cauwer}, \bibinfo{person}{Kenneth~N. Brown}, {and} \bibinfo{person}{Barry
  O'Sullivan}.} \bibinfo{year}{2018}\natexlab{}.
\newblock \showarticletitle{Semi-online task assignment policies for workload
  consolidation in cloud computing systems}.
\newblock \bibinfo{journal}{\emph{Future Gener. Comput. Syst.}}
  \bibinfo{volume}{82} (\bibinfo{year}{2018}), \bibinfo{pages}{89--103}.
\newblock
\urldef\tempurl%
\url{https://doi.org/10.1016/j.future.2017.12.035}
\showDOI{\tempurl}


\bibitem[\protect\citeauthoryear{Arteaga, Cabrera, Xu, Sundararaman, and
  Zhao}{Arteaga et~al\mbox{.}}{2016}]%
        {DBLP:conf/fast/ArteagaCXS016}
\bibfield{author}{\bibinfo{person}{Dulcardo Arteaga}, \bibinfo{person}{Jorge
  Cabrera}, \bibinfo{person}{Jing Xu}, \bibinfo{person}{Swaminathan
  Sundararaman}, {and} \bibinfo{person}{Ming Zhao}.}
  \bibinfo{year}{2016}\natexlab{}.
\newblock \showarticletitle{CloudCache: On-demand Flash Cache Management for
  Cloud Computing}. In \bibinfo{booktitle}{\emph{14th {USENIX} Conference on
  File and Storage Technologies, {FAST} 2016, Santa Clara, CA, USA, February
  22-25, 2016}}, \bibfield{editor}{\bibinfo{person}{Angela~Demke Brown} {and}
  \bibinfo{person}{Florentina~I. Popovici}} (Eds.).
  \bibinfo{publisher}{{USENIX} Association}, \bibinfo{pages}{355--369}.
\newblock
\urldef\tempurl%
\url{https://www.usenix.org/conference/fast16/technical-sessions/presentation/arteaga}
\showURL{%
\tempurl}


\bibitem[\protect\citeauthoryear{Bachiega, Souza, Bruschi, and do~Rocio
  Senger~de Souza}{Bachiega et~al\mbox{.}}{2018}]%
        {DBLP:conf/ic2e/BachiegaSBS18}
\bibfield{author}{\bibinfo{person}{Naylor~G. Bachiega}, \bibinfo{person}{Paulo
  S.~L. Souza}, \bibinfo{person}{Sarita~Mazzini Bruschi}, {and}
  \bibinfo{person}{Simone do~Rocio Senger~de Souza}.}
  \bibinfo{year}{2018}\natexlab{}.
\newblock \showarticletitle{Container-Based Performance Evaluation: {A} Survey
  and Challenges}. In \bibinfo{booktitle}{\emph{2018 {IEEE} International
  Conference on Cloud Engineering, {IC2E} 2018, Orlando, FL, USA, April 17-20,
  2018}}, \bibfield{editor}{\bibinfo{person}{Abhishek Chandra},
  \bibinfo{person}{Jie Li}, \bibinfo{person}{Ying Cai}, {and}
  \bibinfo{person}{Tian Guo}} (Eds.). \bibinfo{publisher}{{IEEE} Computer
  Society}, \bibinfo{pages}{398--403}.
\newblock
\urldef\tempurl%
\url{https://doi.org/10.1109/IC2E.2018.00075}
\showDOI{\tempurl}


\bibitem[\protect\citeauthoryear{{Bacis}, {Brondolin}, and
  {Santambrogio}}{{Bacis} et~al\mbox{.}}{2020}]%
        {blastfunction}
\bibfield{author}{\bibinfo{person}{M. {Bacis}}, \bibinfo{person}{R.
  {Brondolin}}, {and} \bibinfo{person}{M.~D. {Santambrogio}}.}
  \bibinfo{year}{2020}\natexlab{}.
\newblock \showarticletitle{BlastFunction: an FPGA-as-a-Service system for
  Accelerated Serverless Computing}. In \bibinfo{booktitle}{\emph{2020 Design,
  Automation Test in Europe Conference Exhibition (DATE)}}.
  \bibinfo{pages}{852--857}.
\newblock
\urldef\tempurl%
\url{https://doi.org/10.23919/DATE48585.2020.9116333}
\showDOI{\tempurl}


\bibitem[\protect\citeauthoryear{Baldini, Castro, Chang, Cheng, Fink, Ishakian,
  Mitchell, Muthusamy, Rabbah, Slominski, et~al\mbox{.}}{Baldini
  et~al\mbox{.}}{2017a}]%
        {baldini2017serverless}
\bibfield{author}{\bibinfo{person}{Ioana Baldini}, \bibinfo{person}{Paul
  Castro}, \bibinfo{person}{Kerry Chang}, \bibinfo{person}{Perry Cheng},
  \bibinfo{person}{Stephen Fink}, \bibinfo{person}{Vatche Ishakian},
  \bibinfo{person}{Nick Mitchell}, \bibinfo{person}{Vinod Muthusamy},
  \bibinfo{person}{Rodric Rabbah}, \bibinfo{person}{Aleksander Slominski},
  {et~al\mbox{.}}} \bibinfo{year}{2017}\natexlab{a}.
\newblock \showarticletitle{Serverless computing: Current trends and open
  problems}.
\newblock In \bibinfo{booktitle}{\emph{Research Advances in Cloud Computing}}.
  \bibinfo{publisher}{Springer}, \bibinfo{pages}{1--20}.
\newblock


\bibitem[\protect\citeauthoryear{Baldini, Cheng, Fink, and Mitchell}{Baldini
  et~al\mbox{.}}{2017b}]%
        {DBLP:conf/oopsla/BaldiniCFMMRST17}
\bibfield{author}{\bibinfo{person}{Ioana Baldini}, \bibinfo{person}{Perry
  Cheng}, \bibinfo{person}{Stephen~J. Fink}, {and} \bibinfo{person}{Nick
  Mitchell}.} \bibinfo{year}{2017}\natexlab{b}.
\newblock \showarticletitle{The serverless trilemma: function composition for
  serverless computing}. In \bibinfo{booktitle}{\emph{Proceedings of the 2017
  {ACM} {SIGPLAN} International Symposium on New Ideas, New Paradigms, and
  Reflections on Programming and Software, Onward! 2017, Vancouver, BC, Canada,
  October 23 - 27, 2017}}. \bibinfo{publisher}{{ACM}},
  \bibinfo{pages}{89--103}.
\newblock
\urldef\tempurl%
\url{https://doi.org/10.1145/3133850.3133855}
\showDOI{\tempurl}


\bibitem[\protect\citeauthoryear{Balis}{Balis}{2016}]%
        {Hyperflow}
\bibfield{author}{\bibinfo{person}{Bartosz Balis}.}
  \bibinfo{year}{2016}\natexlab{}.
\newblock \showarticletitle{HyperFlow: {A} model of computation, programming
  approach and enactment engine for complex distributed workflows}.
\newblock \bibinfo{journal}{\emph{Future Gener. Comput. Syst.}}
  \bibinfo{volume}{55} (\bibinfo{year}{2016}), \bibinfo{pages}{147--162}.
\newblock
\urldef\tempurl%
\url{https://doi.org/10.1016/j.future.2015.08.015}
\showDOI{\tempurl}


\bibitem[\protect\citeauthoryear{Bargmann and Tropmann{-}Frick}{Bargmann and
  Tropmann{-}Frick}{2019}]%
        {DBLP:conf/sqamia/BargmannT19}
\bibfield{author}{\bibinfo{person}{Christian Bargmann} {and}
  \bibinfo{person}{Marina Tropmann{-}Frick}.} \bibinfo{year}{2019}\natexlab{}.
\newblock \showarticletitle{A Survey On Secure Container Isolation Approaches
  for Multi-Tenant Container Workloads and Serverless Computing}. In
  \bibinfo{booktitle}{\emph{Proceedings of the Eighth Workshop on Software
  Quality Analysis, Monitoring, Improvement, and Applications, {SQAMIA} 2019,
  Ohrid, North Macedonia, September 22-25, 2019}}
  \emph{(\bibinfo{series}{{CEUR} Workshop Proceedings},
  Vol.~\bibinfo{volume}{2508})}, \bibfield{editor}{\bibinfo{person}{Zoran
  Budimac} {and} \bibinfo{person}{Bojana Koteska}} (Eds.).
  \bibinfo{publisher}{CEUR-WS.org}.
\newblock
\urldef\tempurl%
\url{http://ceur-ws.org/Vol-2508/paper-bar.pdf}
\showURL{%
\tempurl}


\bibitem[\protect\citeauthoryear{Barlev, Basil, Kohanim, Peleg, Regev, and
  Shulman{-}Peleg}{Barlev et~al\mbox{.}}{2016}]%
        {DBLP:journals/ibmrd/BarlevBKPRS16}
\bibfield{author}{\bibinfo{person}{S. Barlev}, \bibinfo{person}{Z. Basil},
  \bibinfo{person}{S. Kohanim}, \bibinfo{person}{R. Peleg}, \bibinfo{person}{S.
  Regev}, {and} \bibinfo{person}{Alexandra Shulman{-}Peleg}.}
  \bibinfo{year}{2016}\natexlab{}.
\newblock \showarticletitle{Secure yet usable: Protecting servers and Linux
  containers}.
\newblock \bibinfo{journal}{\emph{{IBM} J. Res. Dev.}} \bibinfo{volume}{60},
  \bibinfo{number}{4} (\bibinfo{year}{2016}), \bibinfo{pages}{12}.
\newblock
\urldef\tempurl%
\url{https://doi.org/10.1147/JRD.2016.2574138}
\showDOI{\tempurl}


\bibitem[\protect\citeauthoryear{Bermbach, Karakaya, and Buchholz}{Bermbach
  et~al\mbox{.}}{2020}]%
        {DBLP:conf/sac/BermbachKB20}
\bibfield{author}{\bibinfo{person}{David Bermbach},
  \bibinfo{person}{Ahmet{-}Serdar Karakaya}, {and} \bibinfo{person}{Simon
  Buchholz}.} \bibinfo{year}{2020}\natexlab{}.
\newblock \showarticletitle{Using application knowledge to reduce cold starts
  in FaaS services}. In \bibinfo{booktitle}{\emph{{SAC} '20: The 35th
  {ACM/SIGAPP} Symposium on Applied Computing, online event, [Brno, Czech
  Republic], March 30 - April 3, 2020}},
  \bibfield{editor}{\bibinfo{person}{Chih{-}Cheng Hung},
  \bibinfo{person}{Tom{\'{a}}s Cern{\'{y}}}, \bibinfo{person}{Dongwan Shin},
  {and} \bibinfo{person}{Alessio Bechini}} (Eds.). \bibinfo{publisher}{{ACM}},
  \bibinfo{pages}{134--143}.
\newblock
\urldef\tempurl%
\url{https://doi.org/10.1145/3341105.3373909}
\showDOI{\tempurl}


\bibitem[\protect\citeauthoryear{Bessai, Youcef, Oulamara, Godart, and
  Nurcan}{Bessai et~al\mbox{.}}{2012}]%
        {DBLP:conf/IEEEcloud/BessaiYOGN12}
\bibfield{author}{\bibinfo{person}{Kahina Bessai}, \bibinfo{person}{Samir
  Youcef}, \bibinfo{person}{Ammar Oulamara}, \bibinfo{person}{Claude Godart},
  {and} \bibinfo{person}{Selmin Nurcan}.} \bibinfo{year}{2012}\natexlab{}.
\newblock \showarticletitle{Bi-criteria Workflow Tasks Allocation and
  Scheduling in Cloud Computing Environments}. In
  \bibinfo{booktitle}{\emph{2012 {IEEE} Fifth International Conference on Cloud
  Computing, Honolulu, HI, USA, June 24-29, 2012}}. \bibinfo{publisher}{{IEEE}
  Computer Society}, \bibinfo{pages}{638--645}.
\newblock
\urldef\tempurl%
\url{https://doi.org/10.1109/CLOUD.2012.83}
\showDOI{\tempurl}


\bibitem[\protect\citeauthoryear{Bila, Dettori, Kanso, Watanabe, and
  Youssef}{Bila et~al\mbox{.}}{2017}]%
        {DBLP:conf/icdcsw/BilaDKWY17}
\bibfield{author}{\bibinfo{person}{Nilton Bila}, \bibinfo{person}{Paolo
  Dettori}, \bibinfo{person}{Ali Kanso}, \bibinfo{person}{Yuji Watanabe}, {and}
  \bibinfo{person}{Alaa Youssef}.} \bibinfo{year}{2017}\natexlab{}.
\newblock \showarticletitle{Leveraging the Serverless Architecture for Securing
  Linux Containers}. In \bibinfo{booktitle}{\emph{37th {IEEE} International
  Conference on Distributed Computing Systems Workshops, {ICDCS} Workshops
  2017, Atlanta, GA, USA, June 5-8, 2017}},
  \bibfield{editor}{\bibinfo{person}{Aibek Musaev},
  \bibinfo{person}{Jo{\~{a}}o~Eduardo Ferreira}, {and} \bibinfo{person}{Teruo
  Higashino}} (Eds.). \bibinfo{publisher}{{IEEE} Computer Society},
  \bibinfo{pages}{401--404}.
\newblock
\urldef\tempurl%
\url{https://doi.org/10.1109/ICDCSW.2017.66}
\showDOI{\tempurl}


\bibitem[\protect\citeauthoryear{Boucher, Kalia, Andersen, and
  Kaminsky}{Boucher et~al\mbox{.}}{2018}]%
        {DBLP:conf/usenix/BoucherKAK18}
\bibfield{author}{\bibinfo{person}{Sol Boucher}, \bibinfo{person}{Anuj Kalia},
  \bibinfo{person}{David~G. Andersen}, {and} \bibinfo{person}{Michael
  Kaminsky}.} \bibinfo{year}{2018}\natexlab{}.
\newblock \showarticletitle{Putting the "Micro" Back in Microservice}. In
  \bibinfo{booktitle}{\emph{2018 {USENIX} Annual Technical Conference, {USENIX}
  {ATC} 2018, Boston, MA, USA, July 11-13, 2018}},
  \bibfield{editor}{\bibinfo{person}{Haryadi~S. Gunawi} {and}
  \bibinfo{person}{Benjamin Reed}} (Eds.). \bibinfo{publisher}{{USENIX}
  Association}, \bibinfo{pages}{645--650}.
\newblock
\urldef\tempurl%
\url{https://www.usenix.org/conference/atc18/presentation/boucher}
\showURL{%
\tempurl}


\bibitem[\protect\citeauthoryear{Boyd}{Boyd}{2021}]%
        {IOpipe}
\bibfield{author}{\bibinfo{person}{Mark Boyd}.}
  \bibinfo{year}{2021}\natexlab{}.
\newblock \bibinfo{title}{"Serverless: IOpipe Launches a Monitoring Tool for
  AWS Lambda"}.
\newblock
\newblock
\urldef\tempurl%
\url{https://thenewstack.io/iopipe-launches-lambda-monitoring-tool-aws-summit/}
\showURL{%
\tempurl}


\bibitem[\protect\citeauthoryear{BUDINSKY}{BUDINSKY}{2021}]%
        {istio}
\bibfield{author}{\bibinfo{person}{FRANK BUDINSKY}.}
  \bibinfo{year}{2021}\natexlab{}.
\newblock \bibinfo{title}{"Canary Deployments using Istio" about the Red-Black
  and the Blue-green deployment}.
\newblock
\newblock
\urldef\tempurl%
\url{https://istio.io/latest/blog/2017/0.1-canary/}
\showURL{%
\tempurl}


\bibitem[\protect\citeauthoryear{Buyya, Srirama, Casale, and Calheiros}{Buyya
  et~al\mbox{.}}{2019}]%
        {DBLP:journals/csur/BuyyaSCCSVGJVNT19}
\bibfield{author}{\bibinfo{person}{Rajkumar Buyya},
  \bibinfo{person}{Satish~Narayana Srirama}, \bibinfo{person}{Giuliano Casale},
  {and} \bibinfo{person}{Rodrigo~N. Calheiros}.}
  \bibinfo{year}{2019}\natexlab{}.
\newblock \showarticletitle{A Manifesto for Future Generation Cloud Computing:
  Research Directions for the Next Decade}.
\newblock \bibinfo{journal}{\emph{{ACM} Comput. Surv.}} \bibinfo{volume}{51},
  \bibinfo{number}{5} (\bibinfo{year}{2019}), \bibinfo{pages}{105:1--105:38}.
\newblock
\urldef\tempurl%
\url{https://doi.org/10.1145/3241737}
\showDOI{\tempurl}


\bibitem[\protect\citeauthoryear{Buyya, Yeo, Venugopal, Broberg, and
  Brandic}{Buyya et~al\mbox{.}}{2009}]%
        {DBLP:journals/fgcs/BuyyaYVBB09}
\bibfield{author}{\bibinfo{person}{Rajkumar Buyya}, \bibinfo{person}{Chee~Shin
  Yeo}, \bibinfo{person}{Srikumar Venugopal}, \bibinfo{person}{James Broberg},
  {and} \bibinfo{person}{Ivona Brandic}.} \bibinfo{year}{2009}\natexlab{}.
\newblock \showarticletitle{Cloud computing and emerging {IT} platforms:
  Vision, hype, and reality for delivering computing as the 5th utility}.
\newblock \bibinfo{journal}{\emph{Future Gener. Comput. Syst.}}
  \bibinfo{volume}{25}, \bibinfo{number}{6} (\bibinfo{year}{2009}),
  \bibinfo{pages}{599--616}.
\newblock
\urldef\tempurl%
\url{https://doi.org/10.1016/j.future.2008.12.001}
\showDOI{\tempurl}


\bibitem[\protect\citeauthoryear{Cadden, Unger, Awad, and Dong}{Cadden
  et~al\mbox{.}}{2020}]%
        {DBLP:conf/eurosys/CaddenUADKA20}
\bibfield{author}{\bibinfo{person}{James Cadden}, \bibinfo{person}{Thomas
  Unger}, \bibinfo{person}{Yara Awad}, {and} \bibinfo{person}{Han Dong}.}
  \bibinfo{year}{2020}\natexlab{}.
\newblock \showarticletitle{{SEUSS:} skip redundant paths to make serverless
  fast}. In \bibinfo{booktitle}{\emph{EuroSys '20: Fifteenth EuroSys Conference
  2020, Heraklion, Greece, April 27-30, 2020}}. \bibinfo{publisher}{{ACM}},
  \bibinfo{pages}{32:1--32:15}.
\newblock
\urldef\tempurl%
\url{https://doi.org/10.1145/3342195.3392698}
\showDOI{\tempurl}


\bibitem[\protect\citeauthoryear{Carver, Zhang, Wang, Anwar, Wu, and
  Cheng}{Carver et~al\mbox{.}}{2020}]%
        {10.1145/3419111.3421286}
\bibfield{author}{\bibinfo{person}{Benjamin Carver}, \bibinfo{person}{Jingyuan
  Zhang}, \bibinfo{person}{Ao Wang}, \bibinfo{person}{Ali Anwar},
  \bibinfo{person}{Panruo Wu}, {and} \bibinfo{person}{Yue Cheng}.}
  \bibinfo{year}{2020}\natexlab{}.
\newblock \showarticletitle{Wukong: A Scalable and Locality-Enhanced Framework
  for Serverless Parallel Computing}. In \bibinfo{booktitle}{\emph{Proceedings
  of the 11th ACM Symposium on Cloud Computing}} (Virtual Event, USA)
  \emph{(\bibinfo{series}{SoCC '20})}. \bibinfo{publisher}{Association for
  Computing Machinery}, \bibinfo{address}{New York, NY, USA},
  \bibinfo{pages}{1–15}.
\newblock
\showISBNx{9781450381376}
\urldef\tempurl%
\url{https://doi.org/10.1145/3419111.3421286}
\showDOI{\tempurl}


\bibitem[\protect\citeauthoryear{Carver, Zhang, Wang, and Cheng}{Carver
  et~al\mbox{.}}{2019}]%
        {DBLP:journals/corr/abs-1910-05896}
\bibfield{author}{\bibinfo{person}{Benjamin Carver}, \bibinfo{person}{Jingyuan
  Zhang}, \bibinfo{person}{Ao Wang}, {and} \bibinfo{person}{Yue Cheng}.}
  \bibinfo{year}{2019}\natexlab{}.
\newblock \showarticletitle{In Search of a Fast and Efficient Serverless {DAG}
  Engine}.
\newblock \bibinfo{journal}{\emph{CoRR}}  \bibinfo{volume}{abs/1910.05896}
  (\bibinfo{year}{2019}).
\newblock
\showeprint[arxiv]{1910.05896}
\urldef\tempurl%
\url{http://arxiv.org/abs/1910.05896}
\showURL{%
\tempurl}


\bibitem[\protect\citeauthoryear{Casas, Taheri, Ranjan, and Zomaya}{Casas
  et~al\mbox{.}}{2017}]%
        {DBLP:journals/tjs/CasasTRZ17}
\bibfield{author}{\bibinfo{person}{Israel Casas}, \bibinfo{person}{Javid
  Taheri}, \bibinfo{person}{Rajiv Ranjan}, {and} \bibinfo{person}{Albert~Y.
  Zomaya}.} \bibinfo{year}{2017}\natexlab{}.
\newblock \showarticletitle{{PSO-DS:} a scheduling engine for scientific
  workflow managers}.
\newblock \bibinfo{journal}{\emph{J. Supercomput.}} \bibinfo{volume}{73},
  \bibinfo{number}{9} (\bibinfo{year}{2017}), \bibinfo{pages}{3924--3947}.
\newblock
\urldef\tempurl%
\url{https://doi.org/10.1007/s11227-017-1992-z}
\showDOI{\tempurl}


\bibitem[\protect\citeauthoryear{Chang, Yang, Yeh, Lin, and Jeng}{Chang
  et~al\mbox{.}}{2017}]%
        {DBLP:conf/globecom/ChangYYLJ17}
\bibfield{author}{\bibinfo{person}{Chia{-}Chen Chang},
  \bibinfo{person}{Shun{-}Ren Yang}, \bibinfo{person}{En{-}Hau Yeh},
  \bibinfo{person}{Phone Lin}, {and} \bibinfo{person}{Jeu{-}Yih Jeng}.}
  \bibinfo{year}{2017}\natexlab{}.
\newblock \showarticletitle{A Kubernetes-Based Monitoring Platform for Dynamic
  Cloud Resource Provisioning}. In \bibinfo{booktitle}{\emph{2017 {IEEE} Global
  Communications Conference, {GLOBECOM} 2017, Singapore, December 4-8, 2017}}.
  \bibinfo{publisher}{{IEEE}}, \bibinfo{pages}{1--6}.
\newblock
\urldef\tempurl%
\url{https://doi.org/10.1109/GLOCOM.2017.8254046}
\showDOI{\tempurl}


\bibitem[\protect\citeauthoryear{Chen and Shen}{Chen and Shen}{2017}]%
        {DBLP:conf/infocom/ChenS17}
\bibfield{author}{\bibinfo{person}{Liuhua Chen} {and} \bibinfo{person}{Haiying
  Shen}.} \bibinfo{year}{2017}\natexlab{}.
\newblock \showarticletitle{Considering resource demand misalignments to reduce
  resource over-provisioning in cloud datacenters}. In
  \bibinfo{booktitle}{\emph{2017 {IEEE} Conference on Computer Communications,
  {INFOCOM} 2017, Atlanta, GA, USA, May 1-4, 2017}}.
  \bibinfo{publisher}{{IEEE}}, \bibinfo{pages}{1--9}.
\newblock
\urldef\tempurl%
\url{https://doi.org/10.1109/INFOCOM.2017.8057084}
\showDOI{\tempurl}


\bibitem[\protect\citeauthoryear{Chen, Shen, and Platt}{Chen
  et~al\mbox{.}}{2016}]%
        {DBLP:conf/icnp/ChenSP16}
\bibfield{author}{\bibinfo{person}{Liuhua Chen}, \bibinfo{person}{Haiying
  Shen}, {and} \bibinfo{person}{Stephen Platt}.}
  \bibinfo{year}{2016}\natexlab{}.
\newblock \showarticletitle{Cache contention aware Virtual Machine placement
  and migration in cloud datacenters}. In \bibinfo{booktitle}{\emph{24th {IEEE}
  International Conference on Network Protocols, {ICNP} 2016, Singapore,
  November 8-11, 2016}}. \bibinfo{publisher}{{IEEE} Computer Society},
  \bibinfo{pages}{1--10}.
\newblock
\urldef\tempurl%
\url{https://doi.org/10.1109/ICNP.2016.7784447}
\showDOI{\tempurl}


\bibitem[\protect\citeauthoryear{Chen, Delimitrou, and Mart{\'{\i}}nez}{Chen
  et~al\mbox{.}}{2019}]%
        {DBLP:conf/asplos/ChenDM19}
\bibfield{author}{\bibinfo{person}{Shuang Chen}, \bibinfo{person}{Christina
  Delimitrou}, {and} \bibinfo{person}{Jos{\'{e}}~F. Mart{\'{\i}}nez}.}
  \bibinfo{year}{2019}\natexlab{}.
\newblock \showarticletitle{{PARTIES:} QoS-Aware Resource Partitioning for
  Multiple Interactive Services}. In \bibinfo{booktitle}{\emph{Proceedings of
  the Twenty-Fourth International Conference on Architectural Support for
  Programming Languages and Operating Systems, {ASPLOS} 2019, Providence, RI,
  USA, April 13-17, 2019}}, \bibfield{editor}{\bibinfo{person}{Iris Bahar},
  \bibinfo{person}{Maurice Herlihy}, \bibinfo{person}{Emmett Witchel}, {and}
  \bibinfo{person}{Alvin~R. Lebeck}} (Eds.). \bibinfo{publisher}{{ACM}},
  \bibinfo{pages}{107--120}.
\newblock
\urldef\tempurl%
\url{https://doi.org/10.1145/3297858.3304005}
\showDOI{\tempurl}


\bibitem[\protect\citeauthoryear{Chen, Chen, Bajaj, He, He, Wong, and
  Chen}{Chen et~al\mbox{.}}{2020}]%
        {DBLP:conf/cidr/ChenCBHHWC20}
\bibfield{author}{\bibinfo{person}{Xinyu Chen}, \bibinfo{person}{Yao Chen},
  \bibinfo{person}{Ronak Bajaj}, \bibinfo{person}{Jiong He},
  \bibinfo{person}{Bingsheng He}, \bibinfo{person}{Weng{-}Fai Wong}, {and}
  \bibinfo{person}{Deming Chen}.} \bibinfo{year}{2020}\natexlab{}.
\newblock \showarticletitle{Is {FPGA} Useful for Hash Joins?}. In
  \bibinfo{booktitle}{\emph{10th Conference on Innovative Data Systems
  Research, {CIDR} 2020, Amsterdam, The Netherlands, January 12-15, 2020,
  Online Proceedings}}. \bibinfo{publisher}{www.cidrdb.org}.
\newblock
\urldef\tempurl%
\url{http://cidrdb.org/cidr2020/papers/p27-chen-cidr20.pdf}
\showURL{%
\tempurl}


\bibitem[\protect\citeauthoryear{Chen, Tan, Chen, He, Wong, and Chen}{Chen
  et~al\mbox{.}}{2021}]%
        {10.1145/3431920.3439290}
\bibfield{author}{\bibinfo{person}{Xinyu Chen}, \bibinfo{person}{Hongshi Tan},
  \bibinfo{person}{Yao Chen}, \bibinfo{person}{Bingsheng He},
  \bibinfo{person}{Weng-Fai Wong}, {and} \bibinfo{person}{Deming Chen}.}
  \bibinfo{year}{2021}\natexlab{}.
\newblock \showarticletitle{ThunderGP: HLS-Based Graph Processing Framework on
  FPGAs}. In \bibinfo{booktitle}{\emph{The 2021 ACM/SIGDA International
  Symposium on Field-Programmable Gate Arrays}} (Virtual Event, USA)
  \emph{(\bibinfo{series}{FPGA '21})}. \bibinfo{publisher}{Association for
  Computing Machinery}, \bibinfo{address}{New York, NY, USA},
  \bibinfo{pages}{69–80}.
\newblock
\showISBNx{9781450382182}
\urldef\tempurl%
\url{https://doi.org/10.1145/3431920.3439290}
\showDOI{\tempurl}


\bibitem[\protect\citeauthoryear{Cortez, Bonde, and Muzio}{Cortez
  et~al\mbox{.}}{2017}]%
        {DBLP:conf/sosp/CortezBMRFB17}
\bibfield{author}{\bibinfo{person}{Eli Cortez}, \bibinfo{person}{Anand Bonde},
  {and} \bibinfo{person}{Alexandre Muzio}.} \bibinfo{year}{2017}\natexlab{}.
\newblock \showarticletitle{Resource Central: Understanding and Predicting
  Workloads for Improved Resource Management in Large Cloud Platforms}. In
  \bibinfo{booktitle}{\emph{Proceedings of the 26th Symposium on Operating
  Systems Principles, Shanghai, China, October 28-31, 2017}}.
  \bibinfo{publisher}{{ACM}}, \bibinfo{pages}{153--167}.
\newblock
\urldef\tempurl%
\url{https://doi.org/10.1145/3132747.3132772}
\showDOI{\tempurl}


\bibitem[\protect\citeauthoryear{CRIU: A utility to checkpoint/restore Linux
  tasks in userspace}{CRIU: A utility to checkpoint/restore Linux tasks in
  userspace}{2021}]%
        {CRIU}
CRIU: A utility to checkpoint/restore Linux tasks in userspace
  \bibinfo{year}{2021}\natexlab{}.
\newblock
\newblock
\urldef\tempurl%
\url{https://github.com/checkpoint-restore/criu}
\showURL{%
\tempurl}


\bibitem[\protect\citeauthoryear{Daw, Bellur, and Kulkarni}{Daw
  et~al\mbox{.}}{2020}]%
        {DBLP:conf/middleware/DawBK20}
\bibfield{author}{\bibinfo{person}{Nilanjan Daw}, \bibinfo{person}{Umesh
  Bellur}, {and} \bibinfo{person}{Purushottam Kulkarni}.}
  \bibinfo{year}{2020}\natexlab{}.
\newblock \showarticletitle{Xanadu: Mitigating cascading cold starts in
  serverless function chain deployments}. In
  \bibinfo{booktitle}{\emph{Middleware '20: 21st International Middleware
  Conference, Delft, The Netherlands, December 7-11, 2020}},
  \bibfield{editor}{\bibinfo{person}{Dilma~Da Silva} {and}
  \bibinfo{person}{R{\"{u}}diger Kapitza}} (Eds.). \bibinfo{publisher}{{ACM}},
  \bibinfo{pages}{356--370}.
\newblock
\urldef\tempurl%
\url{https://doi.org/10.1145/3423211.3425690}
\showDOI{\tempurl}


\bibitem[\protect\citeauthoryear{Docker}{Docker}{2021}]%
        {Docker}
Docker \bibinfo{year}{2021}\natexlab{}.
\newblock
\newblock
\urldef\tempurl%
\url{https://www.docker.com/}
\showURL{%
\tempurl}


\bibitem[\protect\citeauthoryear{Du, Yu, Xia, Zang, Yan, Qin, Wu, and Chen}{Du
  et~al\mbox{.}}{2020}]%
        {DBLP:conf/asplos/DuYXZYQWC20}
\bibfield{author}{\bibinfo{person}{Dong Du}, \bibinfo{person}{Tianyi Yu},
  \bibinfo{person}{Yubin Xia}, \bibinfo{person}{Binyu Zang},
  \bibinfo{person}{Guanglu Yan}, \bibinfo{person}{Chenggang Qin},
  \bibinfo{person}{Qixuan Wu}, {and} \bibinfo{person}{Haibo Chen}.}
  \bibinfo{year}{2020}\natexlab{}.
\newblock \showarticletitle{Catalyzer: Sub-millisecond Startup for Serverless
  Computing with Initialization-less Booting}. In
  \bibinfo{booktitle}{\emph{{ASPLOS} '20: Architectural Support for Programming
  Languages and Operating Systems, Lausanne, Switzerland, March 16-20, 2020}},
  \bibfield{editor}{\bibinfo{person}{James~R. Larus}, \bibinfo{person}{Luis
  Ceze}, {and} \bibinfo{person}{Karin Strauss}} (Eds.).
  \bibinfo{publisher}{{ACM}}, \bibinfo{pages}{467--481}.
\newblock
\urldef\tempurl%
\url{https://doi.org/10.1145/3373376.3378512}
\showDOI{\tempurl}


\bibitem[\protect\citeauthoryear{Elastic Load Balancing Application Load
  Balancers.}{Elastic Load Balancing Application Load Balancers.}{2021}]%
        {StepFunctionstype}
Elastic Load Balancing Application Load Balancers.
  \bibinfo{year}{2021}\natexlab{}.
\newblock
\newblock
\urldef\tempurl%
\url{https://docs.aws.amazon.com/elasticloadbalancing/latest/application/elb-ag.pdf}
\showURL{%
\tempurl}


\bibitem[\protect\citeauthoryear{Execute mode in Fission}{Execute mode in
  Fission}{2021}]%
        {fissionprewarm}
Execute mode in Fission \bibinfo{year}{2021}\natexlab{}.
\newblock
\newblock
\urldef\tempurl%
\url{https://docs.fission.io/docs/usage/executor/}
\showURL{%
\tempurl}


\bibitem[\protect\citeauthoryear{Eyk, Toader, and Talluri}{Eyk
  et~al\mbox{.}}{2018}]%
        {DBLP:journals/internet/EykTTVUI18}
\bibfield{author}{\bibinfo{person}{Erwin~Van Eyk}, \bibinfo{person}{Lucian
  Toader}, {and} \bibinfo{person}{Sacheendra Talluri}.}
  \bibinfo{year}{2018}\natexlab{}.
\newblock \showarticletitle{Serverless is More: From PaaS to Present Cloud
  Computing}.
\newblock \bibinfo{journal}{\emph{{IEEE} Internet Comput.}}
  \bibinfo{volume}{22}, \bibinfo{number}{5} (\bibinfo{year}{2018}),
  \bibinfo{pages}{8--17}.
\newblock
\urldef\tempurl%
\url{https://doi.org/10.1109/MIC.2018.053681358}
\showDOI{\tempurl}


\bibitem[\protect\citeauthoryear{Fission Workflows: Fast, reliable and
  lightweight function composition for serverless functions}{Fission Workflows:
  Fast, reliable and lightweight function composition for serverless
  functions}{2021}]%
        {Fissionwf}
Fission Workflows: Fast, reliable and lightweight function composition for
  serverless functions \bibinfo{year}{2021}\natexlab{}.
\newblock
\newblock
\urldef\tempurl%
\url{https://github.com/fission/fission-workflows}
\showURL{%
\tempurl}


\bibitem[\protect\citeauthoryear{Fouladi, Wahby, Shacklett, Balasubramaniam,
  Zeng, Bhalerao, Sivaraman, Porter, and Winstein}{Fouladi
  et~al\mbox{.}}{2017}]%
        {DBLP:conf/nsdi/FouladiWSBZBSPW17}
\bibfield{author}{\bibinfo{person}{Sadjad Fouladi}, \bibinfo{person}{Riad~S.
  Wahby}, \bibinfo{person}{Brennan Shacklett}, \bibinfo{person}{Karthikeyan
  Balasubramaniam}, \bibinfo{person}{William Zeng}, \bibinfo{person}{Rahul
  Bhalerao}, \bibinfo{person}{Anirudh Sivaraman}, \bibinfo{person}{George
  Porter}, {and} \bibinfo{person}{Keith Winstein}.}
  \bibinfo{year}{2017}\natexlab{}.
\newblock \showarticletitle{Encoding, Fast and Slow: Low-Latency Video
  Processing Using Thousands of Tiny Threads}. In
  \bibinfo{booktitle}{\emph{14th {USENIX} Symposium on Networked Systems Design
  and Implementation, {NSDI} 2017, Boston, MA, USA, March 27-29, 2017}},
  \bibfield{editor}{\bibinfo{person}{Aditya Akella} {and} \bibinfo{person}{Jon
  Howell}} (Eds.). \bibinfo{publisher}{{USENIX} Association},
  \bibinfo{pages}{363--376}.
\newblock
\urldef\tempurl%
\url{https://www.usenix.org/conference/nsdi17/technical-sessions/presentation/fouladi}
\showURL{%
\tempurl}


\bibitem[\protect\citeauthoryear{Foundations}{Foundations}{2021}]%
        {cncfk8s}
\bibfield{author}{\bibinfo{person}{Cloud Native~Computing Foundations}.}
  \bibinfo{year}{2021}\natexlab{}.
\newblock \bibinfo{title}{"Deployment Strategies on Kubernetes", Software
  engineer at Container Solutions}.
\newblock
\newblock
\urldef\tempurl%
\url{https://www.cncf.io/wp-content/uploads/2020/08/CNCF-Presentation-Template-K8s-Deployment.pdf}
\showURL{%
\tempurl}


\bibitem[\protect\citeauthoryear{Google container runtime sandbox}{Google
  container runtime sandbox}{2021}]%
        {gvisor}
Google container runtime sandbox \bibinfo{year}{2021}\natexlab{}.
\newblock
\newblock
\urldef\tempurl%
\url{https://github.com/google/gvisor}
\showURL{%
\tempurl}


\bibitem[\protect\citeauthoryear{Guan, Wan, Choi, Song, and Zhu}{Guan
  et~al\mbox{.}}{2017}]%
        {DBLP:journals/icl/GuanWCSZ17}
\bibfield{author}{\bibinfo{person}{Xinjie Guan}, \bibinfo{person}{Xili Wan},
  \bibinfo{person}{Baek{-}Young Choi}, \bibinfo{person}{Sejun Song}, {and}
  \bibinfo{person}{Jiafeng Zhu}.} \bibinfo{year}{2017}\natexlab{}.
\newblock \showarticletitle{Application Oriented Dynamic Resource Allocation
  for Data Centers Using Docker Containers}.
\newblock \bibinfo{journal}{\emph{{IEEE} Commun. Lett.}} \bibinfo{volume}{21},
  \bibinfo{number}{3} (\bibinfo{year}{2017}), \bibinfo{pages}{504--507}.
\newblock
\urldef\tempurl%
\url{https://doi.org/10.1109/LCOMM.2016.2644658}
\showDOI{\tempurl}


\bibitem[\protect\citeauthoryear{Harter, Salmon, Liu, Arpaci{-}Dusseau, and
  Arpaci{-}Dusseau}{Harter et~al\mbox{.}}{2016}]%
        {DBLP:conf/fast/HarterSLAA16}
\bibfield{author}{\bibinfo{person}{Tyler Harter}, \bibinfo{person}{Brandon
  Salmon}, \bibinfo{person}{Rose Liu}, \bibinfo{person}{Andrea~C.
  Arpaci{-}Dusseau}, {and} \bibinfo{person}{Remzi~H. Arpaci{-}Dusseau}.}
  \bibinfo{year}{2016}\natexlab{}.
\newblock \showarticletitle{Slacker: Fast Distribution with Lazy Docker
  Containers}. In \bibinfo{booktitle}{\emph{14th {USENIX} Conference on File
  and Storage Technologies, {FAST} 2016, Santa Clara, CA, USA, February 22-25,
  2016}}. \bibinfo{publisher}{{USENIX} Association}, \bibinfo{pages}{181--195}.
\newblock
\urldef\tempurl%
\url{https://www.usenix.org/conference/fast16/technical-sessions/presentation/harter}
\showURL{%
\tempurl}


\bibitem[\protect\citeauthoryear{Hassan, Barakat, and Sarhan}{Hassan
  et~al\mbox{.}}{2021}]%
        {DBLP:journals/jcloudc/HassanBS21}
\bibfield{author}{\bibinfo{person}{Hassan~B. Hassan}, \bibinfo{person}{Saman~A.
  Barakat}, {and} \bibinfo{person}{Qusay~I. Sarhan}.}
  \bibinfo{year}{2021}\natexlab{}.
\newblock \showarticletitle{Survey on serverless computing}.
\newblock \bibinfo{journal}{\emph{J. Cloud Comput.}} \bibinfo{volume}{10},
  \bibinfo{number}{1} (\bibinfo{year}{2021}), \bibinfo{pages}{39}.
\newblock
\urldef\tempurl%
\url{https://doi.org/10.1186/s13677-021-00253-7}
\showDOI{\tempurl}


\bibitem[\protect\citeauthoryear{He, Yang, Fang, Lu, Govindaraju, Luo, and
  Sander}{He et~al\mbox{.}}{2008}]%
        {10.1145/1376616.1376670}
\bibfield{author}{\bibinfo{person}{Bingsheng He}, \bibinfo{person}{Ke Yang},
  \bibinfo{person}{Rui Fang}, \bibinfo{person}{Mian Lu}, \bibinfo{person}{Naga
  Govindaraju}, \bibinfo{person}{Qiong Luo}, {and} \bibinfo{person}{Pedro
  Sander}.} \bibinfo{year}{2008}\natexlab{}.
\newblock \showarticletitle{Relational Joins on Graphics Processors}. In
  \bibinfo{booktitle}{\emph{Proceedings of the 2008 ACM SIGMOD International
  Conference on Management of Data}} (Vancouver, Canada)
  \emph{(\bibinfo{series}{SIGMOD '08})}. \bibinfo{publisher}{Association for
  Computing Machinery}, \bibinfo{address}{New York, NY, USA},
  \bibinfo{pages}{511–524}.
\newblock
\showISBNx{9781605581026}
\urldef\tempurl%
\url{https://doi.org/10.1145/1376616.1376670}
\showDOI{\tempurl}


\bibitem[\protect\citeauthoryear{Hellerstein, Faleiro, and
  Gonzalez}{Hellerstein et~al\mbox{.}}{2019}]%
        {DBLP:conf/cidr/HellersteinFGSS19}
\bibfield{author}{\bibinfo{person}{Joseph~M. Hellerstein},
  \bibinfo{person}{Jose~M. Faleiro}, {and} \bibinfo{person}{Joseph Gonzalez}.}
  \bibinfo{year}{2019}\natexlab{}.
\newblock \showarticletitle{Serverless Computing: One Step Forward, Two Steps
  Back}. In \bibinfo{booktitle}{\emph{{CIDR} 2019, 9th Biennial Conference on
  Innovative Data Systems Research, Asilomar, CA, USA, January 13-16, 2019,
  Online Proceedings}}. \bibinfo{publisher}{www.cidrdb.org}.
\newblock
\urldef\tempurl%
\url{http://cidrdb.org/cidr2019/papers/p119-hellerstein-cidr19.pdf}
\showURL{%
\tempurl}


\bibitem[\protect\citeauthoryear{Hendrickson, Sturdevant, Oakes, Harter,
  Venkataramani, Arpaci{-}Dusseau, and Arpaci{-}Dusseau}{Hendrickson
  et~al\mbox{.}}{2016}]%
        {DBLP:journals/usenix-login/HendricksonSOHV16}
\bibfield{author}{\bibinfo{person}{Scott Hendrickson}, \bibinfo{person}{Stephen
  Sturdevant}, \bibinfo{person}{Edward Oakes}, \bibinfo{person}{Tyler Harter},
  \bibinfo{person}{Venkateshwaran Venkataramani}, \bibinfo{person}{Andrea~C.
  Arpaci{-}Dusseau}, {and} \bibinfo{person}{Remzi~H. Arpaci{-}Dusseau}.}
  \bibinfo{year}{2016}\natexlab{}.
\newblock \showarticletitle{Serverless Computation with OpenLambda}.
\newblock \bibinfo{journal}{\emph{login Usenix Mag.}} \bibinfo{volume}{41},
  \bibinfo{number}{4} (\bibinfo{year}{2016}).
\newblock
\urldef\tempurl%
\url{https://www.usenix.org/publications/login/winter2016/hendrickson}
\showURL{%
\tempurl}


\bibitem[\protect\citeauthoryear{Honeycomb: DevOps tool for code
  inspection}{Honeycomb: DevOps tool for code inspection}{2021}]%
        {honeycomb}
Honeycomb: DevOps tool for code inspection \bibinfo{year}{2021}\natexlab{}.
\newblock
\newblock
\urldef\tempurl%
\url{https://www.honeycomb.io/}
\showURL{%
\tempurl}


\bibitem[\protect\citeauthoryear{HoseinyFarahabady, Zomaya, and
  Tari}{HoseinyFarahabady et~al\mbox{.}}{2018}]%
        {DBLP:journals/tpds/HoseinyFarahabady18}
\bibfield{author}{\bibinfo{person}{M.~Reza HoseinyFarahabady},
  \bibinfo{person}{Albert~Y. Zomaya}, {and} \bibinfo{person}{Zahir Tari}.}
  \bibinfo{year}{2018}\natexlab{}.
\newblock \showarticletitle{A Model Predictive Controller for Managing QoS
  Enforcements and Microarchitecture-Level Interferences in a Lambda Platform}.
\newblock \bibinfo{journal}{\emph{{IEEE} Trans. Parallel Distributed Syst.}}
  \bibinfo{volume}{29}, \bibinfo{number}{7} (\bibinfo{year}{2018}),
  \bibinfo{pages}{1442--1455}.
\newblock
\urldef\tempurl%
\url{https://doi.org/10.1109/TPDS.2017.2779502}
\showDOI{\tempurl}


\bibitem[\protect\citeauthoryear{Hyper-V for Windows containers.}{Hyper-V for
  Windows containers.}{2021}]%
        {hyper}
Hyper-V for Windows containers. \bibinfo{year}{2021}\natexlab{}.
\newblock
\newblock
\urldef\tempurl%
\url{https://docs.microsoft.com/en-us/virtualization/windowscontainers/manage-containers/hyperv-container}
\showURL{%
\tempurl}


\bibitem[\protect\citeauthoryear{Imai, Chestna, and Varela}{Imai
  et~al\mbox{.}}{2013}]%
        {DBLP:conf/ucc/ImaiCV13}
\bibfield{author}{\bibinfo{person}{Shigeru Imai}, \bibinfo{person}{Thomas
  Chestna}, {and} \bibinfo{person}{Carlos~A. Varela}.}
  \bibinfo{year}{2013}\natexlab{}.
\newblock \showarticletitle{Accurate Resource Prediction for Hybrid IaaS Clouds
  Using Workload-Tailored Elastic Compute Units}. In
  \bibinfo{booktitle}{\emph{{IEEE/ACM} 6th International Conference on Utility
  and Cloud Computing, {UCC} 2013, Dresden, Germany, December 9-12, 2013}}.
  \bibinfo{publisher}{{IEEE} Computer Society}, \bibinfo{pages}{171--178}.
\newblock
\urldef\tempurl%
\url{https://doi.org/10.1109/UCC.2013.40}
\showDOI{\tempurl}


\bibitem[\protect\citeauthoryear{Imai, Patterson, and Varela}{Imai
  et~al\mbox{.}}{2018}]%
        {DBLP:conf/ccgrid/ImaiPV18}
\bibfield{author}{\bibinfo{person}{Shigeru Imai}, \bibinfo{person}{Stacy
  Patterson}, {and} \bibinfo{person}{Carlos~A. Varela}.}
  \bibinfo{year}{2018}\natexlab{}.
\newblock \showarticletitle{Uncertainty-Aware Elastic Virtual Machine
  Scheduling for Stream Processing Systems}. In \bibinfo{booktitle}{\emph{18th
  {IEEE/ACM} International Symposium on Cluster, Cloud and Grid Computing,
  {CCGRID} 2018, Washington, DC, USA, May 1-4, 2018}},
  \bibfield{editor}{\bibinfo{person}{Esam El{-}Araby},
  \bibinfo{person}{Dhabaleswar~K. Panda}, \bibinfo{person}{Sandra Gesing},
  \bibinfo{person}{Amy~W. Apon}, \bibinfo{person}{Volodymyr~V. Kindratenko},
  \bibinfo{person}{Massimo Cafaro}, {and} \bibinfo{person}{Alfredo Cuzzocrea}}
  (Eds.). \bibinfo{publisher}{{IEEE} Computer Society},
  \bibinfo{pages}{62--71}.
\newblock
\urldef\tempurl%
\url{https://doi.org/10.1109/CCGRID.2018.00021}
\showDOI{\tempurl}


\bibitem[\protect\citeauthoryear{Ivanov and Smolander}{Ivanov and
  Smolander}{2018}]%
        {DBLP:conf/profes/IvanovS18}
\bibfield{author}{\bibinfo{person}{Vitalii Ivanov} {and} \bibinfo{person}{Kari
  Smolander}.} \bibinfo{year}{2018}\natexlab{}.
\newblock \showarticletitle{Implementation of a DevOps Pipeline for Serverless
  Applications}. In \bibinfo{booktitle}{\emph{Product-Focused Software Process
  Improvement - 19th International Conference, {PROFES} 2018, Wolfsburg,
  Germany, November 28-30, 2018, Proceedings}} \emph{(\bibinfo{series}{Lecture
  Notes in Computer Science}, Vol.~\bibinfo{volume}{11271})},
  \bibfield{editor}{\bibinfo{person}{Marco Kuhrmann}, \bibinfo{person}{Kurt
  Schneider}, \bibinfo{person}{Dietmar Pfahl}, \bibinfo{person}{Sousuke
  Amasaki}, \bibinfo{person}{Marcus Ciolkowski}, \bibinfo{person}{Regina
  Hebig}, \bibinfo{person}{Paolo Tell}, \bibinfo{person}{Jil Kl{\"{u}}nder},
  {and} \bibinfo{person}{Steffen K{\"{u}}pper}} (Eds.).
  \bibinfo{publisher}{Springer}, \bibinfo{pages}{48--64}.
\newblock
\urldef\tempurl%
\url{https://doi.org/10.1007/978-3-030-03673-7\_4}
\showDOI{\tempurl}


\bibitem[\protect\citeauthoryear{Jackson and Clynch}{Jackson and
  Clynch}{2018}]%
        {DBLP:conf/ucc/JacksonC18}
\bibfield{author}{\bibinfo{person}{David Jackson} {and} \bibinfo{person}{Gary
  Clynch}.} \bibinfo{year}{2018}\natexlab{}.
\newblock \showarticletitle{An Investigation of the Impact of Language Runtime
  on the Performance and Cost of Serverless Functions}. In
  \bibinfo{booktitle}{\emph{2018 {IEEE/ACM} International Conference on Utility
  and Cloud Computing Companion, {UCC} Companion 2018, Zurich, Switzerland,
  December 17-20, 2018}}, \bibfield{editor}{\bibinfo{person}{Alan Sill} {and}
  \bibinfo{person}{Josef Spillner}} (Eds.). \bibinfo{publisher}{{IEEE}},
  \bibinfo{pages}{154--160}.
\newblock
\urldef\tempurl%
\url{https://doi.org/10.1109/UCC-Companion.2018.00050}
\showDOI{\tempurl}


\bibitem[\protect\citeauthoryear{Jenkins: DevOps CI tool}{Jenkins: DevOps CI
  tool}{2021}]%
        {jenkins}
Jenkins: DevOps CI tool \bibinfo{year}{2021}\natexlab{}.
\newblock
\newblock
\urldef\tempurl%
\url{https://www.jenkins.io/}
\showURL{%
\tempurl}


\bibitem[\protect\citeauthoryear{Jonas, Pu, Venkataraman, Stoica, and
  Recht}{Jonas et~al\mbox{.}}{2017}]%
        {DBLP:conf/cloud/JonasPVSR17}
\bibfield{author}{\bibinfo{person}{Eric Jonas}, \bibinfo{person}{Qifan Pu},
  \bibinfo{person}{Shivaram Venkataraman}, \bibinfo{person}{Ion Stoica}, {and}
  \bibinfo{person}{Benjamin Recht}.} \bibinfo{year}{2017}\natexlab{}.
\newblock \showarticletitle{Occupy the cloud: distributed computing for the
  99{\%}}. In \bibinfo{booktitle}{\emph{Proceedings of the 2017 Symposium on
  Cloud Computing, SoCC 2017, Santa Clara, CA, USA, September 24-27, 2017}}.
  \bibinfo{publisher}{{ACM}}, \bibinfo{pages}{445--451}.
\newblock
\urldef\tempurl%
\url{https://doi.org/10.1145/3127479.3128601}
\showDOI{\tempurl}


\bibitem[\protect\citeauthoryear{Jonas, Schleier{-}Smith, Sreekanti, Tsai,
  Khandelwal, Pu, Shankar, Carreira, Krauth, Yadwadkar, Gonzalez, Popa, Stoica,
  and Patterson}{Jonas et~al\mbox{.}}{2019}]%
        {DBLP:journals/corr/abs-1902-03383}
\bibfield{author}{\bibinfo{person}{Eric Jonas}, \bibinfo{person}{Johann
  Schleier{-}Smith}, \bibinfo{person}{Vikram Sreekanti},
  \bibinfo{person}{Chia{-}che Tsai}, \bibinfo{person}{Anurag Khandelwal},
  \bibinfo{person}{Qifan Pu}, \bibinfo{person}{Vaishaal Shankar},
  \bibinfo{person}{Joao Carreira}, \bibinfo{person}{Karl Krauth},
  \bibinfo{person}{Neeraja~Jayant Yadwadkar}, \bibinfo{person}{Joseph~E.
  Gonzalez}, \bibinfo{person}{Raluca~Ada Popa}, \bibinfo{person}{Ion Stoica},
  {and} \bibinfo{person}{David~A. Patterson}.} \bibinfo{year}{2019}\natexlab{}.
\newblock \showarticletitle{Cloud Programming Simplified: {A} Berkeley View on
  Serverless Computing}.
\newblock \bibinfo{journal}{\emph{CoRR}}  \bibinfo{volume}{abs/1902.03383}
  (\bibinfo{year}{2019}).
\newblock
\showeprint[arxiv]{1902.03383}
\urldef\tempurl%
\url{http://arxiv.org/abs/1902.03383}
\showURL{%
\tempurl}


\bibitem[\protect\citeauthoryear{Kaffes, Yadwadkar, and Kozyrakis}{Kaffes
  et~al\mbox{.}}{2019}]%
        {DBLP:conf/cloud/KaffesYK19}
\bibfield{author}{\bibinfo{person}{Kostis Kaffes}, \bibinfo{person}{Neeraja~J.
  Yadwadkar}, {and} \bibinfo{person}{Christos Kozyrakis}.}
  \bibinfo{year}{2019}\natexlab{}.
\newblock \showarticletitle{Centralized Core-granular Scheduling for Serverless
  Functions}. In \bibinfo{booktitle}{\emph{Proceedings of the {ACM} Symposium
  on Cloud Computing, SoCC 2019, Santa Cruz, CA, USA, November 20-23, 2019}}.
  \bibinfo{publisher}{{ACM}}, \bibinfo{pages}{158--164}.
\newblock
\urldef\tempurl%
\url{https://doi.org/10.1145/3357223.3362709}
\showDOI{\tempurl}


\bibitem[\protect\citeauthoryear{Kata Containers: an open source container
  runtime, building lightweight virtual machines.}{Kata Containers: an open
  source container runtime, building lightweight virtual machines.}{2021}]%
        {kata}
Kata Containers: an open source container runtime, building lightweight virtual
  machines. \bibinfo{year}{2021}\natexlab{}.
\newblock
\newblock
\urldef\tempurl%
\url{https://katacontainers.io/}
\showURL{%
\tempurl}


\bibitem[\protect\citeauthoryear{Keshavarzian, Sharifian, and
  Seyedin}{Keshavarzian et~al\mbox{.}}{2019}]%
        {DBLP:journals/fgcs/KeshavarzianSS19}
\bibfield{author}{\bibinfo{person}{Alireza Keshavarzian},
  \bibinfo{person}{Saeed Sharifian}, {and} \bibinfo{person}{Sanaz Seyedin}.}
  \bibinfo{year}{2019}\natexlab{}.
\newblock \showarticletitle{Modified deep residual network architecture
  deployed on serverless framework of IoT platform based on human activity
  recognition application}.
\newblock \bibinfo{journal}{\emph{Future Gener. Comput. Syst.}}
  \bibinfo{volume}{101} (\bibinfo{year}{2019}), \bibinfo{pages}{14--28}.
\newblock
\urldef\tempurl%
\url{https://doi.org/10.1016/j.future.2019.06.009}
\showDOI{\tempurl}


\bibitem[\protect\citeauthoryear{Khan}{Khan}{2017}]%
        {khan2017key}
\bibfield{author}{\bibinfo{person}{Asif Khan}.}
  \bibinfo{year}{2017}\natexlab{}.
\newblock \showarticletitle{Key characteristics of a container orchestration
  platform to enable a modern application}.
\newblock \bibinfo{journal}{\emph{IEEE cloud Computing}} \bibinfo{volume}{4},
  \bibinfo{number}{5} (\bibinfo{year}{2017}), \bibinfo{pages}{42--48}.
\newblock
\urldef\tempurl%
\url{https://doi.org/10.1109/MCC.2017.4250933}
\showDOI{\tempurl}


\bibitem[\protect\citeauthoryear{Kim, HoseinyFarahabady, Lee, and Zomaya}{Kim
  et~al\mbox{.}}{2020}]%
        {DBLP:journals/tpds/KimFLZ20}
\bibfield{author}{\bibinfo{person}{Young~Ki Kim}, \bibinfo{person}{M.~Reza
  HoseinyFarahabady}, \bibinfo{person}{Young~Choon Lee}, {and}
  \bibinfo{person}{Albert~Y. Zomaya}.} \bibinfo{year}{2020}\natexlab{}.
\newblock \showarticletitle{Automated Fine-Grained {CPU} Cap Control in
  Serverless Computing Platform}.
\newblock \bibinfo{journal}{\emph{{IEEE} Trans. Parallel Distributed Syst.}}
  \bibinfo{volume}{31}, \bibinfo{number}{10} (\bibinfo{year}{2020}),
  \bibinfo{pages}{2289--2301}.
\newblock
\urldef\tempurl%
\url{https://doi.org/10.1109/TPDS.2020.2989771}
\showDOI{\tempurl}


\bibitem[\protect\citeauthoryear{Klimovic, Wang, Stuedi, and Trivedi}{Klimovic
  et~al\mbox{.}}{2018}]%
        {DBLP:conf/osdi/KlimovicWSTPK18}
\bibfield{author}{\bibinfo{person}{Ana Klimovic}, \bibinfo{person}{Yawen Wang},
  \bibinfo{person}{Patrick Stuedi}, {and} \bibinfo{person}{Animesh Trivedi}.}
  \bibinfo{year}{2018}\natexlab{}.
\newblock \showarticletitle{Pocket: Elastic Ephemeral Storage for Serverless
  Analytics}. In \bibinfo{booktitle}{\emph{13th {USENIX} Symposium on Operating
  Systems Design and Implementation, {OSDI} 2018, Carlsbad, CA, USA, October
  8-10, 2018}}, \bibfield{editor}{\bibinfo{person}{Andrea~C. Arpaci{-}Dusseau}
  {and} \bibinfo{person}{Geoff Voelker}} (Eds.). \bibinfo{publisher}{{USENIX}
  Association}, \bibinfo{pages}{427--444}.
\newblock
\urldef\tempurl%
\url{https://www.usenix.org/conference/osdi18/presentation/klimovic}
\showURL{%
\tempurl}


\bibitem[\protect\citeauthoryear{Kocher, Horn, Fogh, Genkin, Gruss, Haas,
  Hamburg, Lipp, Mangard, Prescher, Schwarz, and Yarom}{Kocher
  et~al\mbox{.}}{2020}]%
        {DBLP:journals/cacm/KocherHFGGHHLMP20}
\bibfield{author}{\bibinfo{person}{Paul Kocher}, \bibinfo{person}{Jann Horn},
  \bibinfo{person}{Anders Fogh}, \bibinfo{person}{Daniel Genkin},
  \bibinfo{person}{Daniel Gruss}, \bibinfo{person}{Werner Haas},
  \bibinfo{person}{Mike Hamburg}, \bibinfo{person}{Moritz Lipp},
  \bibinfo{person}{Stefan Mangard}, \bibinfo{person}{Thomas Prescher},
  \bibinfo{person}{Michael Schwarz}, {and} \bibinfo{person}{Yuval Yarom}.}
  \bibinfo{year}{2020}\natexlab{}.
\newblock \showarticletitle{Spectre attacks: exploiting speculative execution}.
\newblock \bibinfo{journal}{\emph{Commun. {ACM}}} \bibinfo{volume}{63},
  \bibinfo{number}{7} (\bibinfo{year}{2020}), \bibinfo{pages}{93--101}.
\newblock
\urldef\tempurl%
\url{https://doi.org/10.1145/3399742}
\showDOI{\tempurl}


\bibitem[\protect\citeauthoryear{Koller and Dawson}{Koller and Dawson}{2021}]%
        {VulnerabilityAdvisor}
\bibfield{author}{\bibinfo{person}{Ricardo Koller} {and} \bibinfo{person}{Alan
  Dawson}.} \bibinfo{year}{2021}\natexlab{}.
\newblock \bibinfo{title}{Vulnerability Advisor – Secure your Dev + Ops
  across containers}.
\newblock
\newblock
\urldef\tempurl%
\url{https://www.ibm.com/blogs/cloud-archive/2016/11/vulnerability-advisor-secure-your-dev-ops-across-containers/}
\showURL{%
\tempurl}


\bibitem[\protect\citeauthoryear{Kubeless}{Kubeless}{2021}]%
        {kubeless}
Kubeless \bibinfo{year}{2021}\natexlab{}.
\newblock
\newblock
\urldef\tempurl%
\url{https://kubeless.io/}
\showURL{%
\tempurl}


\bibitem[\protect\citeauthoryear{Kubernetes CronJob}{Kubernetes
  CronJob}{2021}]%
        {k8s:cronjob}
Kubernetes CronJob \bibinfo{year}{2021}\natexlab{}.
\newblock
\newblock
\urldef\tempurl%
\url{https://kubernetes.io/docs/concepts/workloads/controllers/cron-jobs/}
\showURL{%
\tempurl}


\bibitem[\protect\citeauthoryear{Kwan, Wong, Jacobsen, and Muthusamy}{Kwan
  et~al\mbox{.}}{2019}]%
        {DBLP:conf/icdcs/KwanWJM19}
\bibfield{author}{\bibinfo{person}{Anthony Kwan}, \bibinfo{person}{Jonathon
  Wong}, \bibinfo{person}{Hans{-}Arno Jacobsen}, {and} \bibinfo{person}{Vinod
  Muthusamy}.} \bibinfo{year}{2019}\natexlab{}.
\newblock \showarticletitle{HyScale: Hybrid and Network Scaling of Dockerized
  Microservices in Cloud Data Centres}. In \bibinfo{booktitle}{\emph{39th
  {IEEE} International Conference on Distributed Computing Systems, {ICDCS}
  2019, Dallas, TX, USA, July 7-10, 2019}}. \bibinfo{publisher}{{IEEE}},
  \bibinfo{pages}{80--90}.
\newblock
\urldef\tempurl%
\url{https://doi.org/10.1109/ICDCS.2019.00017}
\showDOI{\tempurl}


\bibitem[\protect\citeauthoryear{Lee, Satyam, and Fox}{Lee
  et~al\mbox{.}}{2018}]%
        {DBLP:conf/IEEEcloud/LeeSF18}
\bibfield{author}{\bibinfo{person}{Hyungro Lee}, \bibinfo{person}{Kumar
  Satyam}, {and} \bibinfo{person}{Geoffrey~C. Fox}.}
  \bibinfo{year}{2018}\natexlab{}.
\newblock \showarticletitle{Evaluation of Production Serverless Computing
  Environments}. In \bibinfo{booktitle}{\emph{11th {IEEE} International
  Conference on Cloud Computing, {CLOUD} 2018, San Francisco, CA, USA, July
  2-7, 2018}}. \bibinfo{publisher}{{IEEE} Computer Society},
  \bibinfo{pages}{442--450}.
\newblock
\urldef\tempurl%
\url{https://doi.org/10.1109/CLOUD.2018.00062}
\showDOI{\tempurl}


\bibitem[\protect\citeauthoryear{Leitner, Wittern, Spillner, and
  Hummer}{Leitner et~al\mbox{.}}{2019}]%
        {DBLP:journals/jss/LeitnerWSH19}
\bibfield{author}{\bibinfo{person}{Philipp Leitner}, \bibinfo{person}{Erik
  Wittern}, \bibinfo{person}{Josef Spillner}, {and} \bibinfo{person}{Waldemar
  Hummer}.} \bibinfo{year}{2019}\natexlab{}.
\newblock \showarticletitle{A mixed-method empirical study of
  Function-as-a-Service software development in industrial practice}.
\newblock \bibinfo{journal}{\emph{J. Syst. Softw.}}  \bibinfo{volume}{149}
  (\bibinfo{year}{2019}), \bibinfo{pages}{340--359}.
\newblock
\urldef\tempurl%
\url{https://doi.org/10.1016/j.jss.2018.12.013}
\showDOI{\tempurl}


\bibitem[\protect\citeauthoryear{Li, Yuan, Du, Ma, Liu, and Hsu}{Li
  et~al\mbox{.}}{2020}]%
        {DBLP:conf/usenix/LiYDMLH20}
\bibfield{author}{\bibinfo{person}{Huiba Li}, \bibinfo{person}{Yifan Yuan},
  \bibinfo{person}{Rui Du}, \bibinfo{person}{Kai Ma}, \bibinfo{person}{Lanzheng
  Liu}, {and} \bibinfo{person}{Windsor Hsu}.} \bibinfo{year}{2020}\natexlab{}.
\newblock \showarticletitle{{DADI:} Block-Level Image Service for Agile and
  Elastic Application Deployment}. In \bibinfo{booktitle}{\emph{2020 {USENIX}
  Annual Technical Conference, {USENIX} {ATC} 2020, July 15-17, 2020}},
  \bibfield{editor}{\bibinfo{person}{Ada Gavrilovska} {and}
  \bibinfo{person}{Erez Zadok}} (Eds.). \bibinfo{publisher}{{USENIX}
  Association}, \bibinfo{pages}{727--740}.
\newblock
\urldef\tempurl%
\url{https://www.usenix.org/conference/atc20/presentation/li-huiba}
\showURL{%
\tempurl}


\bibitem[\protect\citeauthoryear{Li and Kanso}{Li and Kanso}{2015}]%
        {DBLP:conf/ic2e/LiK15}
\bibfield{author}{\bibinfo{person}{Wubin Li} {and} \bibinfo{person}{Ali
  Kanso}.} \bibinfo{year}{2015}\natexlab{}.
\newblock \showarticletitle{Comparing Containers versus Virtual Machines for
  Achieving High Availability}. In \bibinfo{booktitle}{\emph{2015 {IEEE}
  International Conference on Cloud Engineering, {IC2E} 2015, Tempe, AZ, USA,
  March 9-13, 2015}}. \bibinfo{publisher}{{IEEE} Computer Society},
  \bibinfo{pages}{353--358}.
\newblock
\urldef\tempurl%
\url{https://doi.org/10.1109/IC2E.2015.79}
\showDOI{\tempurl}


\bibitem[\protect\citeauthoryear{Lin and Khazaei}{Lin and Khazaei}{2021}]%
        {DBLP:journals/tpds/LinK21}
\bibfield{author}{\bibinfo{person}{Changyuan Lin} {and} \bibinfo{person}{Hamzeh
  Khazaei}.} \bibinfo{year}{2021}\natexlab{}.
\newblock \showarticletitle{Modeling and Optimization of Performance and Cost
  of Serverless Applications}.
\newblock \bibinfo{journal}{\emph{{IEEE} Trans. Parallel Distributed Syst.}}
  \bibinfo{volume}{32}, \bibinfo{number}{3} (\bibinfo{year}{2021}),
  \bibinfo{pages}{615--632}.
\newblock
\urldef\tempurl%
\url{https://doi.org/10.1109/TPDS.2020.3028841}
\showDOI{\tempurl}


\bibitem[\protect\citeauthoryear{{Ling}, {Ma}, {Tian}, and {Hu}}{{Ling}
  et~al\mbox{.}}{2019}]%
        {9071414}
\bibfield{author}{\bibinfo{person}{W. {Ling}}, \bibinfo{person}{L. {Ma}},
  \bibinfo{person}{C. {Tian}}, {and} \bibinfo{person}{Z. {Hu}}.}
  \bibinfo{year}{2019}\natexlab{}.
\newblock \showarticletitle{Pigeon: A Dynamic and Efficient Serverless and FaaS
  Framework for Private Cloud}. In \bibinfo{booktitle}{\emph{2019 International
  Conference on Computational Science and Computational Intelligence (CSCI)}}.
  \bibinfo{pages}{1416--1421}.
\newblock
\urldef\tempurl%
\url{https://doi.org/10.1109/CSCI49370.2019.00265}
\showDOI{\tempurl}


\bibitem[\protect\citeauthoryear{Lion, Chu, Sun, Zhuang, Grcevski, and
  Yuan}{Lion et~al\mbox{.}}{2017}]%
        {DBLP:journals/usenix-login/LionC0ZGY17}
\bibfield{author}{\bibinfo{person}{David Lion}, \bibinfo{person}{Adrian Chu},
  \bibinfo{person}{Hailong Sun}, \bibinfo{person}{Xin Zhuang},
  \bibinfo{person}{Nikola Grcevski}, {and} \bibinfo{person}{Ding Yuan}.}
  \bibinfo{year}{2017}\natexlab{}.
\newblock \showarticletitle{Don't Get Caught in the Cold, Warm Up Your {JVM}}.
\newblock \bibinfo{journal}{\emph{login Usenix Mag.}} \bibinfo{volume}{42},
  \bibinfo{number}{1} (\bibinfo{year}{2017}).
\newblock
\urldef\tempurl%
\url{https://www.usenix.org/publications/login/spring2017/lion}
\showURL{%
\tempurl}


\bibitem[\protect\citeauthoryear{Lipp, Schwarz, Gruss, Prescher, Haas, Horn,
  Mangard, Kocher, Genkin, Yarom, Hamburg, and Strackx}{Lipp
  et~al\mbox{.}}{2020}]%
        {DBLP:journals/cacm/LippSGPHHMKGYHS20}
\bibfield{author}{\bibinfo{person}{Moritz Lipp}, \bibinfo{person}{Michael
  Schwarz}, \bibinfo{person}{Daniel Gruss}, \bibinfo{person}{Thomas Prescher},
  \bibinfo{person}{Werner Haas}, \bibinfo{person}{Jann Horn},
  \bibinfo{person}{Stefan Mangard}, \bibinfo{person}{Paul Kocher},
  \bibinfo{person}{Daniel Genkin}, \bibinfo{person}{Yuval Yarom},
  \bibinfo{person}{Mike Hamburg}, {and} \bibinfo{person}{Raoul Strackx}.}
  \bibinfo{year}{2020}\natexlab{}.
\newblock \showarticletitle{Meltdown: reading kernel memory from user space}.
\newblock \bibinfo{journal}{\emph{Commun. {ACM}}} \bibinfo{volume}{63},
  \bibinfo{number}{6} (\bibinfo{year}{2020}), \bibinfo{pages}{46--56}.
\newblock
\urldef\tempurl%
\url{https://doi.org/10.1145/3357033}
\showDOI{\tempurl}


\bibitem[\protect\citeauthoryear{L\'{o}pez, Arjona, Samp\'{e}, Slominski, and
  Villard}{L\'{o}pez et~al\mbox{.}}{2020}]%
        {triggerflow}
\bibfield{author}{\bibinfo{person}{Pedro~Garc\'{\i}a L\'{o}pez},
  \bibinfo{person}{Aitor Arjona}, \bibinfo{person}{Josep Samp\'{e}},
  \bibinfo{person}{Aleksander Slominski}, {and} \bibinfo{person}{Lionel
  Villard}.} \bibinfo{year}{2020}\natexlab{}.
\newblock \showarticletitle{Triggerflow: Trigger-Based Orchestration of
  Serverless Workflows}. In \bibinfo{booktitle}{\emph{Proceedings of the 14th
  ACM International Conference on Distributed and Event-Based Systems}}
  (Montreal, Quebec, Canada) \emph{(\bibinfo{series}{DEBS '20})}.
  \bibinfo{publisher}{Association for Computing Machinery},
  \bibinfo{address}{New York, NY, USA}, \bibinfo{pages}{3–14}.
\newblock
\showISBNx{9781450380287}
\urldef\tempurl%
\url{https://doi.org/10.1145/3401025.3401731}
\showDOI{\tempurl}


\bibitem[\protect\citeauthoryear{Madhavapeddy, Mortier, Rotsos, Scott, Singh,
  Gazagnaire, Smith, Hand, and Crowcroft}{Madhavapeddy et~al\mbox{.}}{2013}]%
        {DBLP:conf/asplos/MadhavapeddyMRSSGSHC13}
\bibfield{author}{\bibinfo{person}{Anil Madhavapeddy}, \bibinfo{person}{Richard
  Mortier}, \bibinfo{person}{Charalampos Rotsos}, \bibinfo{person}{David~J.
  Scott}, \bibinfo{person}{Balraj Singh}, \bibinfo{person}{Thomas Gazagnaire},
  \bibinfo{person}{Steven Smith}, \bibinfo{person}{Steven Hand}, {and}
  \bibinfo{person}{Jon Crowcroft}.} \bibinfo{year}{2013}\natexlab{}.
\newblock \showarticletitle{Unikernels: library operating systems for the
  cloud}. In \bibinfo{booktitle}{\emph{Architectural Support for Programming
  Languages and Operating Systems, {ASPLOS} '13, Houston, TX, {USA} - March 16
  - 20, 2013}}, \bibfield{editor}{\bibinfo{person}{Vivek Sarkar} {and}
  \bibinfo{person}{Rastislav Bod{\'{\i}}k}} (Eds.). \bibinfo{publisher}{{ACM}},
  \bibinfo{pages}{461--472}.
\newblock
\urldef\tempurl%
\url{https://doi.org/10.1145/2451116.2451167}
\showDOI{\tempurl}


\bibitem[\protect\citeauthoryear{Mahgoub, Shankar, Mitra, Klimovic, Chaterji,
  and Bagchi}{Mahgoub et~al\mbox{.}}{2021}]%
        {DBLP:conf/usenix/MahgoubSMKCB21}
\bibfield{author}{\bibinfo{person}{Ashraf Mahgoub}, \bibinfo{person}{Karthick
  Shankar}, \bibinfo{person}{Subrata Mitra}, \bibinfo{person}{Ana Klimovic},
  \bibinfo{person}{Somali Chaterji}, {and} \bibinfo{person}{Saurabh Bagchi}.}
  \bibinfo{year}{2021}\natexlab{}.
\newblock \showarticletitle{{SONIC:} Application-aware Data Passing for Chained
  Serverless Applications}. In \bibinfo{booktitle}{\emph{2021 {USENIX} Annual
  Technical Conference, {USENIX} {ATC} 2021, July 14-16, 2021}},
  \bibfield{editor}{\bibinfo{person}{Irina Calciu} {and} \bibinfo{person}{Geoff
  Kuenning}} (Eds.). \bibinfo{publisher}{{USENIX} Association},
  \bibinfo{pages}{285--301}.
\newblock
\urldef\tempurl%
\url{https://www.usenix.org/conference/atc21/presentation/mahgoub}
\showURL{%
\tempurl}


\bibitem[\protect\citeauthoryear{Mahmoudi, Lin, Khazaei, and Litoiu}{Mahmoudi
  et~al\mbox{.}}{2019}]%
        {DBLP:conf/cascon/MahmoudiLKL19}
\bibfield{author}{\bibinfo{person}{Nima Mahmoudi}, \bibinfo{person}{Changyuan
  Lin}, \bibinfo{person}{Hamzeh Khazaei}, {and} \bibinfo{person}{Marin
  Litoiu}.} \bibinfo{year}{2019}\natexlab{}.
\newblock \showarticletitle{Optimizing serverless computing: introducing an
  adaptive function placement algorithm}. In
  \bibinfo{booktitle}{\emph{Proceedings of the 29th Annual International
  Conference on Computer Science and Software Engineering, {CASCON} 2019,
  Markham, Ontario, Canada, November 4-6, 2019}},
  \bibfield{editor}{\bibinfo{person}{Tima Pakfetrat},
  \bibinfo{person}{Guy{-}Vincent Jourdan}, \bibinfo{person}{Kostas
  Kontogiannis}, {and} \bibinfo{person}{Robert~F. Enenkel}} (Eds.).
  \bibinfo{publisher}{{ACM}}, \bibinfo{pages}{203--213}.
\newblock
\urldef\tempurl%
\url{https://dl.acm.org/doi/abs/10.5555/3370272.3370294}
\showURL{%
\tempurl}


\bibitem[\protect\citeauthoryear{Malawski, Gajek, Zima, Balis, and
  Figiela}{Malawski et~al\mbox{.}}{2020}]%
        {DBLP:journals/fgcs/MalawskiGZBF20}
\bibfield{author}{\bibinfo{person}{Maciej Malawski}, \bibinfo{person}{Adam
  Gajek}, \bibinfo{person}{Adam Zima}, \bibinfo{person}{Bartosz Balis}, {and}
  \bibinfo{person}{Kamil Figiela}.} \bibinfo{year}{2020}\natexlab{}.
\newblock \showarticletitle{Serverless execution of scientific workflows:
  Experiments with HyperFlow, {AWS} Lambda and Google Cloud Functions}.
\newblock \bibinfo{journal}{\emph{Future Gener. Comput. Syst.}}
  \bibinfo{volume}{110} (\bibinfo{year}{2020}), \bibinfo{pages}{502--514}.
\newblock
\urldef\tempurl%
\url{https://doi.org/10.1016/j.future.2017.10.029}
\showDOI{\tempurl}


\bibitem[\protect\citeauthoryear{Manco, Lupu, Schmidt, Mendes, Kuenzer, Sati,
  Yasukata, Raiciu, and Huici}{Manco et~al\mbox{.}}{2017}]%
        {DBLP:conf/sosp/MancoLSMKSYRH17}
\bibfield{author}{\bibinfo{person}{Filipe Manco}, \bibinfo{person}{Costin
  Lupu}, \bibinfo{person}{Florian Schmidt}, \bibinfo{person}{Jose Mendes},
  \bibinfo{person}{Simon Kuenzer}, \bibinfo{person}{Sumit Sati},
  \bibinfo{person}{Kenichi Yasukata}, \bibinfo{person}{Costin Raiciu}, {and}
  \bibinfo{person}{Felipe Huici}.} \bibinfo{year}{2017}\natexlab{}.
\newblock \showarticletitle{My {VM} is Lighter (and Safer) than your
  Container}. In \bibinfo{booktitle}{\emph{Proceedings of the 26th Symposium on
  Operating Systems Principles, Shanghai, China, October 28-31, 2017}}.
  \bibinfo{publisher}{{ACM}}, \bibinfo{pages}{218--233}.
\newblock
\urldef\tempurl%
\url{https://doi.org/10.1145/3132747.3132763}
\showDOI{\tempurl}


\bibitem[\protect\citeauthoryear{Masdari, ValiKardan, Shahi, and Azar}{Masdari
  et~al\mbox{.}}{2016}]%
        {DBLP:journals/jnca/MasdariVSA16}
\bibfield{author}{\bibinfo{person}{Mohammad Masdari}, \bibinfo{person}{Sima
  ValiKardan}, \bibinfo{person}{Zahra Shahi}, {and}
  \bibinfo{person}{Sonay~Imani Azar}.} \bibinfo{year}{2016}\natexlab{}.
\newblock \showarticletitle{Towards workflow scheduling in cloud computing: {A}
  comprehensive analysis}.
\newblock \bibinfo{journal}{\emph{J. Netw. Comput. Appl.}}
  \bibinfo{volume}{66} (\bibinfo{year}{2016}), \bibinfo{pages}{64--82}.
\newblock
\urldef\tempurl%
\url{https://doi.org/10.1016/j.jnca.2016.01.018}
\showDOI{\tempurl}


\bibitem[\protect\citeauthoryear{Mattetti, Shulman{-}Peleg, Allouche, Corradi,
  Dolev, and Foschini}{Mattetti et~al\mbox{.}}{2015}]%
        {DBLP:conf/cns/MattettiSACDF15}
\bibfield{author}{\bibinfo{person}{Massimiliano Mattetti},
  \bibinfo{person}{Alexandra Shulman{-}Peleg}, \bibinfo{person}{Yair Allouche},
  \bibinfo{person}{Antonio Corradi}, \bibinfo{person}{Shlomi Dolev}, {and}
  \bibinfo{person}{Luca Foschini}.} \bibinfo{year}{2015}\natexlab{}.
\newblock \showarticletitle{Securing the infrastructure and the workloads of
  linux containers}. In \bibinfo{booktitle}{\emph{2015 {IEEE} Conference on
  Communications and Network Security, {CNS} 2015, Florence, Italy, September
  28-30, 2015}}. \bibinfo{publisher}{{IEEE}}, \bibinfo{pages}{559--567}.
\newblock
\urldef\tempurl%
\url{https://doi.org/10.1109/CNS.2015.7346869}
\showDOI{\tempurl}


\bibitem[\protect\citeauthoryear{McDaniel, Herbein, and Taufer}{McDaniel
  et~al\mbox{.}}{2015}]%
        {DBLP:conf/cluster/McDanielHT15}
\bibfield{author}{\bibinfo{person}{Sean McDaniel}, \bibinfo{person}{Stephen
  Herbein}, {and} \bibinfo{person}{Michela Taufer}.}
  \bibinfo{year}{2015}\natexlab{}.
\newblock \showarticletitle{A Two-Tiered Approach to {I/O} Quality of Service
  in Docker Containers}. In \bibinfo{booktitle}{\emph{2015 {IEEE} International
  Conference on Cluster Computing, {CLUSTER} 2015, Chicago, IL, USA, September
  8-11, 2015}}. \bibinfo{publisher}{{IEEE} Computer Society},
  \bibinfo{pages}{490--491}.
\newblock
\urldef\tempurl%
\url{https://doi.org/10.1109/CLUSTER.2015.77}
\showDOI{\tempurl}


\bibitem[\protect\citeauthoryear{McGrath and Brenner}{McGrath and
  Brenner}{2017}]%
        {DBLP:conf/icdcsw/McGrathB17}
\bibfield{author}{\bibinfo{person}{M.~Garrett McGrath} {and}
  \bibinfo{person}{Paul~R. Brenner}.} \bibinfo{year}{2017}\natexlab{}.
\newblock \showarticletitle{Serverless Computing: Design, Implementation, and
  Performance}. In \bibinfo{booktitle}{\emph{37th {IEEE} International
  Conference on Distributed Computing Systems Workshops, {ICDCS} Workshops
  2017, Atlanta, GA, USA, June 5-8, 2017}},
  \bibfield{editor}{\bibinfo{person}{Aibek Musaev},
  \bibinfo{person}{Jo{\~{a}}o~Eduardo Ferreira}, {and} \bibinfo{person}{Teruo
  Higashino}} (Eds.). \bibinfo{publisher}{{IEEE} Computer Society},
  \bibinfo{pages}{405--410}.
\newblock
\urldef\tempurl%
\url{https://doi.org/10.1109/ICDCSW.2017.36}
\showDOI{\tempurl}


\bibitem[\protect\citeauthoryear{mirage-skeleton with simple MirageOS
  applications}{mirage-skeleton with simple MirageOS applications}{2021}]%
        {mirage-skeleton}
mirage-skeleton with simple MirageOS applications
  \bibinfo{year}{2021}\natexlab{}.
\newblock
\newblock
\urldef\tempurl%
\url{https://github.com/mirage/mirage-skeleton}
\showURL{%
\tempurl}


\bibitem[\protect\citeauthoryear{Mnih, Kavukcuoglu, Silver, Graves, Antonoglou,
  Wierstra, and Riedmiller}{Mnih et~al\mbox{.}}{2013}]%
        {DBLP:journals/corr/MnihKSGAWR13}
\bibfield{author}{\bibinfo{person}{Volodymyr Mnih}, \bibinfo{person}{Koray
  Kavukcuoglu}, \bibinfo{person}{David Silver}, \bibinfo{person}{Alex Graves},
  \bibinfo{person}{Ioannis Antonoglou}, \bibinfo{person}{Daan Wierstra}, {and}
  \bibinfo{person}{Martin~A. Riedmiller}.} \bibinfo{year}{2013}\natexlab{}.
\newblock \showarticletitle{Playing Atari with Deep Reinforcement Learning}.
\newblock \bibinfo{journal}{\emph{CoRR}}  \bibinfo{volume}{abs/1312.5602}
  (\bibinfo{year}{2013}).
\newblock
\showeprint[arxiv]{1312.5602}
\urldef\tempurl%
\url{http://arxiv.org/abs/1312.5602}
\showURL{%
\tempurl}


\bibitem[\protect\citeauthoryear{Mohan, Sane, Doshi, and Edupuganti}{Mohan
  et~al\mbox{.}}{2019}]%
        {DBLP:conf/hotcloud/MohanSDENS19}
\bibfield{author}{\bibinfo{person}{Anup Mohan}, \bibinfo{person}{Harshad Sane},
  \bibinfo{person}{Kshitij Doshi}, {and} \bibinfo{person}{Saikrishna
  Edupuganti}.} \bibinfo{year}{2019}\natexlab{}.
\newblock \showarticletitle{Agile Cold Starts for Scalable Serverless}. In
  \bibinfo{booktitle}{\emph{11th {USENIX} Workshop on Hot Topics in Cloud
  Computing, HotCloud 2019, Renton, WA, USA, July 8, 2019}},
  \bibfield{editor}{\bibinfo{person}{Christina Delimitrou} {and}
  \bibinfo{person}{Dan R.~K. Ports}} (Eds.). \bibinfo{publisher}{{USENIX}
  Association}.
\newblock
\urldef\tempurl%
\url{https://www.usenix.org/conference/hotcloud19/presentation/mohan}
\showURL{%
\tempurl}


\bibitem[\protect\citeauthoryear{Naranjo, Risco, {de Alfonso}, Pérez,
  Blanquer, and Moltó}{Naranjo et~al\mbox{.}}{2020}]%
        {NARANJO202032}
\bibfield{author}{\bibinfo{person}{Diana~M. Naranjo},
  \bibinfo{person}{Sebastián Risco}, \bibinfo{person}{Carlos {de Alfonso}},
  \bibinfo{person}{Alfonso Pérez}, \bibinfo{person}{Ignacio Blanquer}, {and}
  \bibinfo{person}{Germán Moltó}.} \bibinfo{year}{2020}\natexlab{}.
\newblock \showarticletitle{Accelerated serverless computing based on GPU
  virtualization}.
\newblock \bibinfo{journal}{\emph{J. Parallel and Distrib. Comput.}}
  \bibinfo{volume}{139} (\bibinfo{year}{2020}), \bibinfo{pages}{32 -- 42}.
\newblock
\showISSN{0743-7315}
\urldef\tempurl%
\url{https://doi.org/10.1016/j.jpdc.2020.01.004}
\showDOI{\tempurl}


\bibitem[\protect\citeauthoryear{Netto, Lung, Correia, Luiz, and
  de~Souza}{Netto et~al\mbox{.}}{2017}]%
        {DBLP:journals/jsa/NettoLCLS17}
\bibfield{author}{\bibinfo{person}{Hylson~Vescovi Netto},
  \bibinfo{person}{Lau~Cheuk Lung}, \bibinfo{person}{Miguel Correia},
  \bibinfo{person}{Aldelir~Fernando Luiz}, {and} \bibinfo{person}{Luciana
  Moreira~S{\'{a}} de Souza}.} \bibinfo{year}{2017}\natexlab{}.
\newblock \showarticletitle{State machine replication in containers managed by
  Kubernetes}.
\newblock \bibinfo{journal}{\emph{J. Syst. Archit.}}  \bibinfo{volume}{73}
  (\bibinfo{year}{2017}), \bibinfo{pages}{53--59}.
\newblock
\urldef\tempurl%
\url{https://doi.org/10.1016/j.sysarc.2016.12.007}
\showDOI{\tempurl}


\bibitem[\protect\citeauthoryear{Nguyen, Shen, Gu, Subbiah, and Wilkes}{Nguyen
  et~al\mbox{.}}{2013}]%
        {DBLP:conf/icac/NguyenSGSW13}
\bibfield{author}{\bibinfo{person}{Hiep Nguyen}, \bibinfo{person}{Zhiming
  Shen}, \bibinfo{person}{Xiaohui Gu}, \bibinfo{person}{Sethuraman Subbiah},
  {and} \bibinfo{person}{John Wilkes}.} \bibinfo{year}{2013}\natexlab{}.
\newblock \showarticletitle{{AGILE:} Elastic Distributed Resource Scaling for
  Infrastructure-as-a-Service}. In \bibinfo{booktitle}{\emph{10th International
  Conference on Autonomic Computing, ICAC'13, San Jose, CA, USA, June 26-28,
  2013}}, \bibfield{editor}{\bibinfo{person}{Jeffrey~O. Kephart},
  \bibinfo{person}{Calton Pu}, {and} \bibinfo{person}{Xiaoyun Zhu}} (Eds.).
  \bibinfo{publisher}{{USENIX} Association}, \bibinfo{pages}{69--82}.
\newblock
\urldef\tempurl%
\url{https://www.usenix.org/conference/icac13/technical-sessions/presentation/nguyen}
\showURL{%
\tempurl}


\bibitem[\protect\citeauthoryear{Oakes, Yang, Zhou, Houck, Harter,
  Arpaci{-}Dusseau, and Arpaci{-}Dusseau}{Oakes et~al\mbox{.}}{2018}]%
        {DBLP:conf/usenix/OakesYZHHAA18}
\bibfield{author}{\bibinfo{person}{Edward Oakes}, \bibinfo{person}{Leon Yang},
  \bibinfo{person}{Dennis Zhou}, \bibinfo{person}{Kevin Houck},
  \bibinfo{person}{Tyler Harter}, \bibinfo{person}{Andrea~C. Arpaci{-}Dusseau},
  {and} \bibinfo{person}{Remzi~H. Arpaci{-}Dusseau}.}
  \bibinfo{year}{2018}\natexlab{}.
\newblock \showarticletitle{{SOCK:} Rapid Task Provisioning with
  Serverless-Optimized Containers}. In \bibinfo{booktitle}{\emph{2018 {USENIX}
  Annual Technical Conference, {USENIX} {ATC} 2018, Boston, MA, USA, July
  11-13, 2018}}, \bibfield{editor}{\bibinfo{person}{Haryadi~S. Gunawi} {and}
  \bibinfo{person}{Benjamin Reed}} (Eds.). \bibinfo{publisher}{{USENIX}
  Association}, \bibinfo{pages}{57--70}.
\newblock
\urldef\tempurl%
\url{https://www.usenix.org/conference/atc18/presentation/oakes}
\showURL{%
\tempurl}


\bibitem[\protect\citeauthoryear{Olivier, Chiba, Lankes, Min, and
  Ravindran}{Olivier et~al\mbox{.}}{2019}]%
        {DBLP:conf/vee/OlivierCLMR19}
\bibfield{author}{\bibinfo{person}{Pierre Olivier}, \bibinfo{person}{Daniel
  Chiba}, \bibinfo{person}{Stefan Lankes}, \bibinfo{person}{Changwoo Min},
  {and} \bibinfo{person}{Binoy Ravindran}.} \bibinfo{year}{2019}\natexlab{}.
\newblock \showarticletitle{A binary-compatible unikernel}. In
  \bibinfo{booktitle}{\emph{Proceedings of the 15th {ACM} {SIGPLAN/SIGOPS}
  International Conference on Virtual Execution Environments, {VEE} 2019,
  Providence, RI, USA, April 14, 2019}},
  \bibfield{editor}{\bibinfo{person}{Jennifer~B. Sartor},
  \bibinfo{person}{Mayur Naik}, {and} \bibinfo{person}{Chris Rossbach}} (Eds.).
  \bibinfo{publisher}{{ACM}}, \bibinfo{pages}{59--73}.
\newblock
\urldef\tempurl%
\url{https://doi.org/10.1145/3313808.3313817}
\showDOI{\tempurl}


\bibitem[\protect\citeauthoryear{OpenWhisk: Serverless functions platform for
  building cloud applications}{OpenWhisk: Serverless functions platform for
  building cloud applications}{2021}]%
        {Openwhisk}
OpenWhisk: Serverless functions platform for building cloud applications
  \bibinfo{year}{2021}\natexlab{}.
\newblock
\newblock
\urldef\tempurl%
\url{https://github.com/apache/openwhisk}
\showURL{%
\tempurl}


\bibitem[\protect\citeauthoryear{Prewarm in Apache OpenWhisk}{Prewarm in Apache
  OpenWhisk}{2021}]%
        {OpenWhiskprewarm}
Prewarm in Apache OpenWhisk \bibinfo{year}{2021}\natexlab{}.
\newblock
\newblock
\urldef\tempurl%
\url{https://github.com/apache/openwhisk/blob/master/docs/actions-python.md}
\showURL{%
\tempurl}


\bibitem[\protect\citeauthoryear{Prewarm in Azure Functions}{Prewarm in Azure
  Functions}{2021}]%
        {azureprewarm}
Prewarm in Azure Functions \bibinfo{year}{2021}\natexlab{}.
\newblock
\newblock
\urldef\tempurl%
\url{https://docs.microsoft.com/en-us/azure/azure-functions/functions-premium-plan}
\showURL{%
\tempurl}


\bibitem[\protect\citeauthoryear{Pu, Venkataraman, and Stoica}{Pu
  et~al\mbox{.}}{2019}]%
        {DBLP:conf/nsdi/PuVS19}
\bibfield{author}{\bibinfo{person}{Qifan Pu}, \bibinfo{person}{Shivaram
  Venkataraman}, {and} \bibinfo{person}{Ion Stoica}.}
  \bibinfo{year}{2019}\natexlab{}.
\newblock \showarticletitle{Shuffling, Fast and Slow: Scalable Analytics on
  Serverless Infrastructure}. In \bibinfo{booktitle}{\emph{16th {USENIX}
  Symposium on Networked Systems Design and Implementation, {NSDI} 2019,
  Boston, MA, February 26-28, 2019}}, \bibfield{editor}{\bibinfo{person}{Jay~R.
  Lorch} {and} \bibinfo{person}{Minlan Yu}} (Eds.).
  \bibinfo{publisher}{{USENIX} Association}, \bibinfo{pages}{193--206}.
\newblock
\urldef\tempurl%
\url{https://www.usenix.org/conference/nsdi19/presentation/pu}
\showURL{%
\tempurl}


\bibitem[\protect\citeauthoryear{Rashmi, Chowdhury, Kosaian, Stoica, and
  Ramchandran}{Rashmi et~al\mbox{.}}{2016}]%
        {DBLP:conf/osdi/RashmiCKSR16}
\bibfield{author}{\bibinfo{person}{K.~V. Rashmi}, \bibinfo{person}{Mosharaf
  Chowdhury}, \bibinfo{person}{Jack Kosaian}, \bibinfo{person}{Ion Stoica},
  {and} \bibinfo{person}{Kannan Ramchandran}.} \bibinfo{year}{2016}\natexlab{}.
\newblock \showarticletitle{EC-Cache: Load-Balanced, Low-Latency Cluster
  Caching with Online Erasure Coding}. In \bibinfo{booktitle}{\emph{12th
  {USENIX} Symposium on Operating Systems Design and Implementation, {OSDI}
  2016, Savannah, GA, USA, November 2-4, 2016}},
  \bibfield{editor}{\bibinfo{person}{Kimberly Keeton} {and}
  \bibinfo{person}{Timothy Roscoe}} (Eds.). \bibinfo{publisher}{{USENIX}
  Association}, \bibinfo{pages}{401--417}.
\newblock


\bibitem[\protect\citeauthoryear{Samp{\'{e}}, Artigas, L{\'{o}}pez, and
  Par{\'{\i}}s}{Samp{\'{e}} et~al\mbox{.}}{2017}]%
        {DBLP:conf/middleware/SampeALP17}
\bibfield{author}{\bibinfo{person}{Josep Samp{\'{e}}},
  \bibinfo{person}{Marc~S{\'{a}}nchez Artigas},
  \bibinfo{person}{Pedro~Garc{\'{\i}}a L{\'{o}}pez}, {and}
  \bibinfo{person}{Gerard Par{\'{\i}}s}.} \bibinfo{year}{2017}\natexlab{}.
\newblock \showarticletitle{Data-driven serverless functions for object
  storage}. In \bibinfo{booktitle}{\emph{Proceedings of the 18th
  {ACM/IFIP/USENIX} Middleware Conference, Las Vegas, NV, USA, December 11 -
  15, 2017}}, \bibfield{editor}{\bibinfo{person}{K.~R. Jayaram},
  \bibinfo{person}{Anshul Gandhi}, \bibinfo{person}{Bettina Kemme}, {and}
  \bibinfo{person}{Peter~R. Pietzuch}} (Eds.). \bibinfo{publisher}{{ACM}},
  \bibinfo{pages}{121--133}.
\newblock
\urldef\tempurl%
\url{https://doi.org/10.1145/3135974.3135980}
\showDOI{\tempurl}


\bibitem[\protect\citeauthoryear{Samp{\'{e}}, L{\'{o}}pez, and
  Artigas}{Samp{\'{e}} et~al\mbox{.}}{2016}]%
        {DBLP:conf/IEEEcloud/SampeLA16}
\bibfield{author}{\bibinfo{person}{Josep Samp{\'{e}}},
  \bibinfo{person}{Pedro~Garc{\'{\i}}a L{\'{o}}pez}, {and}
  \bibinfo{person}{Marc~S{\'{a}}nchez Artigas}.}
  \bibinfo{year}{2016}\natexlab{}.
\newblock \showarticletitle{Vertigo: Programmable Micro-controllers for
  Software-Defined Object Storage}. In \bibinfo{booktitle}{\emph{9th {IEEE}
  International Conference on Cloud Computing, {CLOUD} 2016, San Francisco, CA,
  USA, June 27 - July 2, 2016}}. \bibinfo{publisher}{{IEEE} Computer Society},
  \bibinfo{pages}{180--187}.
\newblock
\urldef\tempurl%
\url{https://doi.org/10.1109/CLOUD.2016.0033}
\showDOI{\tempurl}


\bibitem[\protect\citeauthoryear{Scheuner and Leitner}{Scheuner and
  Leitner}{2020a}]%
        {DBLP:journals/jss/Scheuner020}
\bibfield{author}{\bibinfo{person}{Joel Scheuner} {and}
  \bibinfo{person}{Philipp Leitner}.} \bibinfo{year}{2020}\natexlab{a}.
\newblock \showarticletitle{Function-as-a-Service performance evaluation: {A}
  multivocal literature review}.
\newblock \bibinfo{journal}{\emph{J. Syst. Softw.}}  \bibinfo{volume}{170}
  (\bibinfo{year}{2020}), \bibinfo{pages}{110708}.
\newblock
\urldef\tempurl%
\url{https://doi.org/10.1016/j.jss.2020.110708}
\showDOI{\tempurl}


\bibitem[\protect\citeauthoryear{Scheuner and Leitner}{Scheuner and
  Leitner}{2020b}]%
        {DBLP:journals/corr/abs-2004-03276}
\bibfield{author}{\bibinfo{person}{Joel Scheuner} {and}
  \bibinfo{person}{Philipp Leitner}.} \bibinfo{year}{2020}\natexlab{b}.
\newblock \showarticletitle{The State of Research on Function-as-a-Service
  Performance Evaluation: {A} Multivocal Literature Review}.
\newblock \bibinfo{journal}{\emph{CoRR}}  \bibinfo{volume}{abs/2004.03276}
  (\bibinfo{year}{2020}).
\newblock
\showeprint[arxiv]{2004.03276}
\urldef\tempurl%
\url{https://arxiv.org/abs/2004.03276}
\showURL{%
\tempurl}


\bibitem[\protect\citeauthoryear{Schleier{-}Smith, Sreekanti, Khandelwal,
  Carreira, Yadwadkar, Popa, Gonzalez, Stoica, and Patterson}{Schleier{-}Smith
  et~al\mbox{.}}{2021}]%
        {DBLP:journals/cacm/Schleier-SmithS21}
\bibfield{author}{\bibinfo{person}{Johann Schleier{-}Smith},
  \bibinfo{person}{Vikram Sreekanti}, \bibinfo{person}{Anurag Khandelwal},
  \bibinfo{person}{Joao Carreira}, \bibinfo{person}{Neeraja~Jayant Yadwadkar},
  \bibinfo{person}{Raluca~Ada Popa}, \bibinfo{person}{Joseph~E. Gonzalez},
  \bibinfo{person}{Ion Stoica}, {and} \bibinfo{person}{David~A. Patterson}.}
  \bibinfo{year}{2021}\natexlab{}.
\newblock \showarticletitle{What serverless computing is and should become: the
  next phase of cloud computing}.
\newblock \bibinfo{journal}{\emph{Commun. {ACM}}} \bibinfo{volume}{64},
  \bibinfo{number}{5} (\bibinfo{year}{2021}), \bibinfo{pages}{76--84}.
\newblock
\urldef\tempurl%
\url{https://doi.org/10.1145/3406011}
\showDOI{\tempurl}


\bibitem[\protect\citeauthoryear{Schmidt}{Schmidt}{2017}]%
        {DBLP:conf/sigcomm/Schmidt17}
\bibfield{author}{\bibinfo{person}{Florian Schmidt}.}
  \bibinfo{year}{2017}\natexlab{}.
\newblock \showarticletitle{uniprof: {A} Unikernel Stack Profiler}. In
  \bibinfo{booktitle}{\emph{Posters and Demos Proceedings of the Conference of
  the {ACM} Special Interest Group on Data Communication, {SIGCOMM} 2017, Los
  Angeles, CA, USA, August 21-25, 2017}}. \bibinfo{publisher}{{ACM}},
  \bibinfo{pages}{31--33}.
\newblock
\urldef\tempurl%
\url{https://doi.org/10.1145/3123878.3131976}
\showDOI{\tempurl}


\bibitem[\protect\citeauthoryear{Schwarz, Lipp, Moghimi, Bulck, Stecklina,
  Prescher, and Gruss}{Schwarz et~al\mbox{.}}{2019}]%
        {DBLP:conf/ccs/0001LMBS0G19}
\bibfield{author}{\bibinfo{person}{Michael Schwarz}, \bibinfo{person}{Moritz
  Lipp}, \bibinfo{person}{Daniel Moghimi}, \bibinfo{person}{Jo~Van Bulck},
  \bibinfo{person}{Julian Stecklina}, \bibinfo{person}{Thomas Prescher}, {and}
  \bibinfo{person}{Daniel Gruss}.} \bibinfo{year}{2019}\natexlab{}.
\newblock \showarticletitle{ZombieLoad: Cross-Privilege-Boundary Data
  Sampling}. In \bibinfo{booktitle}{\emph{Proceedings of the 2019 {ACM}
  {SIGSAC} Conference on Computer and Communications Security, {CCS} 2019,
  London, UK, November 11-15, 2019}},
  \bibfield{editor}{\bibinfo{person}{Lorenzo Cavallaro},
  \bibinfo{person}{Johannes Kinder}, \bibinfo{person}{XiaoFeng Wang}, {and}
  \bibinfo{person}{Jonathan Katz}} (Eds.). \bibinfo{publisher}{{ACM}},
  \bibinfo{pages}{753--768}.
\newblock
\urldef\tempurl%
\url{https://doi.org/10.1145/3319535.3354252}
\showDOI{\tempurl}


\bibitem[\protect\citeauthoryear{Setty, Su, and Lorch}{Setty
  et~al\mbox{.}}{2016}]%
        {DBLP:conf/osdi/SettySLZCPR16}
\bibfield{author}{\bibinfo{person}{Srinath T.~V. Setty},
  \bibinfo{person}{Chunzhi Su}, {and} \bibinfo{person}{Jacob~R. Lorch}.}
  \bibinfo{year}{2016}\natexlab{}.
\newblock \showarticletitle{Realizing the Fault-Tolerance Promise of Cloud
  Storage Using Locks with Intent}. In \bibinfo{booktitle}{\emph{12th {USENIX}
  Symposium on Operating Systems Design and Implementation, {OSDI} 2016,
  Savannah, GA, USA, November 2-4, 2016}},
  \bibfield{editor}{\bibinfo{person}{Kimberly Keeton} {and}
  \bibinfo{person}{Timothy Roscoe}} (Eds.). \bibinfo{publisher}{{USENIX}
  Association}, \bibinfo{pages}{501--516}.
\newblock
\urldef\tempurl%
\url{https://www.usenix.org/conference/osdi16/technical-sessions/presentation/setty}
\showURL{%
\tempurl}


\bibitem[\protect\citeauthoryear{Shafiei, Khonsari, and Mousavi}{Shafiei
  et~al\mbox{.}}{2021}]%
        {shafiei2021serverless}
\bibfield{author}{\bibinfo{person}{Hossein Shafiei}, \bibinfo{person}{Ahmad
  Khonsari}, {and} \bibinfo{person}{Payam Mousavi}.}
  \bibinfo{year}{2021}\natexlab{}.
\newblock \bibinfo{title}{Serverless Computing: A Survey of Opportunities,
  Challenges and Applications}.
\newblock
\newblock
\showeprint[arxiv]{1911.01296}~[cs.NI]


\bibitem[\protect\citeauthoryear{Shahrad, Balkind, and Wentzlaff}{Shahrad
  et~al\mbox{.}}{2019}]%
        {DBLP:conf/micro/ShahradBW19}
\bibfield{author}{\bibinfo{person}{Mohammad Shahrad}, \bibinfo{person}{Jonathan
  Balkind}, {and} \bibinfo{person}{David Wentzlaff}.}
  \bibinfo{year}{2019}\natexlab{}.
\newblock \showarticletitle{Architectural Implications of Function-as-a-Service
  Computing}. In \bibinfo{booktitle}{\emph{Proceedings of the 52nd Annual
  {IEEE/ACM} International Symposium on Microarchitecture, {MICRO} 2019,
  Columbus, OH, USA, October 12-16, 2019}}. \bibinfo{publisher}{{ACM}},
  \bibinfo{pages}{1063--1075}.
\newblock
\urldef\tempurl%
\url{https://doi.org/10.1145/3352460.3358296}
\showDOI{\tempurl}


\bibitem[\protect\citeauthoryear{Shahrad, Fonseca, Goiri, and Chaudhry}{Shahrad
  et~al\mbox{.}}{2020}]%
        {DBLP:conf/usenix/ShahradFGCBCLTR20}
\bibfield{author}{\bibinfo{person}{Mohammad Shahrad}, \bibinfo{person}{Rodrigo
  Fonseca}, \bibinfo{person}{I{\~{n}}igo Goiri}, {and} \bibinfo{person}{Gohar
  Chaudhry}.} \bibinfo{year}{2020}\natexlab{}.
\newblock \showarticletitle{Serverless in the Wild: Characterizing and
  Optimizing the Serverless Workload at a Large Cloud Provider}. In
  \bibinfo{booktitle}{\emph{2020 {USENIX} Annual Technical Conference, {USENIX}
  {ATC} 2020, July 15-17, 2020}}, \bibfield{editor}{\bibinfo{person}{Ada
  Gavrilovska} {and} \bibinfo{person}{Erez Zadok}} (Eds.).
  \bibinfo{publisher}{{USENIX} Association}, \bibinfo{pages}{205--218}.
\newblock
\urldef\tempurl%
\url{https://www.usenix.org/conference/atc20/presentation/shahrad}
\showURL{%
\tempurl}


\bibitem[\protect\citeauthoryear{Shankar, Krauth, and Pu}{Shankar
  et~al\mbox{.}}{2018}]%
        {DBLP:journals/corr/abs-1810-09679}
\bibfield{author}{\bibinfo{person}{Vaishaal Shankar}, \bibinfo{person}{Karl
  Krauth}, {and} \bibinfo{person}{Qifan Pu}.} \bibinfo{year}{2018}\natexlab{}.
\newblock \showarticletitle{numpywren: serverless linear algebra}.
\newblock \bibinfo{journal}{\emph{CoRR}}  \bibinfo{volume}{abs/1810.09679}
  (\bibinfo{year}{2018}).
\newblock
\showeprint[arxiv]{1810.09679}
\urldef\tempurl%
\url{http://arxiv.org/abs/1810.09679}
\showURL{%
\tempurl}


\bibitem[\protect\citeauthoryear{Singhvi, Khalid, Akella, and Banerjee}{Singhvi
  et~al\mbox{.}}{2020}]%
        {DBLP:conf/cloud/SinghviKAB20}
\bibfield{author}{\bibinfo{person}{Arjun Singhvi}, \bibinfo{person}{Junaid
  Khalid}, \bibinfo{person}{Aditya Akella}, {and} \bibinfo{person}{Sujata
  Banerjee}.} \bibinfo{year}{2020}\natexlab{}.
\newblock \showarticletitle{{SNF:} serverless network functions}. In
  \bibinfo{booktitle}{\emph{SoCC '20: {ACM} Symposium on Cloud Computing,
  Virtual Event, USA, October 19-21, 2020}},
  \bibfield{editor}{\bibinfo{person}{Rodrigo Fonseca},
  \bibinfo{person}{Christina Delimitrou}, {and} \bibinfo{person}{Beng~Chin
  Ooi}} (Eds.). \bibinfo{publisher}{{ACM}}, \bibinfo{pages}{296--310}.
\newblock
\urldef\tempurl%
\url{https://doi.org/10.1145/3419111.3421295}
\showDOI{\tempurl}


\bibitem[\protect\citeauthoryear{Skarlatos, Darbaz, Gopireddy, and
  Kim}{Skarlatos et~al\mbox{.}}{2020}]%
        {DBLP:conf/isca/SkarlatosDGKT20}
\bibfield{author}{\bibinfo{person}{Dimitrios Skarlatos}, \bibinfo{person}{Umur
  Darbaz}, \bibinfo{person}{Bhargava Gopireddy}, {and}
  \bibinfo{person}{Nam~Sung Kim}.} \bibinfo{year}{2020}\natexlab{}.
\newblock \showarticletitle{BabelFish: Fusing Address Translations for
  Containers}. In \bibinfo{booktitle}{\emph{47th {ACM/IEEE} Annual
  International Symposium on Computer Architecture, {ISCA} 2020, Valencia,
  Spain, May 30 - June 3, 2020}}. \bibinfo{publisher}{{IEEE}},
  \bibinfo{pages}{501--514}.
\newblock
\urldef\tempurl%
\url{https://doi.org/10.1109/ISCA45697.2020.00049}
\showDOI{\tempurl}


\bibitem[\protect\citeauthoryear{Sonarqube: Code quality and security
  platform}{Sonarqube: Code quality and security platform}{2021}]%
        {sonarqube}
Sonarqube: Code quality and security platform \bibinfo{year}{2021}\natexlab{}.
\newblock
\newblock
\urldef\tempurl%
\url{https://www.sonarqube.org/}
\showURL{%
\tempurl}


\bibitem[\protect\citeauthoryear{Sparta: A Go framework for AWS Lambda
  microservices}{Sparta: A Go framework for AWS Lambda microservices}{2021}]%
        {Sparta}
Sparta: A Go framework for AWS Lambda microservices
  \bibinfo{year}{2021}\natexlab{}.
\newblock
\newblock
\urldef\tempurl%
\url{http://gosparta.io/}
\showURL{%
\tempurl}


\bibitem[\protect\citeauthoryear{Sreekanti, Wu, and Chhatrapati}{Sreekanti
  et~al\mbox{.}}{2020a}]%
        {DBLP:conf/eurosys/SreekantiWCGHF20}
\bibfield{author}{\bibinfo{person}{Vikram Sreekanti},
  \bibinfo{person}{Chenggang Wu}, {and} \bibinfo{person}{Saurav Chhatrapati}.}
  \bibinfo{year}{2020}\natexlab{a}.
\newblock \showarticletitle{A fault-tolerance shim for serverless computing}.
  In \bibinfo{booktitle}{\emph{EuroSys '20: Fifteenth EuroSys Conference 2020,
  Heraklion, Greece, April 27-30, 2020}}. \bibinfo{publisher}{{ACM}},
  \bibinfo{pages}{15:1--15:15}.
\newblock
\urldef\tempurl%
\url{https://doi.org/10.1145/3342195.3387535}
\showDOI{\tempurl}


\bibitem[\protect\citeauthoryear{Sreekanti, Wu, Lin, and
  Schleier{-}Smith}{Sreekanti et~al\mbox{.}}{2020b}]%
        {DBLP:journals/pvldb/SreekantiWLSGHT20}
\bibfield{author}{\bibinfo{person}{Vikram Sreekanti},
  \bibinfo{person}{Chenggang Wu}, \bibinfo{person}{Xiayue~Charles Lin}, {and}
  \bibinfo{person}{Johann Schleier{-}Smith}.} \bibinfo{year}{2020}\natexlab{b}.
\newblock \showarticletitle{Cloudburst: Stateful Functions-as-a-Service}.
\newblock \bibinfo{journal}{\emph{Proc. {VLDB} Endow.}} \bibinfo{volume}{13},
  \bibinfo{number}{11} (\bibinfo{year}{2020}), \bibinfo{pages}{2438--2452}.
\newblock
\urldef\tempurl%
\url{http://www.vldb.org/pvldb/vol13/p2438-sreekanti.pdf}
\showURL{%
\tempurl}


\bibitem[\protect\citeauthoryear{Srirama and Ostovar}{Srirama and
  Ostovar}{2018}]%
        {DBLP:journals/ijcc/SriramaO18}
\bibfield{author}{\bibinfo{person}{Satish~Narayana Srirama} {and}
  \bibinfo{person}{Alireza Ostovar}.} \bibinfo{year}{2018}\natexlab{}.
\newblock \showarticletitle{Optimal cloud resource provisioning for
  auto-scaling enterprise applications}.
\newblock \bibinfo{journal}{\emph{Int. J. Cloud Comput.}} \bibinfo{volume}{7},
  \bibinfo{number}{2} (\bibinfo{year}{2018}), \bibinfo{pages}{129--162}.
\newblock
\urldef\tempurl%
\url{https://doi.org/10.1504/IJCC.2018.10014880}
\showDOI{\tempurl}


\bibitem[\protect\citeauthoryear{Suresh and Gandhi}{Suresh and Gandhi}{2019}]%
        {DBLP:conf/middleware/SureshG19}
\bibfield{author}{\bibinfo{person}{Amoghavarsha Suresh} {and}
  \bibinfo{person}{Anshul Gandhi}.} \bibinfo{year}{2019}\natexlab{}.
\newblock \showarticletitle{FnSched: An Efficient Scheduler for Serverless
  Functions}. In \bibinfo{booktitle}{\emph{Proceedings of the 5th International
  Workshop on Serverless Computing, WOSC@Middleware 2019, Davis, CA, USA,
  December 09-13, 2019}}. \bibinfo{publisher}{{ACM}}, \bibinfo{pages}{19--24}.
\newblock
\urldef\tempurl%
\url{https://doi.org/10.1145/3366623.3368136}
\showDOI{\tempurl}


\bibitem[\protect\citeauthoryear{Tak, Isci, Duri, Bila, Nadgowda, and
  Doran}{Tak et~al\mbox{.}}{2017}]%
        {tak2017understanding}
\bibfield{author}{\bibinfo{person}{Byungchul Tak}, \bibinfo{person}{Canturk
  Isci}, \bibinfo{person}{Sastry Duri}, \bibinfo{person}{Nilton Bila},
  \bibinfo{person}{Shripad Nadgowda}, {and} \bibinfo{person}{James Doran}.}
  \bibinfo{year}{2017}\natexlab{}.
\newblock \showarticletitle{Understanding security implications of using
  containers in the cloud}. In \bibinfo{booktitle}{\emph{2017 $\{$USENIX$\}$
  Annual Technical Conference ($\{$USENIX$\}$$\{$ATC$\}$ 17)}}.
  \bibinfo{pages}{313--319}.
\newblock


\bibitem[\protect\citeauthoryear{Tariq, Pahl, Nimmagadda, Rozner, and
  Lanka}{Tariq et~al\mbox{.}}{2020}]%
        {DBLP:conf/cloud/TariqPNRL20}
\bibfield{author}{\bibinfo{person}{Ali Tariq}, \bibinfo{person}{Austin Pahl},
  \bibinfo{person}{Sharat Nimmagadda}, \bibinfo{person}{Eric Rozner}, {and}
  \bibinfo{person}{Siddharth Lanka}.} \bibinfo{year}{2020}\natexlab{}.
\newblock \showarticletitle{Sequoia: enabling quality-of-service in serverless
  computing}. In \bibinfo{booktitle}{\emph{SoCC '20: {ACM} Symposium on Cloud
  Computing, Virtual Event, USA, October 19-21, 2020}},
  \bibfield{editor}{\bibinfo{person}{Rodrigo Fonseca},
  \bibinfo{person}{Christina Delimitrou}, {and} \bibinfo{person}{Beng~Chin
  Ooi}} (Eds.). \bibinfo{publisher}{{ACM}}, \bibinfo{pages}{311--327}.
\newblock
\urldef\tempurl%
\url{https://doi.org/10.1145/3419111.3421306}
\showDOI{\tempurl}


\bibitem[\protect\citeauthoryear{Thalheim, Bhatotia, Fonseca, and
  Kasikci}{Thalheim et~al\mbox{.}}{2018}]%
        {DBLP:conf/usenix/ThalheimBFK18}
\bibfield{author}{\bibinfo{person}{J{\"{o}}rg Thalheim},
  \bibinfo{person}{Pramod Bhatotia}, \bibinfo{person}{Pedro Fonseca}, {and}
  \bibinfo{person}{Baris Kasikci}.} \bibinfo{year}{2018}\natexlab{}.
\newblock \showarticletitle{Cntr: Lightweight {OS} Containers}. In
  \bibinfo{booktitle}{\emph{2018 {USENIX} Annual Technical Conference, {USENIX}
  {ATC} 2018, Boston, MA, USA, July 11-13, 2018}},
  \bibfield{editor}{\bibinfo{person}{Haryadi~S. Gunawi} {and}
  \bibinfo{person}{Benjamin Reed}} (Eds.). \bibinfo{publisher}{{USENIX}
  Association}, \bibinfo{pages}{199--212}.
\newblock
\urldef\tempurl%
\url{https://www.usenix.org/conference/atc18/presentation/thalheim}
\showURL{%
\tempurl}


\bibitem[\protect\citeauthoryear{Tinedo, L{\'{o}}pez, Artigas, and
  Samp{\'{e}}}{Tinedo et~al\mbox{.}}{2016}]%
        {DBLP:journals/internet/TinedoLASMRNNCO16}
\bibfield{author}{\bibinfo{person}{Ra{\'{u}}l~Gracia Tinedo},
  \bibinfo{person}{Pedro~Garc{\'{\i}}a L{\'{o}}pez},
  \bibinfo{person}{Marc~S{\'{a}}nchez Artigas}, {and} \bibinfo{person}{Josep
  Samp{\'{e}}}.} \bibinfo{year}{2016}\natexlab{}.
\newblock \showarticletitle{IOStack: Software-Defined Object Storage}.
\newblock \bibinfo{journal}{\emph{{IEEE} Internet Comput.}}
  \bibinfo{volume}{20}, \bibinfo{number}{3} (\bibinfo{year}{2016}),
  \bibinfo{pages}{10--18}.
\newblock
\urldef\tempurl%
\url{https://doi.org/10.1109/MIC.2016.46}
\showDOI{\tempurl}


\bibitem[\protect\citeauthoryear{Tinedo, Samp{\'{e}}, and
  Zamora{-}G{\'{o}}mez}{Tinedo et~al\mbox{.}}{2017}]%
        {DBLP:conf/fast/TinedoSZALMR17}
\bibfield{author}{\bibinfo{person}{Ra{\'{u}}l~Gracia Tinedo},
  \bibinfo{person}{Josep Samp{\'{e}}}, {and} \bibinfo{person}{Edgar
  Zamora{-}G{\'{o}}mez}.} \bibinfo{year}{2017}\natexlab{}.
\newblock \showarticletitle{Crystal: Software-Defined Storage for Multi-Tenant
  Object Stores}. In \bibinfo{booktitle}{\emph{15th {USENIX} Conference on File
  and Storage Technologies, {FAST} 2017, Santa Clara, CA, USA, February 27 -
  March 2, 2017}}, \bibfield{editor}{\bibinfo{person}{Geoff Kuenning} {and}
  \bibinfo{person}{Carl~A. Waldspurger}} (Eds.). \bibinfo{publisher}{{USENIX}
  Association}, \bibinfo{pages}{243--256}.
\newblock
\urldef\tempurl%
\url{https://www.usenix.org/conference/fast17/technical-sessions/presentation/gracia-tinedo}
\showURL{%
\tempurl}


\bibitem[\protect\citeauthoryear{Toka, Dobreff, Fodor, and Sonkoly}{Toka
  et~al\mbox{.}}{2020}]%
        {DBLP:conf/ccgrid/TokaDFS20}
\bibfield{author}{\bibinfo{person}{L{\'{a}}szl{\'{o}} Toka},
  \bibinfo{person}{Gergely Dobreff}, \bibinfo{person}{Bal{\'{a}}zs Fodor},
  {and} \bibinfo{person}{Bal{\'{a}}zs Sonkoly}.}
  \bibinfo{year}{2020}\natexlab{}.
\newblock \showarticletitle{Adaptive AI-based auto-scaling for Kubernetes}. In
  \bibinfo{booktitle}{\emph{20th {IEEE/ACM} International Symposium on Cluster,
  Cloud and Internet Computing, {CCGRID} 2020, Melbourne, Australia, May 11-14,
  2020}}. \bibinfo{publisher}{{IEEE}}, \bibinfo{pages}{599--608}.
\newblock
\urldef\tempurl%
\url{https://doi.org/10.1109/CCGrid49817.2020.00-33}
\showDOI{\tempurl}


\bibitem[\protect\citeauthoryear{Using custom Docker images as the action
  runtime in OpenWhisk}{Using custom Docker images as the action runtime in
  OpenWhisk}{2021}]%
        {OpenWhiskruntime}
Using custom Docker images as the action runtime in OpenWhisk
  \bibinfo{year}{2021}\natexlab{}.
\newblock
\newblock
\urldef\tempurl%
\url{https://github.com/apache/openwhisk/blob/master/docs/actions-docker.md}
\showURL{%
\tempurl}


\bibitem[\protect\citeauthoryear{Verbitski, Gupta, and Saha}{Verbitski
  et~al\mbox{.}}{2017}]%
        {DBLP:conf/sigmod/VerbitskiGSBGMK17}
\bibfield{author}{\bibinfo{person}{Alexandre Verbitski},
  \bibinfo{person}{Anurag Gupta}, {and} \bibinfo{person}{Debanjan Saha}.}
  \bibinfo{year}{2017}\natexlab{}.
\newblock \showarticletitle{Amazon Aurora: Design Considerations for High
  Throughput Cloud-Native Relational Databases}. In
  \bibinfo{booktitle}{\emph{Proceedings of the 2017 {ACM} International
  Conference on Management of Data, {SIGMOD} Conference 2017, Chicago, IL, USA,
  May 14-19, 2017}}, \bibfield{editor}{\bibinfo{person}{Semih Salihoglu},
  \bibinfo{person}{Wenchao Zhou}, \bibinfo{person}{Rada Chirkova},
  \bibinfo{person}{Jun Yang}, {and} \bibinfo{person}{Dan Suciu}} (Eds.).
  \bibinfo{publisher}{{ACM}}, \bibinfo{pages}{1041--1052}.
\newblock
\urldef\tempurl%
\url{https://doi.org/10.1145/3035918.3056101}
\showDOI{\tempurl}


\bibitem[\protect\citeauthoryear{Viil and Srirama}{Viil and Srirama}{2018}]%
        {DBLP:journals/tjs/ViilS18}
\bibfield{author}{\bibinfo{person}{Jaagup Viil} {and}
  \bibinfo{person}{Satish~Narayana Srirama}.} \bibinfo{year}{2018}\natexlab{}.
\newblock \showarticletitle{Framework for automated partitioning and execution
  of scientific workflows in the cloud}.
\newblock \bibinfo{journal}{\emph{J. Supercomput.}} \bibinfo{volume}{74},
  \bibinfo{number}{6} (\bibinfo{year}{2018}), \bibinfo{pages}{2656--2683}.
\newblock
\urldef\tempurl%
\url{https://doi.org/10.1007/s11227-018-2296-7}
\showDOI{\tempurl}


\bibitem[\protect\citeauthoryear{Wajahat, Gandhi, Karve, and Kochut}{Wajahat
  et~al\mbox{.}}{2016}]%
        {DBLP:conf/green/WajahatGKK16}
\bibfield{author}{\bibinfo{person}{Muhammad Wajahat}, \bibinfo{person}{Anshul
  Gandhi}, \bibinfo{person}{Alexei~A. Karve}, {and} \bibinfo{person}{Andrzej
  Kochut}.} \bibinfo{year}{2016}\natexlab{}.
\newblock \showarticletitle{Using machine learning for black-box autoscaling}.
  In \bibinfo{booktitle}{\emph{Seventh International Green and Sustainable
  Computing Conference, {IGSC} 2016, Hangzhou, China, November 7-9, 2016}}.
  \bibinfo{publisher}{{IEEE} Computer Society}, \bibinfo{pages}{1--8}.
\newblock
\urldef\tempurl%
\url{https://doi.org/10.1109/IGCC.2016.7892598}
\showDOI{\tempurl}


\bibitem[\protect\citeauthoryear{Wang, Zhang, Ma, Anwar, Rupprecht, Skourtis,
  Tarasov, Yan, and Cheng}{Wang et~al\mbox{.}}{2020}]%
        {DBLP:conf/fast/WangZMARST0C20}
\bibfield{author}{\bibinfo{person}{Ao Wang}, \bibinfo{person}{Jingyuan Zhang},
  \bibinfo{person}{Xiaolong Ma}, \bibinfo{person}{Ali Anwar},
  \bibinfo{person}{Lukas Rupprecht}, \bibinfo{person}{Dimitrios Skourtis},
  \bibinfo{person}{Vasily Tarasov}, \bibinfo{person}{Feng Yan}, {and}
  \bibinfo{person}{Yue Cheng}.} \bibinfo{year}{2020}\natexlab{}.
\newblock \showarticletitle{InfiniCache: Exploiting Ephemeral Serverless
  Functions to Build a Cost-Effective Memory Cache}. In
  \bibinfo{booktitle}{\emph{18th {USENIX} Conference on File and Storage
  Technologies, {FAST} 2020, Santa Clara, CA, USA, February 24-27, 2020}},
  \bibfield{editor}{\bibinfo{person}{Sam~H. Noh} {and} \bibinfo{person}{Brent
  Welch}} (Eds.). \bibinfo{publisher}{{USENIX} Association},
  \bibinfo{pages}{267--281}.
\newblock
\urldef\tempurl%
\url{https://www.usenix.org/conference/fast20/presentation/wang-ao}
\showURL{%
\tempurl}


\bibitem[\protect\citeauthoryear{Wang, Niu, and Li}{Wang
  et~al\mbox{.}}{2019b}]%
        {DBLP:conf/infocom/WangNL19}
\bibfield{author}{\bibinfo{person}{Hao Wang}, \bibinfo{person}{Di Niu}, {and}
  \bibinfo{person}{Baochun Li}.} \bibinfo{year}{2019}\natexlab{b}.
\newblock \showarticletitle{Distributed Machine Learning with a Serverless
  Architecture}. In \bibinfo{booktitle}{\emph{2019 {IEEE} Conference on
  Computer Communications, {INFOCOM} 2019, Paris, France, April 29 - May 2,
  2019}}. \bibinfo{publisher}{{IEEE}}, \bibinfo{pages}{1288--1296}.
\newblock
\urldef\tempurl%
\url{https://doi.org/10.1109/INFOCOM.2019.8737391}
\showDOI{\tempurl}


\bibitem[\protect\citeauthoryear{Wang, Ho, and Wu}{Wang
  et~al\mbox{.}}{[n.d.]}]%
        {wang2019a}
\bibfield{author}{\bibinfo{person}{Kai-Ting~Amy Wang}, \bibinfo{person}{Rayson
  Ho}, {and} \bibinfo{person}{Peng Wu}.} \bibinfo{year}{[n.d.]}\natexlab{}.
\newblock \showarticletitle{Replayable {{Execution Optimized}} for {{Page
  Sharing}} for a {{Managed Runtime Environment}}}. In
  \bibinfo{booktitle}{\emph{Proceedings of the {{Fourteenth EuroSys
  Conference}} 2019}} ({New York, NY, USA}, 2019)
  \emph{(\bibinfo{series}{{{EuroSys}} '19})}. \bibinfo{publisher}{{Association
  for Computing Machinery}}.
\newblock
\showISBNx{978-1-4503-6281-8}
\urldef\tempurl%
\url{https://doi.org/10.1145/3302424.3303978}
\showDOI{\tempurl}


\bibitem[\protect\citeauthoryear{Wang, Li, Zhang, Ristenpart, and Swift}{Wang
  et~al\mbox{.}}{2018}]%
        {DBLP:conf/usenix/WangLZRS18}
\bibfield{author}{\bibinfo{person}{Liang Wang}, \bibinfo{person}{Mengyuan Li},
  \bibinfo{person}{Yinqian Zhang}, \bibinfo{person}{Thomas Ristenpart}, {and}
  \bibinfo{person}{Michael~M. Swift}.} \bibinfo{year}{2018}\natexlab{}.
\newblock \showarticletitle{Peeking Behind the Curtains of Serverless
  Platforms}. In \bibinfo{booktitle}{\emph{2018 {USENIX} Annual Technical
  Conference, {USENIX} {ATC} 2018, Boston, MA, USA, July 11-13, 2018}},
  \bibfield{editor}{\bibinfo{person}{Haryadi~S. Gunawi} {and}
  \bibinfo{person}{Benjamin Reed}} (Eds.). \bibinfo{publisher}{{USENIX}
  Association}, \bibinfo{pages}{133--146}.
\newblock
\urldef\tempurl%
\url{https://www.usenix.org/conference/atc18/presentation/wang-liang}
\showURL{%
\tempurl}


\bibitem[\protect\citeauthoryear{Wang, Liagouris, and Nishihara}{Wang
  et~al\mbox{.}}{2019a}]%
        {DBLP:conf/sosp/WangLNMMTS19}
\bibfield{author}{\bibinfo{person}{Stephanie Wang}, \bibinfo{person}{John
  Liagouris}, {and} \bibinfo{person}{Robert Nishihara}.}
  \bibinfo{year}{2019}\natexlab{a}.
\newblock \showarticletitle{Lineage stash: fault tolerance off the critical
  path}. In \bibinfo{booktitle}{\emph{Proceedings of the 27th {ACM} Symposium
  on Operating Systems Principles, {SOSP} 2019, Huntsville, ON, Canada, October
  27-30, 2019}}, \bibfield{editor}{\bibinfo{person}{Tim Brecht} {and}
  \bibinfo{person}{Carey Williamson}} (Eds.). \bibinfo{publisher}{{ACM}},
  \bibinfo{pages}{338--352}.
\newblock
\urldef\tempurl%
\url{https://doi.org/10.1145/3341301.3359653}
\showDOI{\tempurl}


\bibitem[\protect\citeauthoryear{Wires and Warfield}{Wires and
  Warfield}{2017}]%
        {DBLP:conf/fast/WiresW17}
\bibfield{author}{\bibinfo{person}{Jake Wires} {and} \bibinfo{person}{Andrew
  Warfield}.} \bibinfo{year}{2017}\natexlab{}.
\newblock \showarticletitle{Mirador: An Active Control Plane for Datacenter
  Storage}. In \bibinfo{booktitle}{\emph{15th {USENIX} Conference on File and
  Storage Technologies, {FAST} 2017, Santa Clara, CA, USA, February 27 - March
  2, 2017}}, \bibfield{editor}{\bibinfo{person}{Geoff Kuenning} {and}
  \bibinfo{person}{Carl~A. Waldspurger}} (Eds.). \bibinfo{publisher}{{USENIX}
  Association}, \bibinfo{pages}{213--228}.
\newblock
\urldef\tempurl%
\url{https://www.usenix.org/conference/fast17/technical-sessions/presentation/wires}
\showURL{%
\tempurl}


\bibitem[\protect\citeauthoryear{Wu, Mi, and Xia}{Wu et~al\mbox{.}}{2020}]%
        {9183650}
\bibfield{author}{\bibinfo{person}{Mingyu Wu}, \bibinfo{person}{Zeyu Mi}, {and}
  \bibinfo{person}{Yubin Xia}.} \bibinfo{year}{2020}\natexlab{}.
\newblock \showarticletitle{A Survey on Serverless Computing and Its
  Implications for JointCloud Computing}. In \bibinfo{booktitle}{\emph{2020
  IEEE International Conference on Joint Cloud Computing}}.
  \bibinfo{pages}{94--101}.
\newblock
\urldef\tempurl%
\url{https://doi.org/10.1109/JCC49151.2020.00023}
\showDOI{\tempurl}


\bibitem[\protect\citeauthoryear{Xie, Feng, Li, and Long}{Xie
  et~al\mbox{.}}{2016}]%
        {DBLP:journals/fgcs/XieF0L16}
\bibfield{author}{\bibinfo{person}{Yulai Xie}, \bibinfo{person}{Dan Feng},
  \bibinfo{person}{Yan Li}, {and} \bibinfo{person}{Darrell D.~E. Long}.}
  \bibinfo{year}{2016}\natexlab{}.
\newblock \showarticletitle{Oasis: An active storage framework for object
  storage platform}.
\newblock \bibinfo{journal}{\emph{Future Gener. Comput. Syst.}}
  \bibinfo{volume}{56} (\bibinfo{year}{2016}), \bibinfo{pages}{746--758}.
\newblock
\urldef\tempurl%
\url{https://doi.org/10.1016/j.future.2015.08.011}
\showDOI{\tempurl}


\bibitem[\protect\citeauthoryear{Xu, Zhang, Geng, Wu, and Ma}{Xu
  et~al\mbox{.}}{2019}]%
        {DBLP:conf/icpads/XuZGWM19}
\bibfield{author}{\bibinfo{person}{Zhengjun Xu}, \bibinfo{person}{Haitao
  Zhang}, \bibinfo{person}{Xin Geng}, \bibinfo{person}{Qiong Wu}, {and}
  \bibinfo{person}{Huadong Ma}.} \bibinfo{year}{2019}\natexlab{}.
\newblock \showarticletitle{Adaptive Function Launching Acceleration in
  Serverless Computing Platforms}. In \bibinfo{booktitle}{\emph{25th {IEEE}
  International Conference on Parallel and Distributed Systems, {ICPADS} 2019,
  Tianjin, China, December 4-6, 2019}}. \bibinfo{publisher}{{IEEE}},
  \bibinfo{pages}{9--16}.
\newblock
\urldef\tempurl%
\url{https://doi.org/10.1109/ICPADS47876.2019.00011}
\showDOI{\tempurl}


\bibitem[\protect\citeauthoryear{Ye, Wu, Wang, Zhou, Si, Jiang, and Zomaya}{Ye
  et~al\mbox{.}}{2015}]%
        {DBLP:journals/tpds/YeWWZSJZ15}
\bibfield{author}{\bibinfo{person}{Kejiang Ye}, \bibinfo{person}{Zhaohui Wu},
  \bibinfo{person}{Chen Wang}, \bibinfo{person}{Bing~Bing Zhou},
  \bibinfo{person}{Weisheng Si}, \bibinfo{person}{Xiaohong Jiang}, {and}
  \bibinfo{person}{Albert~Y. Zomaya}.} \bibinfo{year}{2015}\natexlab{}.
\newblock \showarticletitle{Profiling-Based Workload Consolidation and
  Migration in Virtualized Data Centers}.
\newblock \bibinfo{journal}{\emph{{IEEE} Trans. Parallel Distributed Syst.}}
  \bibinfo{volume}{26}, \bibinfo{number}{3} (\bibinfo{year}{2015}),
  \bibinfo{pages}{878--890}.
\newblock
\urldef\tempurl%
\url{https://doi.org/10.1109/TPDS.2014.2313335}
\showDOI{\tempurl}


\bibitem[\protect\citeauthoryear{Yu, Liu, and Du}{Yu et~al\mbox{.}}{2020}]%
        {DBLP:conf/cloud/YuLDXZLYQ020}
\bibfield{author}{\bibinfo{person}{Tianyi Yu}, \bibinfo{person}{Qingyuan Liu},
  {and} \bibinfo{person}{Dong Du}.} \bibinfo{year}{2020}\natexlab{}.
\newblock \showarticletitle{Characterizing serverless platforms with
  serverlessbench}. In \bibinfo{booktitle}{\emph{SoCC '20: {ACM} Symposium on
  Cloud Computing, Virtual Event, USA, October 19-21, 2020}},
  \bibfield{editor}{\bibinfo{person}{Rodrigo Fonseca},
  \bibinfo{person}{Christina Delimitrou}, {and} \bibinfo{person}{Beng~Chin
  Ooi}} (Eds.). \bibinfo{publisher}{{ACM}}, \bibinfo{pages}{30--44}.
\newblock
\urldef\tempurl%
\url{https://doi.org/10.1145/3419111.3421280}
\showDOI{\tempurl}


\bibitem[\protect\citeauthoryear{Yu, Huang, Wang, Zhang, and Letaief}{Yu
  et~al\mbox{.}}{2018}]%
        {DBLP:conf/sc/YuHWZL18}
\bibfield{author}{\bibinfo{person}{Yinghao Yu}, \bibinfo{person}{Renfei Huang},
  \bibinfo{person}{Wei Wang}, \bibinfo{person}{Jun Zhang}, {and}
  \bibinfo{person}{Khaled~Ben Letaief}.} \bibinfo{year}{2018}\natexlab{}.
\newblock \showarticletitle{SP-cache: load-balanced, redundancy-free cluster
  caching with selective partition}. In \bibinfo{booktitle}{\emph{Proceedings
  of the International Conference for High Performance Computing, Networking,
  Storage, and Analysis, {SC} 2018, Dallas, TX, USA, November 11-16, 2018}}.
  \bibinfo{publisher}{{IEEE} / {ACM}}, \bibinfo{pages}{1:1--1:13}.
\newblock
\urldef\tempurl%
\url{http://dl.acm.org/citation.cfm?id=3291658}
\showURL{%
\tempurl}


\bibitem[\protect\citeauthoryear{Zhang, Yu, Wang, and Yan}{Zhang
  et~al\mbox{.}}{2019b}]%
        {mark}
\bibfield{author}{\bibinfo{person}{Chengliang Zhang}, \bibinfo{person}{Minchen
  Yu}, \bibinfo{person}{Wei Wang}, {and} \bibinfo{person}{Feng Yan}.}
  \bibinfo{year}{2019}\natexlab{b}.
\newblock \showarticletitle{MArk: Exploiting Cloud Services for Cost-Effective,
  SLO-Aware Machine Learning Inference Serving}. In
  \bibinfo{booktitle}{\emph{2019 {USENIX} Annual Technical Conference ({USENIX}
  {ATC} 19)}}. \bibinfo{publisher}{{USENIX} Association},
  \bibinfo{address}{Renton, WA}, \bibinfo{pages}{1049--1062}.
\newblock
\showISBNx{978-1-939133-03-8}
\urldef\tempurl%
\url{https://www.usenix.org/conference/atc19/presentation/zhang-chengliang}
\showURL{%
\tempurl}


\bibitem[\protect\citeauthoryear{Zhang, Cardoza, and Chen}{Zhang
  et~al\mbox{.}}{2020a}]%
        {DBLP:conf/osdi/ZhangCCAL20}
\bibfield{author}{\bibinfo{person}{Haoran Zhang}, \bibinfo{person}{Adney
  Cardoza}, {and} \bibinfo{person}{Peter~Baile Chen}.}
  \bibinfo{year}{2020}\natexlab{a}.
\newblock \showarticletitle{Fault-tolerant and transactional stateful
  serverless workflows}. In \bibinfo{booktitle}{\emph{14th {USENIX} Symposium
  on Operating Systems Design and Implementation, {OSDI} 2020, Virtual Event,
  November 4-6, 2020}}. \bibinfo{publisher}{{USENIX} Association},
  \bibinfo{pages}{1187--1204}.
\newblock


\bibitem[\protect\citeauthoryear{Zhang, Xie, Li, and Stutsman}{Zhang
  et~al\mbox{.}}{2019a}]%
        {DBLP:conf/cloud/Zhang00S19}
\bibfield{author}{\bibinfo{person}{Tian Zhang}, \bibinfo{person}{Dong Xie},
  \bibinfo{person}{Feifei Li}, {and} \bibinfo{person}{Ryan Stutsman}.}
  \bibinfo{year}{2019}\natexlab{a}.
\newblock \showarticletitle{Narrowing the Gap Between Serverless and its State
  with Storage Functions}. In \bibinfo{booktitle}{\emph{Proceedings of the
  {ACM} Symposium on Cloud Computing, SoCC 2019, Santa Cruz, CA, USA, November
  20-23, 2019}}. \bibinfo{publisher}{{ACM}}, \bibinfo{pages}{1--12}.
\newblock
\urldef\tempurl%
\url{https://doi.org/10.1145/3357223.3362723}
\showDOI{\tempurl}


\bibitem[\protect\citeauthoryear{Zhang, Fang, Panda, and Shenker}{Zhang
  et~al\mbox{.}}{2020b}]%
        {DBLP:conf/cloud/ZhangFPS20}
\bibfield{author}{\bibinfo{person}{Wen Zhang}, \bibinfo{person}{Vivian Fang},
  \bibinfo{person}{Aurojit Panda}, {and} \bibinfo{person}{Scott Shenker}.}
  \bibinfo{year}{2020}\natexlab{b}.
\newblock \showarticletitle{Kappa: a programming framework for serverless
  computing}. In \bibinfo{booktitle}{\emph{SoCC '20: {ACM} Symposium on Cloud
  Computing, Virtual Event, USA, October 19-21, 2020}},
  \bibfield{editor}{\bibinfo{person}{Rodrigo Fonseca},
  \bibinfo{person}{Christina Delimitrou}, {and} \bibinfo{person}{Beng~Chin
  Ooi}} (Eds.). \bibinfo{publisher}{{ACM}}, \bibinfo{pages}{328--343}.
\newblock
\urldef\tempurl%
\url{https://doi.org/10.1145/3419111.3421277}
\showDOI{\tempurl}


\bibitem[\protect\citeauthoryear{Zheng and Peng}{Zheng and Peng}{2019}]%
        {DBLP:conf/IEEEcloud/ZhengP19}
\bibfield{author}{\bibinfo{person}{Ge Zheng} {and} \bibinfo{person}{Yang
  Peng}.} \bibinfo{year}{2019}\natexlab{}.
\newblock \showarticletitle{GlobalFlow: {A} Cross-Region Orchestration Service
  for Serverless Computing Services}. In \bibinfo{booktitle}{\emph{12th {IEEE}
  International Conference on Cloud Computing, {CLOUD} 2019, Milan, Italy, July
  8-13, 2019}}, \bibfield{editor}{\bibinfo{person}{Elisa Bertino},
  \bibinfo{person}{Carl~K. Chang}, {and} \bibinfo{person}{Peter Chen}} (Eds.).
  \bibinfo{publisher}{{IEEE}}, \bibinfo{pages}{508--510}.
\newblock
\urldef\tempurl%
\url{https://doi.org/10.1109/CLOUD.2019.00093}
\showDOI{\tempurl}


\bibitem[\protect\citeauthoryear{Zheng, Tynes, Gorelick, Mao, Cheng, and
  Hou}{Zheng et~al\mbox{.}}{2019}]%
        {DBLP:conf/icpp/ZhengTGMCH19}
\bibfield{author}{\bibinfo{person}{Wenjia Zheng}, \bibinfo{person}{Michael
  Tynes}, \bibinfo{person}{Henry Gorelick}, \bibinfo{person}{Ying Mao},
  \bibinfo{person}{Long Cheng}, {and} \bibinfo{person}{Yantian Hou}.}
  \bibinfo{year}{2019}\natexlab{}.
\newblock \showarticletitle{FlowCon: Elastic Flow Configuration for
  Containerized Deep Learning Applications}. In
  \bibinfo{booktitle}{\emph{Proceedings of the 48th International Conference on
  Parallel Processing, {ICPP} 2019, Kyoto, Japan, August 05-08, 2019}}.
  \bibinfo{publisher}{{ACM}}, \bibinfo{pages}{87:1--87:10}.
\newblock
\urldef\tempurl%
\url{https://doi.org/10.1145/3337821.3337868}
\showDOI{\tempurl}


\bibitem[\protect\citeauthoryear{Zhong and He}{Zhong and He}{2014}]%
        {10.1109/TPDS.2013.111}
\bibfield{author}{\bibinfo{person}{Jianlong Zhong} {and}
  \bibinfo{person}{Bingsheng He}.} \bibinfo{year}{2014}\natexlab{}.
\newblock \showarticletitle{Medusa: Simplified Graph Processing on GPUs}.
\newblock \bibinfo{journal}{\emph{IEEE Trans. Parallel Distrib. Syst.}}
  \bibinfo{volume}{25}, \bibinfo{number}{6} (\bibinfo{date}{June}
  \bibinfo{year}{2014}), \bibinfo{pages}{1543–1552}.
\newblock
\showISSN{1045-9219}
\urldef\tempurl%
\url{https://doi.org/10.1109/TPDS.2013.111}
\showDOI{\tempurl}


\end{thebibliography}

\appendix

\end{document}